%% file: main.tex
\documentclass[prb,twocolumn,notitlepage,superscriptaddress,nofootinbib]{revtex4-2}
\usepackage{times}
\usepackage{graphicx}
\usepackage{float}
\usepackage{latexsym,amsmath,amssymb,bm,euscript}
\usepackage{color}
\usepackage{subfigure}
\usepackage{epstopdf}
\usepackage[colorlinks=true,linkcolor=blue,citecolor=blue,urlcolor=blue]{hyperref}
\usepackage{hyperref}
\usepackage{ulem}
\usepackage{cleveref}

\usepackage{tikz}
\usetikzlibrary{decorations.pathreplacing}

\usepackage{ytableau}

\newcommand{\be}{\begin{equation}}
\newcommand{\ee}{\end{equation}}
\newcommand{\bea}{\begin{eqnarray}}
\newcommand{\eea}{\end{eqnarray}}

\newcommand{\eq}[1]{\begin{equation} #1 \end{equation}}

\newcommand{\eqa}[1]{\begin{align}\begin{split} #1 \end{split}\end{align}}

\newcommand{\bi}{\begin{itemize}}
\newcommand{\ei}{\end{itemize}}

\newcommand{\bpm}{\begin{pmatrix}}
\newcommand{\epm}{\end{pmatrix}}
\newcommand{\eps}{\epsilon}

\renewcommand{\th}{\theta}

\newcommand{\lp}{\left(}
\newcommand{\rp}{\right)}

\usepackage{amsmath,amsfonts,amssymb,amsthm,epsfig,array}
\usepackage{dsfont}
\usepackage{slashed}
\usepackage{graphics}
\usepackage{float}
\usepackage{verbatim}
\usepackage{color}
\usepackage{tabularx}

\usepackage{braket}

\newcommand{\bsl}[1]{\boldsymbol{#1}}
\newcommand{\mbf}[1]{\boldsymbol{#1}}

\newcommand{\ii}{\mathrm{i}}

\usepackage{environ}
\NewEnviron{eqs}{%
\begin{equation}\begin{split}
\BODY
\end{split}\end{equation}
}

\let\oldAA\AA
\renewcommand{\AA}{\text{\normalfont\oldAA}}

\newcommand{\V}{\mathcal{V}}

\newcommand{\K}{\text{K}}

\usepackage{cleveref}
\crefname{appendix}{App.}{Apps.}
\crefname{equation}{Eq.}{Eqs.}
\crefname{figure}{Fig.}{Figs.}
\crefname{table}{Tab.}{Tabs.}
\crefname{section}{Sec.}{Secs.}
\creflabelformat{appendix}{#2#1#3}

\usepackage{rotating}

\begin{document}

\title{Multi-Band Exact Diagonalization and an Iteration Approach to Hunt For Fractional Chern Insulators in Rhombohedral Multilayer Graphene}

\author{Heqiu Li}
\affiliation{Donostia International Physics Center, P. Manuel de Lardizabal 4, 20018 Donostia-San Sebastian, Spain}

\author{B. Andrei Bernevig}
\affiliation{Department of Physics, Princeton University, Princeton, New Jersey 08544, USA}
\affiliation{Donostia International Physics Center, P. Manuel de Lardizabal 4, 20018 Donostia-San Sebastian, Spain}
\affiliation{IKERBASQUE, Basque Foundation for Science, Bilbao, Spain}

\author{Nicolas Regnault}
\affiliation{Center for Computational Quantum Physics, Flatiron Institute, 162 5th Avenue, New York, NY 10010, USA}
\affiliation{Department of Physics, Princeton University, Princeton, New Jersey 08544, USA}
\affiliation{Laboratoire de Physique de l'Ecole normale sup\'{e}rieure, ENS, Universit\'{e} PSL, CNRS, Sorbonne Universit\'{e}, Universit\'{e} Paris-Diderot, Sorbonne Paris Cit\'{e}, 75005 Paris, France}

\date{\today}

\begin{abstract}

We perform a multi-band exact diagonalization (ED) study of rhombohedral pentalayer graphene twisted on hexagonal boron nitride with a focus on fractional Chern insulators (FCI) in systems with weak \text{moir\'e} gaps, complementing the results of [Yu et al., arXiv:2407.13770]. We consider both the charge-neutrality (CN) and average (AVE) interaction schemes. Saliently and surprisingly, we now find using the particle entanglement spectrum that the FCI at filling factor 1/3 in the CN scheme predicted by single-(Hartree-Fock) band ED is unstable towards a transition to charge density wave once a small fraction of electrons is allowed to occupy the higher bands. Meanwhile, the FCI at filling 2/3 in the AVE scheme remains more robust under similar band mixing until being suppressed when increasing band mixing. To tackle truncation errors that arise from including multiple bands in larger system sizes, we propose an ED iteration method that iteratively optimizes the single-particle basis so that the particles in the ground state should reside mainly in the lowest band. Nevertheless, we find that the FCI gap remains absent after convergence when the mixing with higher bands is considered. These findings highlight the delicate sensitivity of FCIs to multi-band effects and the shortcoming of all of the current models to explain the experimental emergence of such phases.

\end{abstract}


\maketitle

\section{Introduction}

Fractional Chern insulators (FCIs)~\cite{PhysRevLett.106.236804,Sheng2011,FCI-PhysRevX.1.021014,PhysRevLett.106.236802,checkerboard,parameswaran_fractional_2013,bergholtz_topological_2013} are lattice analogs of fractional quantum Hall states that are realized without an external magnetic field. In some \text{moir\'e} superlattices, the small bandwidth and strong correlations make them a particularly promising platform to realize FCIs. Over the past few years, extensive theoretical~\cite{Macdonaldtwist2011,PhysRevB.89.205414-2014,AndreiMacdonald2020,PhysRevLett.127.147203-2021,PhysRevResearch.2.023237,PhysRevB.99.075127,PhysRevLett.122.086402,PhysRevResearch.2.023238,PhysRevLett.124.106803,liu_gate-tunable_2020,Wilhelm2021FCITBG,Li2021fci,PhysRevB.107.L201109,Morales-Duran_Wei_Shi_MacDonald_2023,PhysRevB.110.035130-2024,Morales-Duran2024fe,PhysRevLett.134.116501} and experimental~\cite{Young2018FCIBLGMoire,Xie_Pierce2021,Cai2023f,Zeng2023e,Park2023f,Xu2023i,Lu2024PGexp,LuLong2025,PhysRevX.15.011045} efforts have focused on \text{moir\'e} materials. A significant breakthrough was achieved recently with the observation of zero-field FCIs in twisted MoTe$_2$ bilayers~\cite{Cai2023f,Zeng2023e,Park2023f,Xu2023i} and rhombohedral n-layer graphene on aligned hexagonal boron nitride (RnG/hBN), including pentalayer~\cite{Lu2024PGexp,LuLong2025}, tetralayer~\cite{Choi_tetralayer2024} and hexalayer~\cite{Xiaobohexa2025} systems. These findings demonstrate that fractional quantum Hall-like physics can arise purely from the interplay of strong correlations and band topology. The experimental observation of FCIs in \text{moir\'e} systems has generated significant theoretical interest both in rhombohedral multilayer graphene~\cite{PhysRevLett.133.206502_2024,Zhou_Yang_Zhang_2023,MFCI2,MFCI3,MFCIIV,PhysRevLett.133.206503_2024,Guo_Lu_Xie_Liu_2023,PhysRevB.110.115146-2024,PhysRevB.109.L241115-2024} and twisted MoTe$_2$~\cite{PhysRevLett.131.136502,PhysRevLett.131.136501,Hu_Xiao_Ran_2023,Zhang2023MoTe2,YuMFCI02024,Xiao2023tMoTe2abinitio,Morales-Duran_Wei_Shi_MacDonald_2023,MFCI1,reddy2023fractional,Reddy2023GlobalPDFCI,XuLiZhang2024,Zaletel2023tMoTe2FCI,Fu2023BandMixingFCItMoTe2,Fengcheng2023tMoTe2HFnum1,Li2024tmdtbg,XuLiZhang2024} to faithfully capture the experimental systems.

Many approaches for FCIs consider a single topological flat band well-isolated from all other bands. But these new \text{moir\'e} systems make this assumption break apart. For twisted MoTe$_2$, several theoretical~\cite{YuMFCI02024,XuLiZhang2024,Fu2023BandMixingFCItMoTe2} studies pointed out the role of band mixing  to correctly understand the different aspects of phase diagram. This issue becomes even more critical in cases such as R5G/hBN. This latest example features additional nearly degenerate conduction bands and extremely small single-particle gaps~\cite{Zhou_Yang_Zhang_2023,MFCI2,MFCI3,MFCIIV,PhysRevLett.133.206503_2024}. In current models,  due to layer polarization of the wavefunctions of this system, the displacement field  pushes conduction electrons away from the \text{moir\'e} pattern, which significantly suppresses the \text{moir\'e} potential on low-lying conduction bands, leaving only a tiny gap (typically around 0.1 meV) at the Brillouin zone edge. Such a small energy separation in the current single-particle models - which for all practical purposes can be considered zero - leads to strong band-mixing effects once interactions are considered, which can drastically alter the stability of both integer and fractional Chern phases~\cite{MFCI3,MFCIIV}. Despite this complication, early analyses~\cite{Guo_Lu_Xie_Liu_2023,Zhou_Yang_Zhang_2023,PhysRevLett.133.206502_2024,PhysRevLett.133.206503_2024} followed a single-band approach: one first applies Hartree-Fock (HF) at integer filling to produce an isolated Chern band, then performs exact diagonalization (ED) by projecting the interaction onto that single HF band. This procedure yields FCIs in theory, but has several shortcomings. By construction it neglects mixing with other low-energy bands - although this mixing is \emph{always} stronger than the Hartree-Fock gap - and it poses non-trivial challenges related to the double counting of the interaction accounted for both in the HF and the ED.

 In Ref.~\cite{MFCIIV}, the authors went beyond pure single-band projection by incorporating additional conduction bands into ED, thereby allowing electrons to move between bands and testing the stability of FCIs under more realistic conditions. Due to the computational complexity from including additional bands, they proposed to cap the number of particles in the remote bands. Using the one-body basis derived from the HF at integer filling, this method provides a perturbative approach to band mixing on top of the single HF band. For both filling factors $1/3$ and $2/3$, and irrespective of how interaction with filled valence bands was taken into account, they showed that all FCI phases were unstable under band mixing. 

In this article, we complement this previous study using the same setup, but with new methods. First, we provide additional results by considering the particle entanglement spectrum (PES)~\cite{PhysRevLett.101.010504-2008,PhysRevLett.106.100405,FCI-PhysRevX.1.021014}. Surprisingly, we find that the nature of the phase at filling $\nu=1/3$ in the charge-neutrality (CN) scheme~\cite{MFCI3,MFCIIV,Zhou_Yang_Zhang_2023,PhysRevLett.133.206503_2024,PhysRevLett.133.206502_2024} used to include interactions, is more subtle than previously thought. In the limit of a single HF Chern band, a clear FCI phase is observed in ED. The energy gap on top of the ground state collapses under strong enough band mixing, whereas a finite energy gap exists when band mixing is weak, leading to the conclusion that the FCI phase is stable in a given range of band mixing. However, PES gives a strikingly contrasted picture where, in reality, a small amount of band mixing actually drives the system into a charge density wave (CDW).

Despite relying on a truncated many-body Hilbert space by limiting the occupation of remote bands, multi-band ED is still severely limited and thus can be plagued by finite size effects. To overcome this limitation, we propose an iterative procedure called the ED iteration method, intended to refine the single-particle basis to maximize the weight of the ground state manifold on the lowest band. This method, inspired by the case of product states in momentum space, is based on the single-body basis transformation obtained from diagonalizing the momentum resolved density matrix. Despite this further optimization, we find that the FCI remains absent when the particles are allowed to occupy the higher bands, indicating its intrinsic sensitivity to multi-band effects and hinting at the failure of current models and/or sets of parameters to explain the experimental observations.

The remainder of this paper is organized as follows. In Sec.~\ref{Sec_modelm}, we introduce the single-particle model for twisted rhombohedral pentalayer graphene on aligned hBN substrate. We also review the interaction schemes and introduce the HF basis in which the multi-band ED is performed. In Sec.~\ref{Sec_beforeiter}, we present our numerical results at fractional fillings in complement to Ref.~\cite{MFCIIV}, including the immediate collapse of PES gap at FCI counting and the transition to CDW under minimal band mixing in the CN scheme at 1/3 filling. In Sec.~\ref{Sec_itermethod}, we introduce the ED iteration method, discussing cases where it is an exact method and how we define its convergence. Finally in Sec.~\ref{Sec_afteriter}, we apply the ED iteration method to two representative cases and explore its impact on the low energy physics, leaving the systematic study for all system sizes, filling factors and interaction scheme to App.~\ref{app_allfig}.

\section{Overview of the model}
\label{Sec_modelm}

In this section, we provide an overview of both the single-particle model of R5G/hBN~\cite{MFCI2,MFCI3,MFCIIV,PhysRevB.110.115146-2024,Zhou_Yang_Zhang_2023,Guo_Lu_Xie_Liu_2023,PhysRevLett.133.206503_2024,PhysRevB.110.115146-2024,PhysRevLett.133.206502_2024} and the two different interaction schemes that have been considered in the literature. More details can be found in Appendices~\ref{Sec_h0} and~\ref{Sec_appscheme}. We follow the notations and presentations of Refs.~[\onlinecite{MFCI2,MFCI3,MFCIIV}].

\subsection{Single-particle model}

\begin{figure}
\centering
\includegraphics[width=3.0 in]{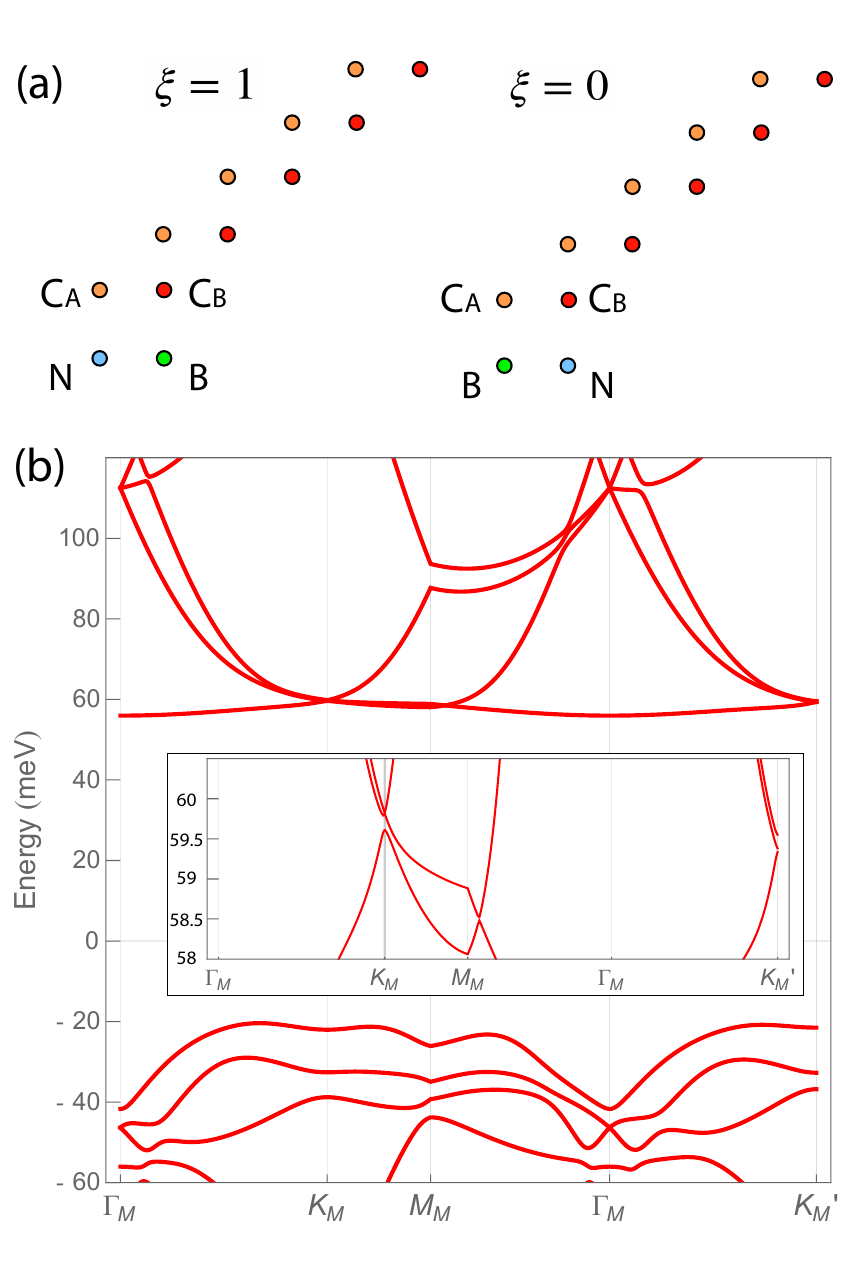}
\caption{ (a) Schematics of the two stacking configurations $\xi=1$ and $\xi=0$. $\text{C}_\text{A}$ and $\text{C}_\text{B}$ are carbon sites. N and B denote nitrogen and boron sites respectively. We focus on $\xi=1$ in this work. (b) Band structure of R5G/hBN with twist angle $\theta=0.77^\circ$ and $\xi=1$ under perpendicular electric field with interlayer potential $V=28$ meV. The electric field opens a gap between conduction and valence bands, with conduction (valence) bands polarized in the top (bottom) layer when $V>0$. The three conduction bands are very close to each other at the $K$ point of the MBZ, as shown in the inset, hence the spectrum cannot be approximated by an isolated flat band.  }
\label{fig_bandbare}
\end{figure}

The single-particle model of rhombohedral pentalayer graphene twisted against hBN consists of the pristine R5G Hamiltonian $H_{R5G}^\eta$ under a perpendicular electric field and the \text{moir\'e} potential term $H_{\text{moir\'e},\xi}$ induced by the hBN near the bottom graphene layer:
\bea
H_{0,\eta}&=&H_{R5G}^\eta+H_{\text{moir\'e}}.\ \ 
\eea
The creation operator in the continuum model for R5G is $ c^\dagger_{ \eta,\bsl{k},\bsl G, l,\sigma, s} $, where $\bsl{k}$ is the momentum inside the \text{moir\'e} Brillouin zone (MBZ), $\bsl G$ is the \text{moir\'e} reciprocal lattice vector, $l=0,1,...,4$ is the layer index, $\sigma=A,B$ represents the sublattice, $\eta=\pm \K$ labels the valley, and $s=\uparrow,\downarrow$ is the spin index. The hBN is near the bottom layer with $l=0$. There are two distinct stackings depending on whether the carbon \(A\) site aligns with boron or nitrogen site. We use a variable $\xi=0,1$ to distinguish these two stackings as shown in Fig.~\ref{fig_bandbare}(a). 

In this work, we focus on the twist angle $\theta=0.77^\circ$ and choose $\xi=1$, which has a stable Chern insulator at filling 1 in HF calculations~\cite{MFCI3,MFCIIV,Guo_Lu_Xie_Liu_2023}. The detailed formulation of the Hamiltonian can be found in Appendix \ref{Sec_h0}. The band structure of the non-interacting Hamiltonian is shown in Fig.~\ref{fig_bandbare}(b). A perpendicular electric field with an interlayer potential of $V=28$ meV is applied (corresponding to the typical experimental displacement field at which the FCI is observed), opening a gap between conduction and valence bands, with conduction (valence) bands polarized in the top (bottom) layer. The lowest conduction band is relatively flat and is very close to the two higher conduction bands at the boundary of the MBZ, with a small direct gap of $\sim 0.1$ meV. Therefore, when the lowest conduction band is partially filled, the system \emph{cannot} be approximated by an isolated flat band.

\subsection{Interaction schemes}\label{sec:interactionschemes}

Next, we discuss the interacting Hamiltonian in the CN~\cite{MFCI3,MFCIIV,Zhou_Yang_Zhang_2023,PhysRevLett.133.206503_2024,PhysRevLett.133.206502_2024} and average (AVE)~\cite{MFCI3,MFCIIV,TBG3,2021PhRvB.103t5414L,Parker2021fieldtunedFCITBG,Calugaru2021TSTG,Christos2022TSTG,wang2023kekule,kwan2023strongcoupling} schemes (see Appendix \ref{Sec_appscheme} for more details). In the CN scheme, the interaction is normal-ordered with respect to all the valence bands below the Fermi level at charge neutrality. This is equivalent to neglecting all contributions to the interaction from the occupied valence bands, including the charge density background formed by those valence bands. The Hamiltonian in the CN scheme reads:
\bea
H_{\text{CN}} &=& \sum_{\eta} H_{0,\eta} + \frac{1}{2\mathcal{V}}\sum_{\mbf{q}, \mbf{G}} V(\mbf{q}+\mbf{G}) : \hat \rho_{\mbf{q}+\mbf{G}}\hat \rho_{-\mbf{q}-\mbf{G}}: \nonumber\\ 
\eea
The notation $:\hat{O}:$ refers to the normal ordering that places all annihilation (creation) operators on the right for conduction (valence) electrons in $\hat{O}$ while keeping track of the fermionic minus signs. Note that this differs from the conventional normal-ordering operation with respect to the vacuum, which places all annihilation operators to the right regardless of whether the operator is in the conduction or valence band. With this normal-ordering convention, the interaction in the CN scheme annihilates many-body states at charge neutrality, in which all valence bands are occupied, and none of the conduction bands are occupied. The density operator is
\bea
\hat \rho_{\mbf{q}+\mbf{G}} &=& \sum_{\mbf{k} mn \eta s} M_{mn}^{\eta}(\mbf{k},\mbf{q}+\mbf{G}) c^\dag_{\eta,\mbf{k}+\mbf{q}, m,s} c_{\eta,\mbf{k}, n,s},
\eea
where $m,n$ are summed over all conduction and valence bands. The form factor is given by 
\eq{
M^{\eta}_{mn}(\bsl{k},\bsl{q}+\bsl{G}) = \sum_{\bsl{G}'l \sigma} \left[U^{\eta}_{m}(\bsl{k}+\bsl{q}+\bsl{G})\right]_{\bsl{G}'l\sigma}^* \left[U^{\eta}_{n}(\bsl{k})\right]_{\bsl{G}'l\sigma},
}
where $U_n^\eta(\bsl k)$ is the eigenvector of the $n$-th band of the non-interacting Hamiltonian $H_{0,\eta}$, and the creation operator in the band basis reads
\bea
c^\dagger_{\eta,\bsl{k},n ,s} &=&  \sum_{\bsl{G}l\sigma}c^\dagger_{\eta,\bsl{k},\bsl{G},l,\sigma , s} \left[ U_{n}^{\eta}(\bsl k) \right]_{\bsl{G}l\sigma}.
\eea
The interaction is taken to be the dual-gate screened Coulomb interaction:
\be
V(\mbf{q}) = \frac{e^2}{2 \eps} \frac{\tanh |\mbf{q}| d_{sc}}{|\mbf{q}|},
\ee
with the dielectric constant chosen as $\epsilon=5\epsilon_0$, and the gate-to-sample distance set to $d_{sc}=10\text{nm}$.

The average (AVE) scheme differs from the CN scheme by additional single-particle terms proportional to the Coulomb interaction strength. The Hamiltonian in the AVE scheme is
\bea
H_{\text{AVE}} &=& \sum_{\eta} H_{0,\eta} + \frac{1}{2\mathcal{V}}\sum_{\mbf{q}, \mbf{G}} V(\mbf{q}+\mbf{G}) \, \delta \hat \rho_{\mbf{q}+\mbf{G}}\delta\hat \rho_{-\mbf{q}-\mbf{G}},\ \ 
\eea
with
\bea
\delta \hat \rho_{\mbf{q}+\mbf{G}} &=& \!\!\!\!\sum_{\mbf{k} mn \eta s} \!\!\!\! M_{mn}^{\eta}(\mbf{k},\mbf{q}+\mbf{G})(c^\dag_{\eta,\mbf{k}+\mbf{q},m,s}c_{\eta,\mbf{k},n,s} - \frac{1}{2}\delta_{\mbf{q},0} \delta_{mn}). \nonumber\\
\label{Haveall}
\eea
Here, the summation over bands includes all valence and conduction bands. The difference between the AVE and CN schemes is a one-body background term:
\bea
&&\sum_{\eta} H_{b}^{\eta}=H_{\text{AVE}} - H_{\text{CN}} \nonumber\\
&=&  \sum_{\mbf{q} \mbf{G}} \frac{V(\mbf{q}+\mbf{G})}{2\mathcal{V}} ( \delta \hat \rho_{\mbf{q}+\mbf{G}}\delta\hat \rho_{-\mbf{q}-\mbf{G}}\ -\  \!:\!\hat \rho_{\mbf{q}+\mbf{G}}\hat \rho_{-\mbf{q}-\mbf{G}}\!:).\ 
\label{Hbdiff}
\eea
This term accounts for the influence of the background charge density on the electrons in the active conduction band, which is absent in the CN scheme (see Appendix \ref{Sec_appscheme}). 

Throughout this work, we consider many-body states where all valence bands are fully occupied. Motivated by the band structure in Fig.~\ref{fig_bandbare}(b), we restrict our computations to the lowest $n_{\text{act}}$ conduction bands, denoted as the active bands, and take $n_{\text{act}}=3$ unless otherwise mentioned. We also assume that the active bands are spin- and valley-polarized; hence, we fix the spin to $\uparrow$ and the valley index to $K$ in electron operators from now on. Without loss of generality, the Hamiltonian in the CN scheme, after this simplification, becomes:
\bea
\label{eq:H_CN}
H_{\text{CN}} &=& H_0 + \frac{1}{2\V} \sum_{\bsl{k}_1 \bsl{k}_2 \bsl{q} }  \sum_{n_1 n_2 n_3 n_4}  V^{\bsl K,\bsl K}_{n_1 n_2 n_3 n_4}(\bsl{k}_1,\bsl{k}_2,\bsl{q}) \nonumber\\
&& \times c^\dagger_{K,\bsl{k}_1+\bsl{q},n_1,\uparrow } c^\dagger_{K,\bsl{k}_2-\bsl{q},n_2,\uparrow }  c_{K,\bsl{k}_2,n_3 ,\uparrow}  c_{K,\bsl{k}_1,n_4,\uparrow } \ 
\eea
\bea
V_{n_1 n_2 n_3 n_4}^{\eta_1 \eta_2}(\bsl{k}_1,\bsl{k}_2,\bsl{q}) &=&\sum_{\bsl{G}} V(\bsl{q}+\bsl{G}) M_{n_1 n_4}^{ \eta_1}(\bsl{k}_1,\bsl{q}+\bsl{G})  \nonumber\\
&&\times M_{n_2 n_3}^{ \eta_2}(\bsl{k}_2,-\bsl{q}-\bsl{G}).
\eea
Here, the summation over bands is restricted to the active bands. The Hamiltonian in the AVE scheme has an additional background term $H_b$:
\be
H_{\text{AVE}}=H_{\text{CN}}+H_b,
\ee
where $H_b$ is obtained by restricting the background term $H_b^K$ in Eq.~\eqref{Hbdiff} to the three active bands. The explicit expression for this term is provided in Eq.~\eqref{hbglong} in Appendix \ref{Sec_appscheme}.

\subsection{Hartree-Fock basis and truncation of the Hilbert space}\label{Subsec_HFtruncation}

The low energy conduction bands in R5G/hBN are nearly gapless, posing a challenge for the ED computation. Unlike simple models with a single isolated narrow Chern band, in which an FCI can be observed in ED even for small system sizes~\cite{FCI-PhysRevX.1.021014}, band-mixing become significant in systems with a nearly gapless non-interacting band structure. We perform three-band ED computations to capture the effect of band-mixing. The large number of bands significantly increases the dimension of the Hilbert space. To make the ED computation manageable, we apply a truncation to the Hilbert space by limiting the allowed number of particles in the higher bands  Ref.~\cite{MFCIIV}.

We represent the truncation of the Hilbert space by the two truncation parameters $\{N_{\text{band1}},N_{\text{band2}}\}$, which means we only consider many-body states with no more than $N_{\text{band1}}$ particles in band 1, no more than $N_{\text{band2}}$ particles in band 2, while allowing any number of particles in the lowest band 0. Larger truncation parameters $\{N_{\text{band1}},N_{\text{band2}}\}$ indicate stronger band-mixing, and $\{0,0\}$ represents the single-band limit in which the ED computation is fully within the lowest HF band. This truncation efficiently reduces the size of the Hilbert space. For example, for a system with 18 unit cells and 12 particles, the Hilbert space dimension per momentum sector for the three-band system is $2\times 10^{10}$ without truncation (beyond current computational resources), whereas the dimension for a truncated Hilbert space with $\{2,1\}$ is only $9\times 10^6$.

When performing ED computation in the Hilbert space spanned by the active bands, we have the freedom to choose a new one-body basis obtained as a linear combination of the original conduction bands. This basis degree of freedom does not affect the ED spectrum if no truncation is applied to the band occupation, as the change of basis is simply a unitary transformation. However, this equivalence between different bases breaks down when we consider a truncated many-body Hilbert space. For large system sizes, the ED computation is feasible only at small truncation numbers. In this case, the ED result remains accurate if the low-energy states contain most of the particles in band "0" (we will detail below how this band is chosen) and only a few particles in the higher bands. Therefore, the choice of basis becomes important for the ED computation in the truncated Hilbert space. 

To search for FCIs, we first choose the basis to be the eigenstates of the self-consistent Hartree-Fock (HF) Hamiltonian at integer filling, in which the lowest band has Chern number 1 (Ch=1). We call this new basis the HF basis and refer to the original basis, consisting of eigenstates of the kinetic Hamiltonian, as the bare basis. The HF Hamiltonian in the AVE scheme is written as~\cite{MFCI3,MFCIIV}:
\begin{widetext}
\bea
\label{eq:hartreefockapp}
H_{\text{HF}}[P] &=& H_0 + H_b + H_H[P]+ H_F[P], \\
H_H[P]&=&  \sum_{\mbf{k}mn\mbf{G}} \frac{V(\mbf{G})}{\V} M_{mn}^K(\mbf{k},\mbf{G})\lp\sum_{\mbf{k}'m'n'} M^{K*}_{m'n'}(\mbf{k}',\mbf{G}) P_{m' n'}(\mbf{k}') \rp c^\dag_{K,\mbf{k},m,\uparrow}c_{K,\mbf{k},n,\uparrow}, \\
H_F[P]&=& - \sum_{\mbf{k}mn} \sum_{\mbf{q} \mbf{G} m' n'}\frac{V(\mbf{q}+\mbf{G})}{\V} M^{K*}_{n'm}(\mbf{k},\mbf{q}+\mbf{G}) P^*_{m' n'}(\mbf{k}+\mbf{q}) M_{m' n}^K(\mbf{k},\mbf{q}+\mbf{G}) c^\dag_{K,\mbf{k},m,\uparrow}c_{K,\mbf{k},n,\uparrow}.
\eea
\end{widetext}
Here, the summation over bands is restricted to the active bands. The self-consistent order parameter is given by $P_{mn}(\bsl{k}) = \langle c^\dag_{K,\mbf{k},m,\uparrow}c_{K,\mbf{k},n,\uparrow}  \rangle $, while $H_H[P]$ and $H_F[P]$ are the Hartree and Fock terms generated by the interaction. The HF Hamiltonian in the CN scheme is obtained by dropping the $H_b$ term. 

We performed self-consistent HF calculations that involve the lowest $n_{\text{act}}$ conduction bands at filling $\nu=1$. We choose $n_{\text{act}}=5$ for the CN scheme and $n_{\text{act}}=3$ for the AVE scheme, which yields a self-consistent HF Hamiltonian whose lowest HF band has Ch=1 as reported in Ref.~\cite{MFCIIV}. Let $\tilde{U}_{n \alpha}(\mbf{k})$ be the $\alpha$th eigenvector of $H^{\text{HF}}[P]$ at momentum $\mbf{k}$, where $\alpha=0,1,2$ is ordered by increasing energy of the HF bands. {If the number of HF bands is larger than three, we only keep the lowest three HF bands and set the particle number in the higher bands to zero.} Then, the creation operator $\gamma^\dagger_{\bsl k,\alpha}$ in the HF basis is related to that in the bare basis, $c^\dagger_{K,\bsl k,n,\uparrow}$, by:
\bea
\gamma^\dag_{\mbf{k},\alpha} &=& \sum_{n} c^\dag_{K,\bsl{k}, n,\uparrow}  \tilde{U}_{n \alpha}(\mbf{k}).
\label{eq:HF_basis}
\eea 
When generating the HF basis, we start from a dense momentum mesh (e.g., $18\times 18$) to ensure that the HF ground state at $\nu=1$ has Chern number Ch=1~\cite{MFCI3,MFCIIV}. Then, for the ED computation, we extract a subset of momentum points from this dense momentum mesh, and only keep the three lowest HF band at each momentum.

\begin{figure}
\centering
\includegraphics[width=3.4 in]{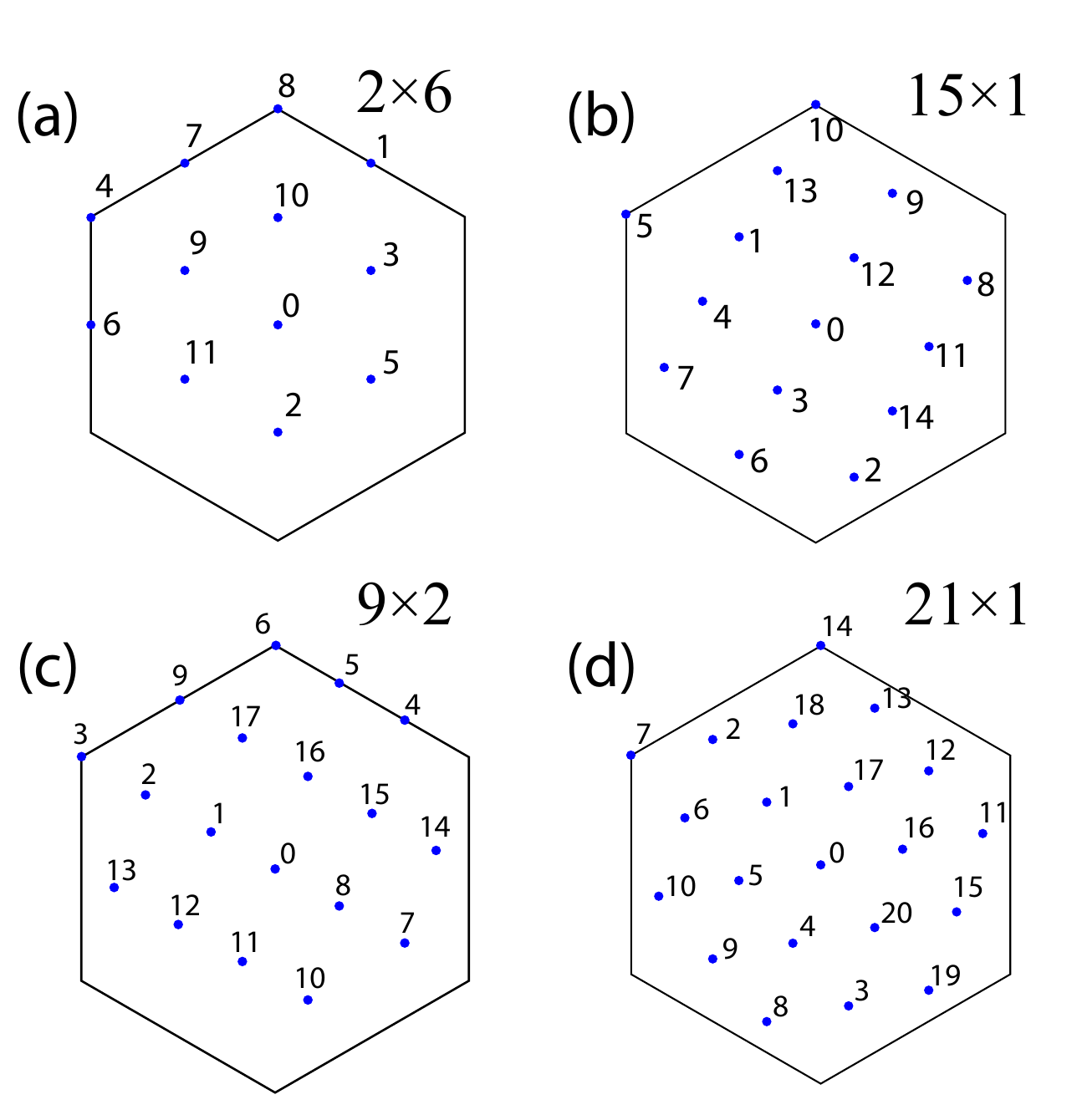}
\caption{ Different momentum meshes used in ED, with $(N_1,N_2,\tilde n_{11},\tilde n_{12},\tilde n_{21},\tilde n_{22})$ chosen as (a) $(2,6,1,0,1,1)$, (b) $(15,1,1,-5,0,1)$, (c) $(9,2,1,-2,0,1)$, and (d) $(21,1,1,-5,0,1)$, respectively. The momentum points are labeled by $k_1+k_2N_1$. The meshes in (a) and (d) preserve the threefold rotation symmetry.}
\label{fig_kmesh}
\end{figure}

The momentum mesh for ED is chosen as follows. Let $\bsl b_{M1}$ and $\bsl b_{M2}$ be the shortest \text{moir\'e} reciprocal lattice vectors, with $\bsl b_{M2}=R(\frac{2\pi}{6})\bsl b_{M1}$, where $R(\theta)$ is the rotation matrix that rotates vectors by an angle $\theta$. The momentum points in the Brillouin zone of an $N_1\times N_2$ system are given by
\eq{
\bsl k=\frac{k_1}{N_1}\bsl f_1+\frac{k_2}{N_2}\bsl f_2, 
}
where $k_i=0,1,...,N_i-1$. The reciprocal lattice vectors $\bsl f_1$ and $\bsl f_2$ are defined as
\eq{
\bsl f_1=\tilde n_{11}\bsl b_{M1}+\tilde n_{12}\bsl b_{M2}, \quad
\bsl f_2=\tilde n_{21}\bsl b_{M1}+\tilde n_{22}\bsl b_{M2}, 
}
where $\tilde n_{ij}$ are integers. Each distinct choice of $N_1, N_2, \tilde n_{ij}$ corresponds to a different momentum mesh. {We choose the same momentum meshes as Ref.~\cite{MFCIIV} as illustrated in Fig.~\ref{fig_kmesh}. With these choices, the momentum $K$ and $K'$ points are contained in the mesh and the three FCI states have distinct momenta, which reduces the level repulsion between FCI states. }

\section{Numerical results in the HF basis}
\label{Sec_beforeiter}

In this section, we present the multi-band ED results in the HF basis obtained from HF computations for the CN and AVE schemes. We focus on the FCI at 1/3 and 2/3 filling and study its stability under band mixing by considering the three lowest HF bands. For pedagogical purpose, we primarily focus on results for $9\times 2$ systems with a momentum mesh shown in Fig.~\ref{fig_kmesh}(c). The results for other system sizes, including $21\times 1$ and $15\times 1$, are available in Appendix~\ref{app:additionalED}.

\subsection{CN scheme at 1/3 filling}\label{subsec:CNonethird}

\begin{figure*}
\centering
\includegraphics[width=\textwidth]{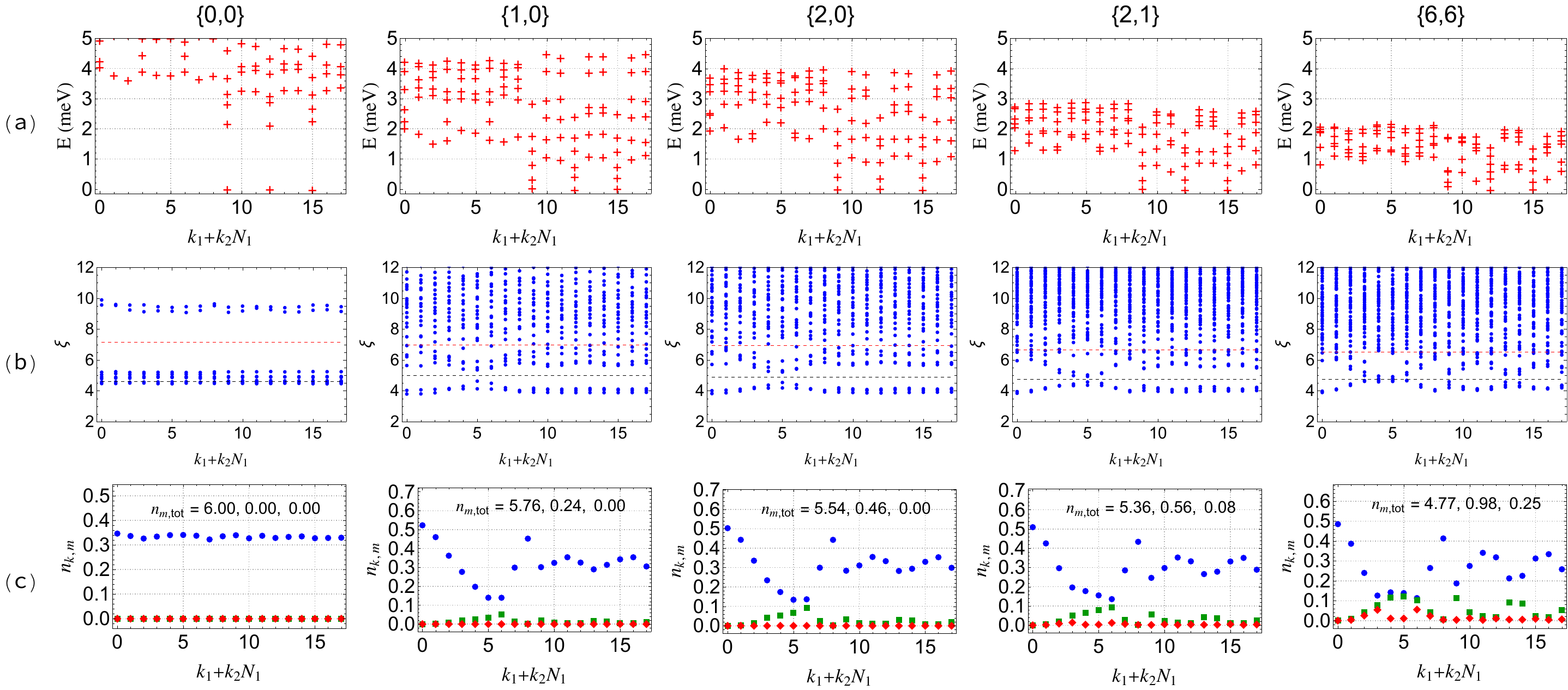}
\caption{ Energy spectrum, PES, and occupation number of the $9\times 2$ system at 1/3 filling in the CN scheme with $V=28$ meV. Different columns represent ED results computed at different truncation parameters $\{N_{\text{band1}},N_{\text{band2}}\}$. (a) Energy spectrum. The energies are globally shifted by a constant so that the ground state energy is set to zero. (b) PES of the three lowest states at FCI momenta $9,12,15$. There are 117 states below the red line, which marks the FCI counting, and 45 states below the black line, which marks the CDW counting. Note that the size of the many-body Hilbert space in each momentum sector of the PES increases when increasing $N_{\text{band1}}$ or $N_{\text{band2}}$, leading to an exponentially larger number of entanglement levels. (c) Occupation number $n_{\bsl k,m}$ as defined in Eq.~\eqref{eq:bandoccupationperk}. Blue, green, and red colors correspond to $m=0,1,2$. The inset shows the total occupation number $n_{m,\text{tot}}$ for band $m=0,1,2$ respectively (see Eq.~\ref{eq:totalbandoccupation}).  }
\label{fig_CN9b2n6i0ch}
\end{figure*}

The spectra at 1/3 filling in the HF basis in the CN scheme with different truncation parameters are shown in Fig.~\ref{fig_CN9b2n6i0ch}(a). We begin with the single-band results at truncation $\{0,0\}$, where all particles occupy the lowest HF band with Chern number 1. As was shown in previous works~\cite{PhysRevLett.133.206502_2024,MFCIIV,Zhou_Yang_Zhang_2023,Guo_Lu_Xie_Liu_2023,PhysRevB.110.115146-2024}, the single band calculation is heavily biased towards FCIs, since FCIs are known to exist in projected Chern bands such as the one obtained here from HF. For example, the ED spectrum at truncation $\{0,0\}$ has threefold almost degenerate ground state at the FCI momenta $k_1+k_2N_1=9,12,15$, as expect for a Laughlin-like state on such a geometry~\cite{Bernevig2012} , which are separated from excited states by a gap of $\sim 1$ meV. Upon band mixing controlled by the parameters $\{N_{\text{band1}},N_{\text{band2}}\}$ and as pointed out in Ref.~\cite{MFCIIV}, the approximate threefold degeneracy of the ground state seems to survive for a range of mixing until being washed out at large mixing.

However, 3-fold degeneracy is not limited to FCI states. 
To further confirm the emergence of an FCI and rule out the possibility of competing states such as a CDW, which also exhibits gapped ground states at the same momenta on a such a $9 \times 2$ geometry, we compute the particle entanglement spectrum (PES)~\cite{PhysRevLett.101.010504-2008,PhysRevLett.106.100405,FCI-PhysRevX.1.021014} from the three ground states. This has not yet been tested in prior calculations, and as we shall see, brings about big surprises.  To compute the PES, we divide the $N$ particles into two subsets with $N_A$ and $N_B$ particles and compute the eigenvalues $e^{-\xi_i}$ of the reduced density matrix $\rho_A$, obtained by tracing out subsystem $B$:
\be
\rho_A=\text{Tr}_B\left[\frac{1}{N_{\text{GS}}}\mathcal{P}_{\text{GS}}\right],\label{eq:reduceddensitymatrix}
\ee
with
\be
\mathcal{P}_{\text{GS}}=\sum_{a\in \{\text{GS}\}} | \Psi_a \rangle \langle \Psi_a|\label{eq:GSprojector}
\ee
the projection to the ground state manifold $\{\text{GS}\}$, i.e., the $N_{\text{GS}}=3$ ground states for $\nu=1/3$. We choose $\text{GS}$ to contain the three lowest many-body states in the momentum sectors in which FCI could appear. For a given momentum mesh and filling fraction, if the allowed FCI momenta are distinct, then $\text{GS}$ contains the lowest energy state at each of the three momentum sectors. If the three FCI states have the same momentum, then $\text{GS}$ contains the lowest three states in that momentum sector. Note that later we will encounter cases where the momenta of the three absolute lowest energy states differ from FCI momenta. In those cases we still denote $\text{GS}$ as the three lowest states at FCI momenta. For FCI states at 1/3 filling in a BZ torus with $N_k=N_1 N_2$ momentum points, the PES spectrum is gapped, and the total number of states ${\cal N}_{\rm FCI}$ below the gap is given by~\cite{FCI-PhysRevX.1.021014}
\be
{\cal N}_{\rm FCI}(N_k,N_A)=\frac{N_k(N_k-2N_A-1)!}{N_A!(N_k-3N_A)!}.
\ee
For the $9\times 2$ system at 1/3 filling with $N_A=2$, the FCI counting predicts ${\cal N}_{\rm FCI}(18,2)=117$. We plot the PES with $N_A=2$ in Fig.~\ref{fig_CN9b2n6i0ch}(b) and use a horizontal red line to mark the threshold below which there are 117 states. The PES at truncation $\{0,0\}$ is gapped at the FCI counting, in agreement with previous calculations~\cite{PhysRevLett.133.206502_2024,PhysRevLett.133.206503_2024,PhysRevB.110.115146-2024}. {An entanglement gap is also observed for other values of $N_A$ such as $N_A=3$. These observations confirm the Laughlin-like nature of the ground state when solely considering the lowest HF Chern band. }

Surprisingly, the ED computation at finite truncation levels reveals a significant change in the ground state. We find that the band mixing controlled by the truncation parameters $\{N_{\text{band1}},N_{\text{band2}}\}$ can induce a transition from an FCI to a CDW in the CN scheme at 1/3 filling.
Considering band mixing provides a striking different picture. At truncation $\{1,0\}$, where one particle is allowed to populate the first remote HF band, the three lowest states remain at the FCI momenta with a significant decrease of the energy gap. A set of 3 other states at same momenta, live in close vicinity to the 3 ground-states, begging the question of whether the ground state is still FCI or has changed. We compute the PES and surprisingly find that the gap at the FCI counting, labeled by the red line, closes, while another PES gap opens. Below this new gap indicated by the black line, the number of entanglement level matches the CDW counting~\cite{bernevig2012thintoruslimitfractionaltopological}

\be
{\cal N}_{\text{CDW}}(N,N_A)=3\frac{N!}{N_A!(N-N_A)!}\label{eq:countingCDW}
\ee

Indeed, we find ${\cal N}_{\text{CDW}}(6, 2)=45$ states below the gap. We provide a full discussion of $K$-CDW and the PES derivation for such a phase in App.~\ref{app_CDWktot}. The closing of the FCI gap in the PES and the opening of a CDW gap suggest that the ground states transition to a CDW already for the smallest deviation to the single HF band calculation, namely the truncation $\{1,0\}$ - which for $6$ particle in $18$ momenta, seems to influence the result dramatically. The CDW state has far less low energy levels in its  entanglement spectrum than an FCI has, as it is-  essentially- a product state,  whereas an FCI is not. This transition from an FCI to a CDW due to band mixing is also observed in other system sizes, e.g., as shown in Fig.~\ref{fig_CN21b1n7} of Appendix~\ref{app_allfig} for a 21-site system. Note that for $9\times 2$ systems, we can perform a full three-band computation without truncation at 1/3 filling, which corresponds to  $ \{N_{\text{band1}},N_{\text{band2}}\}=\{6,6\}$. In that case, the PES with $N_A=2$ is not gapped at either the FCI or CDW counting, concomitant to the absence of any gap in the energy spectrum.

We also compute the occupation number $n_{\bsl k,m}$ of the ground states, defined as
\bea
n_{\bsl k,m}&=&\frac{1}{N_{\text{GS}}}{\text Tr} \left[\gamma^\dagger_{\bsl k,m}\gamma_{\bsl k,m} \mathcal{P}_{\text{GS}} \right],\label{eq:bandoccupationperk}
\eea
where $N_{\text{GS}}$ is the number of ground states. We also define the total occupation number per momentum $n_{\bsl k}$ and per band $n_{m,\text{tot}}$ as:
\bea
n_{\bsl k}&=&\sum_m n_{\bsl k,m},\ \ n_{m,\text{tot}}=\sum_{\bsl k}n_{\bsl k,m}.\label{eq:totalbandoccupation}
\eea
Here, $\gamma^\dagger_{\bsl k,m}$ is the creation operator for band $m$, where $m=0,1,2$ labels the three HF bands, and $N_{\text{GS}}$ is the number of states in the set of ground states $\{\text{GS}\}$. We choose $\{\text{GS}\}$ to contain the three lowest states at the FCI momenta with $N_{\text{GS}}=3$. The plots of $n_{\bsl k,m}$ are shown in Fig.~\ref{fig_CN9b2n6i0ch}(c). The occupation number distribution $n_{\bsl k,m}$ at $\{0,0\}$ is relatively uniform, consistent with the ground states being FCIs. However, when including band mixing, there is a significant decrease of $n_{\bsl k}$ at the $K'$ point with $k_1+k_2N_1=6$ and the other momenta at the boundary of MBZ hexagon, making the distribution of occupation number extremely nonuniform. As opposed to the thin-torus limit~\cite{bernevig2012thintoruslimitfractionaltopological} where the distribution of $n_{\bsl k}$ is uniform, here $n_{\bsl k}$ is non-uniform. But as shown in App.~\ref{app_CDWPES}, a $K$-CDW could also have a uniform $n_{\bsl k}$ and thus its non-uniformness cannot be used as a signature.

\begin{figure}
\centering
\includegraphics[width=3.4 in]{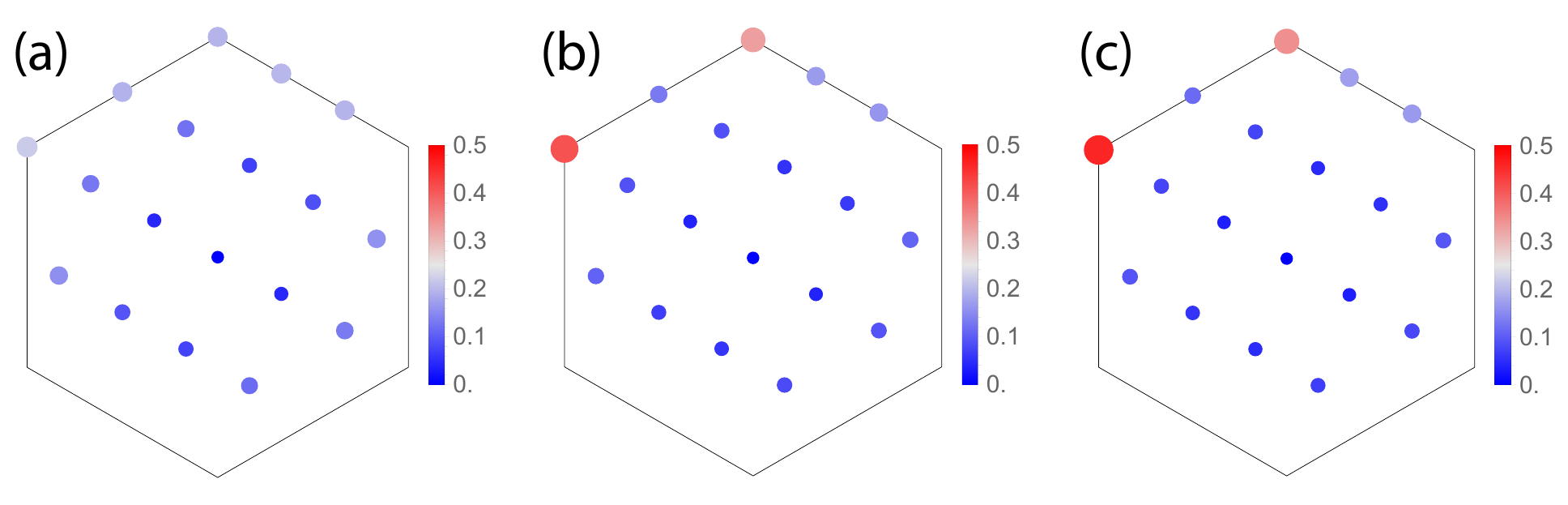}
\caption{ Structure factor $S(\bsl q)$ of the ground state in the $9\times 2$ system at 1/3 filling in the CN scheme in the HF basis at different truncation parameters: $\{0,0\}$, $\{1,0\}$, $\{2,0\}$ in (a), (b), and (c), respectively. The hexagon represents the MBZ. With small band-mixing, the correlation function is peaked at the $K$, $K'$ points, indicating a CDW transition.}
\label{fig_corCN9b2N6m}
\end{figure}

To reliably confirm the emergence of a $K$-CDW at small band mixing, we compute the density correlation function and structure factor $S(\bsl q)$~\cite{Wilhelm2021FCITBG}:
\bea
S(\bsl q)&=&\frac{1}{N_k}(\langle\hat\rho_{\bsl q}\hat{\rho}_{-\bsl q})\rangle - \langle\hat\rho_{\bsl q}\rangle\langle\hat\rho_{-\bsl q}\rangle ).
\eea
Details on the computation of $S(\bsl q)$ are provided in Appendix \ref{Sec_cor}. The plots of $S(\bsl q)$ at $\{0,0\},\{1,0\},\{2,0\}$ are shown in Fig.~\ref{fig_corCN9b2N6m}. $S(\bsl q)$ is peaked at the $K,K'$ points of the MBZ for $\{1,0\}$ and $\{2,0\}$ but not for $\{0,0\}$, signaling a transition from an FCI at the single-band limit $\{0,0\}$ to a CDW with ordering vector $K$ under small band-mixing at $\{1,0\}$ and $\{2,0\}$. {The ordering vector $K$ of the CDW is also consistent with the fact that there are three quasi-degenerate ground states with total momenta that differ by $\pm K$. A more detailed discussion on the total momentum of CDW states is given in App.~\ref{app_CDWktot}.}

\subsection{CN scheme at 2/3 filling }\label{subsec:CNtwothird}
\begin{figure*}
\centering
\includegraphics[width=\textwidth]{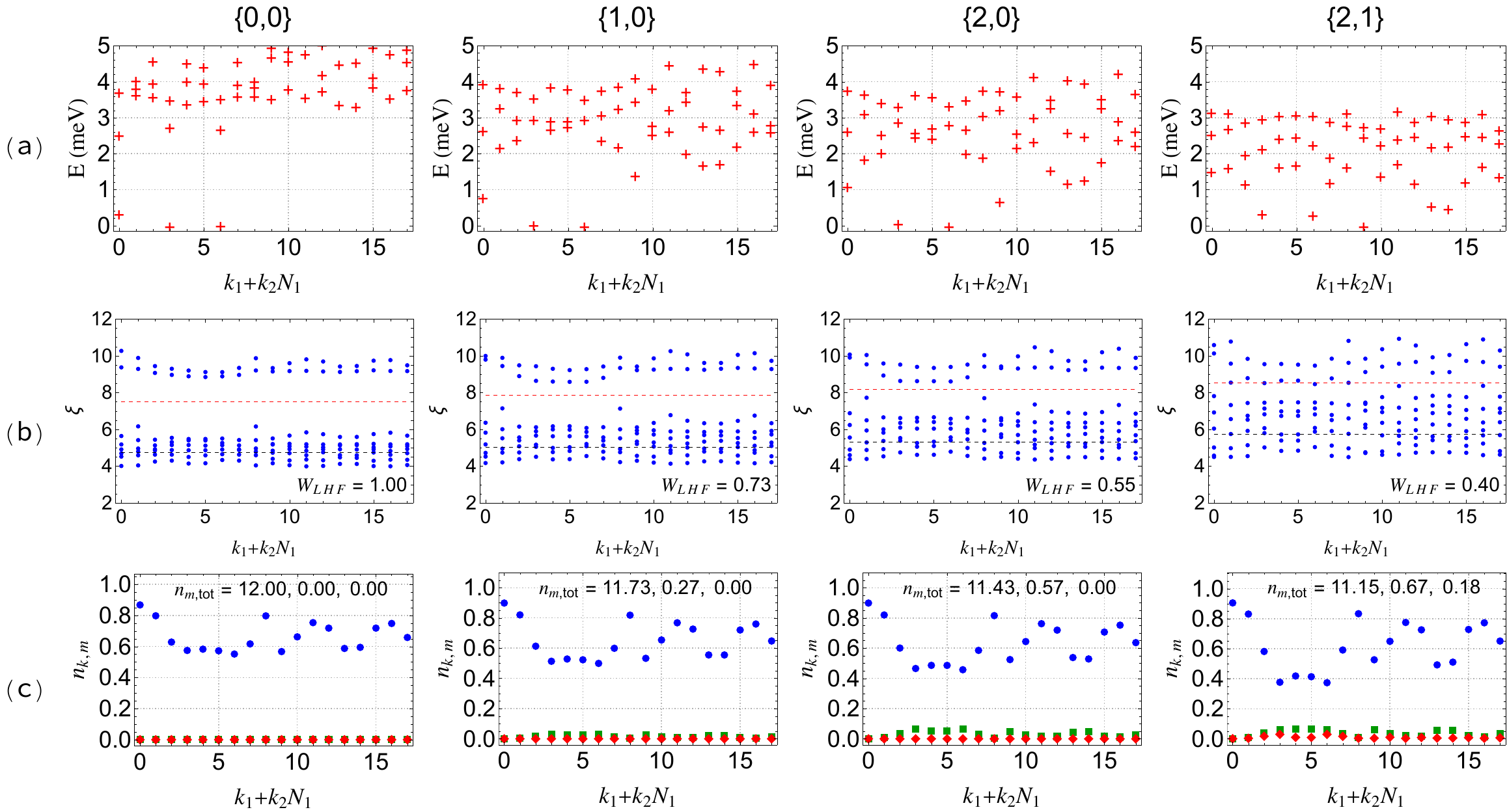}
\caption{ Energy spectrum, PES, and occupation number of the $9\times 2$ system at 2/3 filling in the CN scheme with $V=28$ meV. Different columns represent ED results computed at different truncation parameters. (a) Energy spectrum. The energies are globally shifted by a constant so that the ground state energy is set to zero. (b) PES of the three lowest states at FCI momenta $0,3,6$ after single-band projection and PH transform. There are 117 states below the red line, which marks the FCI counting, and 45 states below the black line, which marks the CDW counting. (c) Occupation number $n_{\bsl k,m}$ as defined in Eq.~\eqref{eq:bandoccupationperk}. Blue, green, and red colors correspond to $m=0,1,2$. The inset shows the total occupation number $n_{m,\text{tot}}$ for band $m=0,1,2$ respectively (see Eq.~\ref{eq:totalbandoccupation}).}
\label{fig_CN9b2n12i0ch}
\end{figure*}

At 2/3 filling in the CN scheme, we observe the collapse of the FCI energy gap even for moderate band mixing, but, unlike at 1/3 filling, there is no intermediate CDW phase at small band mixing. The energy spectrum, PES, and $n_{\bsl k,m}$ of the $9\times 2$ system at 2/3 filling in the HF basis are shown in Fig.~\ref{fig_CN9b2n12i0ch}. The PES and $n_{\bsl k}$ are always computed from the lowest state at the three FCI momenta $k_1+k_2N_1=0,3,6$, regardless of whether the three lowest energy states globally lie at the FCI momenta. The energy spectrum shows that the lowest three many-body states are at the FCI momenta only with zero or small band-mixing at $\{0,0\}$ and $\{1,0\}$, in agreement with previous computations in Ref.~\cite{MFCIIV}. Therefore, with finite band mixing, the ground state cannot be an FCI. 

\begin{figure}
\centering
\includegraphics[width=3.4 in]{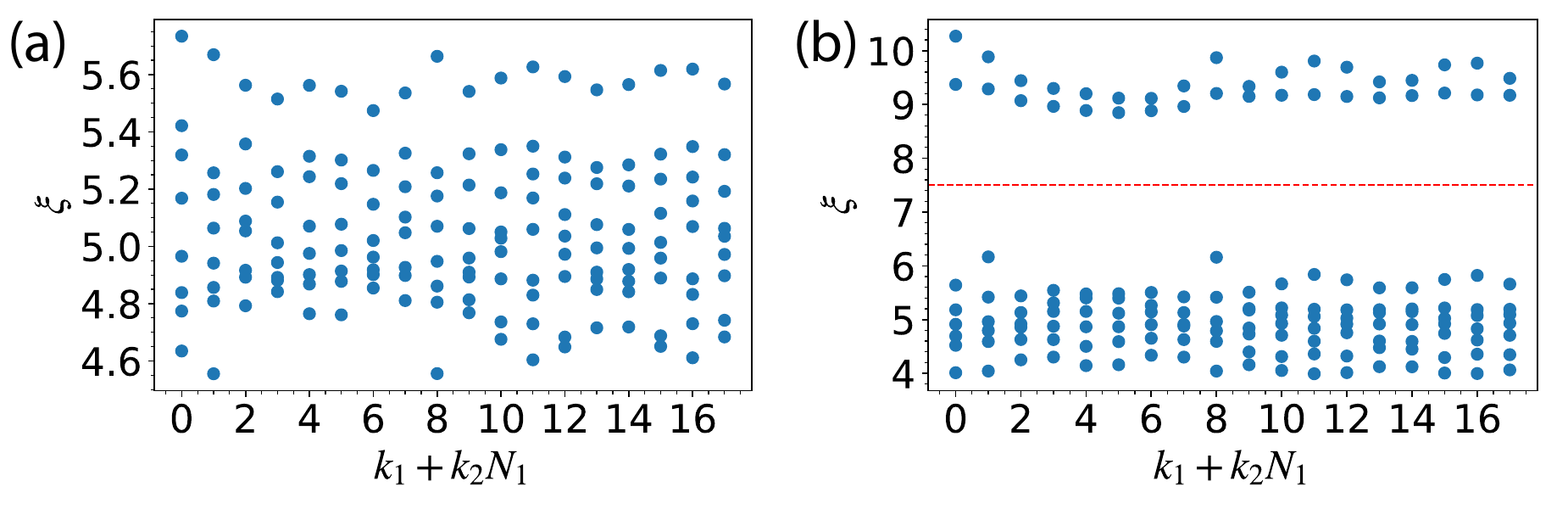}
\caption{ PES of the three ground states of the $9\times 2$ system at 2/3 filling in the CN scheme in the HF basis at $\{0,0\}$. (a) PES computed directly from the three ground states at 2/3 filling. No clear gap is visible in the PES. (b) PES computed for the same states after the PH transformation. An entanglement gap emerges in the PES, with 117 states below the red line, consistent with the FCI counting at 1/3 filling.}
\label{fig_PESnoph}
\end{figure}

To probe the nature of the ground states in the lowest HF band, i.e., at truncation parameter $\{0,0\}$, we compute the PES in Fig.~\ref{fig_CN9b2n12i0ch}(b). In general, for many-body states at 2/3 filling, the PES usually does not provide any insightful information. Indeed, in a single band setup with a filling factor greater than 1/2, no direct connection from the PES to the quasihole physics has been made in the literature so far. Usually, in the literature \cite{PhysRevLett.133.206502_2024}, to bypass this limitation, one first apply a particle-hole (PH) transformation to the FCI state at 2/3 filling, leading to an FCI state at 1/3 filling, whose PES can be computed. Note that such approach would be rigorous in a Landau level setup where the PH symmetry is exact and has  already been used to study the PES at $\nu=2/3$ for the CN scheme in the HF band in Ref.~\cite{PhysRevLett.133.206502_2024}. We compare the PES of the three ground states at truncation $\{0,0\}$ with and without the PH transformation in Fig.~\ref{fig_PESnoph}. As expected, no clear gap in the PES is visible without the PH transformation. However, after applying the PH transformation, a well-defined gap emerges with 117 states below it, consistent with the FCI counting at 1/3 filling and with the results reported in Ref.~\cite{PhysRevLett.133.206502_2024}, and confirming the FCI nature of the ground state at $\nu=2/3$ in the HF band. 

Away from the single HF band calculation, obtaining insight from PES requires one extra step. Performing the PH transformation with three bands is involved would not result in a system at filling factor $1/3$ but $\nu=3-2/3$. Thus prior to the PH transformation, we project each state on to the lowest HF band. Note that the projected states are not normalized and thus neither is the reduced density matrix leading to an irrelevant global shift of the PES. In Fig.~\ref{fig_CN9b2n12i0ch}(b), we thus apply first the projection onto the lowest HF band then the PH transformation when computing the PES for states at 2/3 filling. When computing the PES, we always use the lowest three states at FCI momenta, no matter whether they are the absolute lowest energy states (as discussed in Sec.~\ref{subsec:CNonethird}). We also provide in each PES plot, the weight ${\cal W}_{\text{LHF}}$ of the ground state manifold on the lowest HF band 
\be
{\cal W}_{\text{LHF}}=\frac{1}{N_{\text{GS}}}{\rm Tr}\left[  \mathcal{P}_{\text{GS}}\mathcal{P}_{\text{LHF}} \right] \label{eq:weightHF}
\ee
where $\mathcal{P}_{\text{LHF}}$ stands for the projection operator onto the lowest HF band.  When increasing band mixing in Fig.~\ref{fig_CN9b2n12i0ch}(b), we observe a PES gap separating the FCI counting from the other entanglement levels as long as the weight of the GS manifold is predominantly in the lowest HF band. This is also reflected in the occupation numbers as shown in Fig.~\ref{fig_CN9b2n12i0ch}(c). When the band mixing is too large, no clear entanglement gap is found and the energy gap collapses, indicating the absence of an intermediate CDW phase as opposed to $\nu=1/3$.

\subsection{AVE scheme}\label{subsec:AveHFBasis}

As discussed in Sec.~\ref{sec:interactionschemes}, the AVE scheme includes the influence of the background charge density on the electrons in the active conduction bands, leading to distinct features from the CN scheme in the energy spectrum, in particular a greater robustness to band mixing for the FCI phase at $\nu=2/3$ as shown in Ref.~\cite{MFCIIV} (even though it still  does not survive for strong enough mixing). Since this previous work provides a comprehensive study of the energy spectra, here we will mostly focus on the PES. 

\begin{figure*}
\centering
\includegraphics[width=6.8 in]{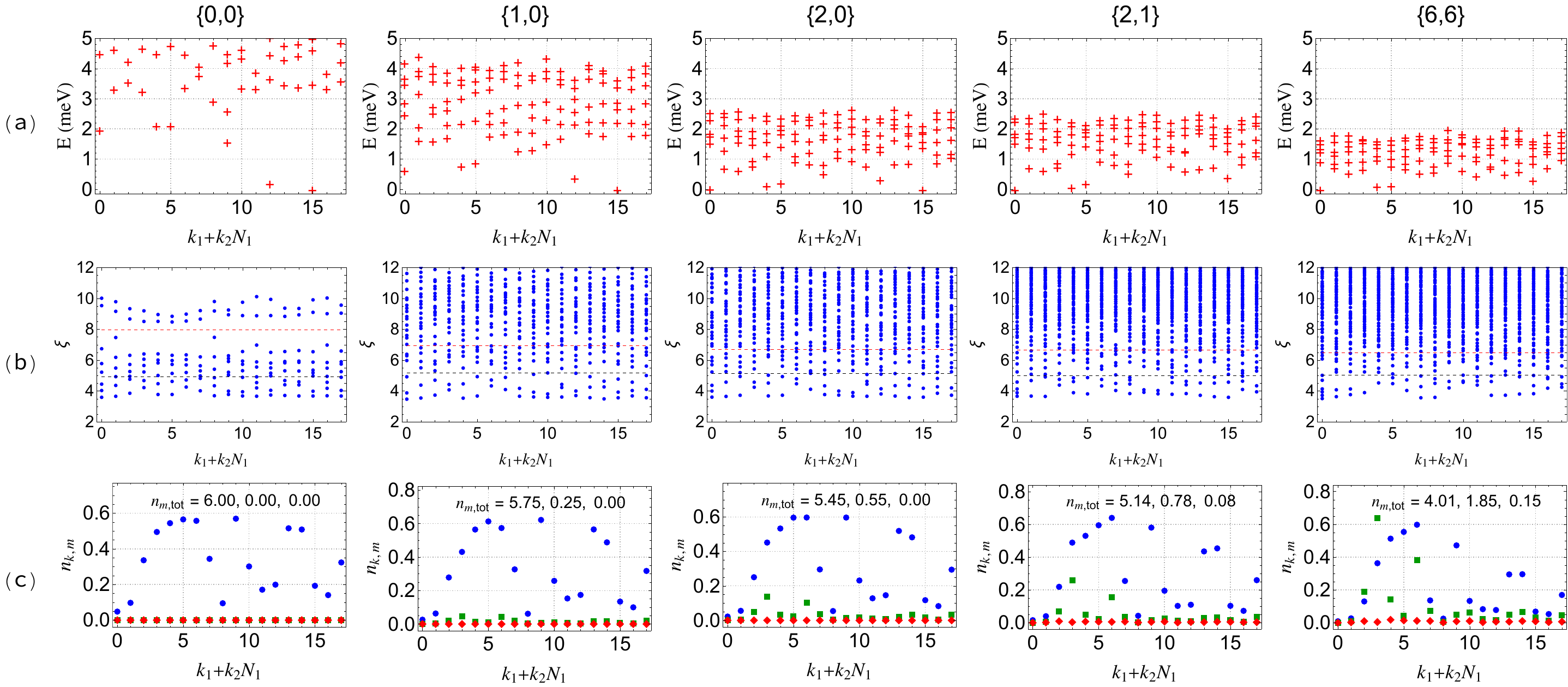}
\caption{ Energy spectrum, PES, and occupation number of the $9\times 2$ system at 1/3 filling in the AVE scheme with $V=22$ meV. Different columns represent ED results computed at different truncation parameters. (a) Energy spectrum. The energies are globally shifted by a constant so that the ground state energy is set to zero. (b) PES of the three lowest states at FCI momenta $9,12,15$. There are 117 states below the red line, which marks the FCI counting, and 45 states below the black line, which marks the CDW counting. (c) Occupation number $n_{\bsl k,m}$ as defined in Eq.~\eqref{eq:bandoccupationperk}. Blue, green, and red colors correspond to $m=0,1,2$. The inset shows the total occupation number $n_{m,\text{tot}}$ for band $m=0,1,2$ respectively (see Eq.~\ref{eq:totalbandoccupation}). }
\label{fig_AVE9b2n6i0ch}
\end{figure*}

\begin{figure*}
\centering
\includegraphics[width=\textwidth]{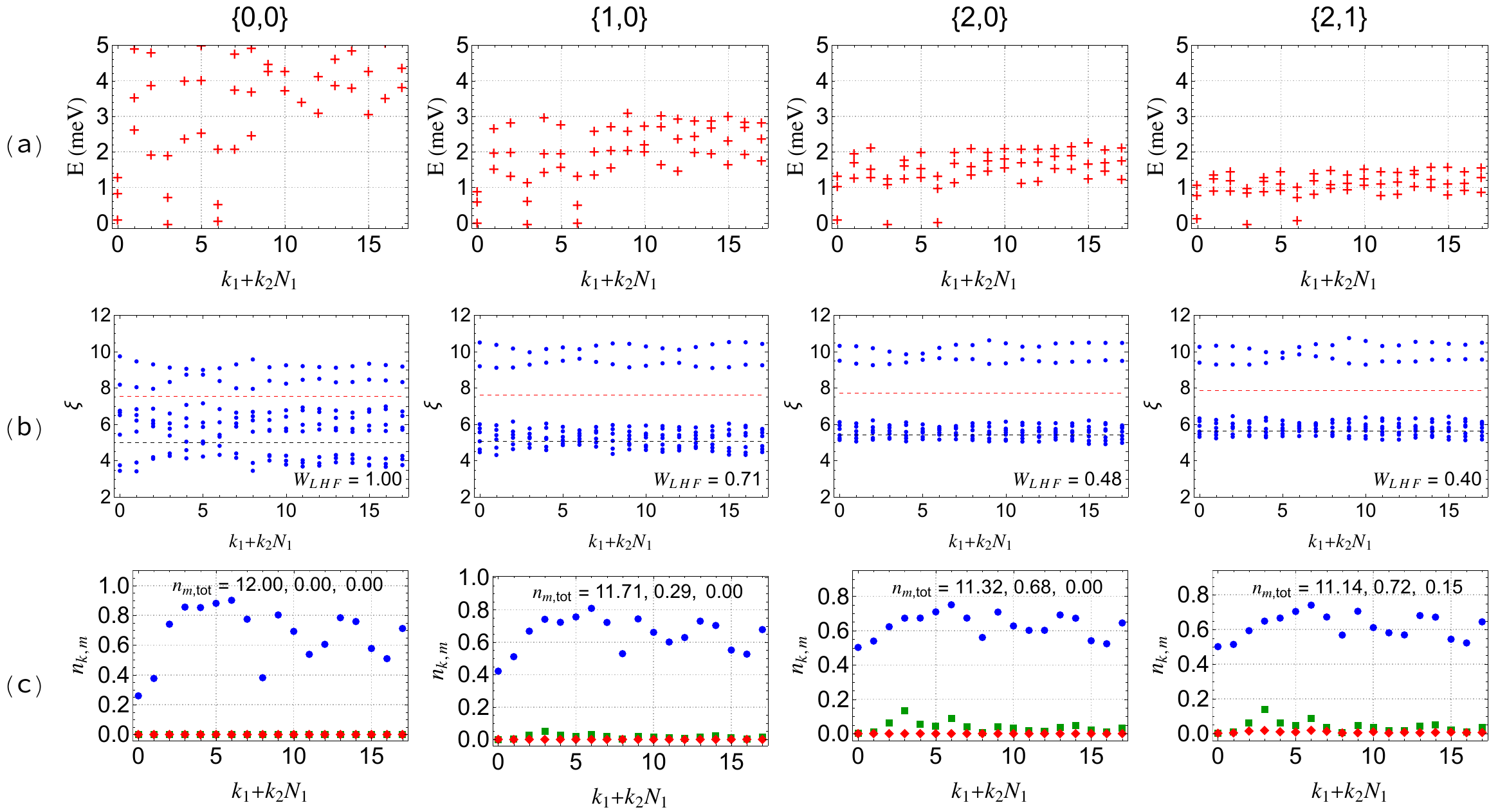}
\caption{ Energy spectrum, PES, and occupation number of the $9\times 2$ system at 2/3 filling in the AVE scheme with $V=22$ meV. Different columns represent ED results computed at different truncation parameters. (a) Energy spectrum. The energies are globally shifted by a constant so that the ground state energy is set to zero. (b) PES of the three lowest states at FCI momenta $0,3,6$ after single-band projection and PH transform. There are 117 states below the red line, which marks the FCI counting, and 45 states below the black line, which marks the CDW counting. (c) Occupation number $n_{\bsl k,m}$ as defined in Eq.~\eqref{eq:bandoccupationperk}. Blue, green, and red colors correspond to $m=0,1,2$. The inset shows the total occupation number $n_{m,\text{tot}}$ for band $m=0,1,2$ respectively (see Eq.~\ref{eq:totalbandoccupation}). }
\label{fig_AVE9b2n12i0ch}
\end{figure*}

The ED results in the HF basis for the $9\times 2$ system at 1/3 and 2/3 filling are shown in Figs.~\ref{fig_AVE9b2n6i0ch} and~\ref{fig_AVE9b2n12i0ch}, respectively. At 1/3 filling, in contrast to the CN scheme, which exhibits a transition from an FCI to a CDW under small band-mixing in the PES, no signature of an FCI or CDW is observed in the AVE scheme as soon as we move away for the single HF band calculation.  Note that the lowest three many-body states coincide with the FCI momenta at $k_1+k_2N_1=9,12,15$ only in the single-band limit $\{0,0\}$, and are at wrong momenta for an FCI for \emph{any} amount of fluctuations. Thus the PES that are shown are for the lowest states in the FCI momenta; these are not the ground-states of the system, but excited states. Nonetheless, even for the excited states, we do not see an FCI in the PES with any fluctuations.

At 2/3 filling, the lowest three states remain at the FCI momenta in the presence of weak band-mixing. When focusing on the momentum sectors where the FCI should appear, the PES in the AVE scheme at 2/3 filling exhibits a clear gap at the FCI counting at moderate band-mixing and even when the weight on the lowest HF band is smaller than 0.5. Actually, as far as we can go in the band truncation parameters and irrespective how low is ${\cal W}_{\text{LHF}}$, the PES of the three lowest energy states in the FCI momentum sector still exhibits clear entanglement gap (see App.~\ref{app_allfig}). Note that these states are \emph{not} the absolute ground states for large band mixing, as discussed in Ref.~\cite{MFCIIV}. However, the previously observed robustness of the FCI in the AVE scheme at 2/3 filling against moderate band-mixing compared to either 1/3 or 2/3 filling in the CN scheme is confirmed by our PES results.

\section{Optimizing the one-body basis by ED iteration}
\label{Sec_itermethod}

Performing exact diagonalizations in multiple active bands requires to choose a specific one-body basis to express the matrix elements. As discussed in Sec.~\ref{Subsec_HFtruncation}, in absence of truncation of the band occupations, any one-body basis obtained by unitary transformation of the band basis, such as the HF basis, leads to the same many-body energy spectrum. Such a property breaks down when we apply a truncation of the remote bands occupation. One could and has wondered if a basis choice could help converging faster (with respect to the truncation parameters, or with adding fluctuations) to the full fledged multiple band calculation, providing access for larger system sizes that would be otherwise computationally too costly even for moderate band mixing. A more attainable goal could just be to maximize the {\it average occupation of the low energy states on the lowest band} to make the small truncation calculations more relevant.  

In absence of truncation, finding a basis transformation maximizing the average occupation of the ground states defining $\{\text{GS}\}$ onto one band at every $\bsl k$ point can be obtained as follows. Let us define the one-body density matrix $\rho(\bsl k)$ computed for the states in $\{\text{GS}\}$
\bea
\rho_{mn}(\bsl k)&=&\frac{1}{N_{\text{GS}}}{\text{Tr}} \left[d^\dagger_{\bsl k,n}d_{\bsl k,m} \mathcal{P}_{\text{GS}} \right].\label{eq:defrho}
\eea 
Here $d_{\bsl k,n}^\dagger$ denotes the creation operator in a generic basis, $n$ being dubbed the band index but is strictly speaking only a label for the available one body states at fixed $k$ if we do not work in the band basis.
We define a unitary transformation $U$ such that the transformed density matrix 
\bea
\Lambda_{m,n}(\bsl k)&=&\left( U^\dagger(\bsl k) \rho(\bsl k) U(\bsl k)  \right)_{mn}.
\label{eq:deflamdbarho}
\eea
has a maximal entry $\Lambda_{0,0}(\bsl k)$. Since $\rho(\bsl k)$ is a positive definite matrix, we can show that $\Lambda(\bsl k)$ has to be the diagonal matrix made out of the eigenvalues of $\rho(\bsl k)$ with $\Lambda_{0,0}(\bsl k)$ being the largest eigenvalue (we will assume that $\Lambda_{0,0}(\bsl k)\ge \Lambda_{1,1}(\bsl k)\ge... \ge 0$). {To see this, we can investigate an arbitrary basis where the density matrix is non-diagonal, which is denoted as $\rho'(\bsl k)$. Then $\rho'(\bsl k)$ can be diagonalized by $\rho'(\bsl k)=U'(\bsl k)\Lambda(\bsl k)U'^\dagger(\bsl k)$ for some unitary matrix $U'(\bsl k)$. Then we have}
\bea
\rho'_{0,0}(\bsl k)&=&\sum_{j=0}^2\Lambda_{j,j}(\bsl k)|U'_{0,j}(\bsl k)|^2\\
&\le&\sum_{j=0}^2\Lambda_{0,0}(\bsl k)|U'_{0,j}(\bsl k)|^2 \\
&=&\Lambda_{0,0}(\bsl k),
\eea 
{where the last line follows from $U'(\bsl k)$ being a unitary matrix. Therefore, the 0,0 entry of density matrix is maximized in the basis where the density matrix is diagonal, and using $U(\bsl k)$- which diagonalizes the density matrix- to define the new basis would guarantee that the average occupation for band $0$ at each $\bsl k$ is maximal.}

From there, we devise a self-consistent procedure, dubbed {\it ED iteration}, that aims at maximizing the many-body ground state occupation on the lowest band at every $\bsl k$ point in the truncated Hilbert space. A generic Hamiltonian with both single-particle terms and two-body interactions can be written as
\bea
\hat H&=\sum_{\bsl k,n_1,n_2}&h_{n_1,n_2}(\bsl k)d^\dagger_{\bsl k,n_1}d_{\bsl k,n_2}\nonumber\\
& + \sum_{\{\bsl k,n\}} & V_{\bsl k_i,n_i,\bsl k_j,n_j,\bsl k_{j'},n_{j'},\bsl k_{i'},n_{i'}} \label{Hgene}\\
&&\times d^\dagger_{\bsl k_i,n_i}d^\dagger_{\bsl k_j,n_j}d_{\bsl k_{j'},n_{j'}}d_{\bsl k_{i'},n_{i'}}  \delta_{\bsl k_i+\bsl k_j,\bsl k_{i'}+\bsl k_{j'}}.\nonumber
\eea
Here, the summation $\{\bsl k,n\}$ runs over all momenta in the MBZ and all active bands. Note that the one-body term $h_{n_1,n_2}$ is not assumed to be diagonal. As pointed out previously, we use the terminology ``band'' for the index $n$ to connect with name used for the Hamiltonian before any iteration. As indicated by Eq.\eqref{Hgene}, the kinetic part of the Hamiltonian does not have to be diagonal (and will not be in general during the procedure). The truncated Hilbert space allows at most $N_{\text{band } n}$ particles in band $n$ for $n>0$, while the number of particles in band 0 is unrestricted. The ED computation in this truncated Hilbert space yields the energy spectrum, from which we define a collection of ground states $\{\text{GS}\}$. Typically at 1/3 or 2/3 filling, $\{\text{GS}\}$ will consist of the three FCI states. We compute the  one-body density matrix $\rho(\bsl k)$ defined in Eq.~\eqref{eq:defrho} and diagonalize it according to Eq.~\eqref{eq:deflamdbarho}. Next at each $\bsl k$ point, we use $U(\bsl k)$ to define a new basis $\tilde d^\dagger_{\bsl k,n}$
\be
\tilde d^\dagger_{\bsl k,n}=\sum_{n'}d^\dagger_{\bsl k,n'} U_{n'n}(\bsl k).
\label{tilded}
\ee
The Hamiltonian written in the new basis $\{\tilde d^\dagger_{\bsl k,n}\}$ takes the form:
\bea
\hat H&=\sum_{\bsl k,n_1,n_2}&\tilde{h}_{n_1,n_2}(\bsl k)\tilde d^\dagger_{\bsl k,n_1}\tilde d_{\bsl k,n_2}\nonumber\\
& + \sum_{\{\bsl k,n\}} & \tilde{V}_{\bsl k_i,n_i,\bsl k_j,n_j,\bsl k_{j'},n_{j'},\bsl k_{i'},n_{i'}} \label{hnewbasis}\\
&&\times \tilde d^\dagger_{\bsl k_i,n_i}\tilde d^\dagger_{\bsl k_j,n_j}\tilde d_{\bsl k_{j'},n_{j'}}\tilde d_{\bsl k_{i'},n_{i'}}  \delta_{\bsl k_i+\bsl k_j,\bsl k_{i'}+\bsl k_{j'}}.\nonumber
\eea
where
\bea
&&\tilde h_{n_1,n_2}(\bsl k)=\left(U^\dagger(\bsl k) h(\bsl k) U(\bsl k) \right)_{n_1,n_2}, \label{eq:htilde}
\eea
and
\bea
&&\tilde V_{\bsl k_1,n_1,\bsl k_2,n_2,\bsl k_3,n_3,\bsl k_4,n_4} \nonumber\\
&=&\sum_{n'_1,n'_2,n'_3,n'_4} V_{\bsl k_1,n'_1,\bsl k_2,n'_2,\bsl k_3,n'_3,\bsl k_4,n'_4} \label{eq:Vtilde}\\
&&\quad\times U_{n_1',n_1}(\bsl k_1)U_{n_2',n_2}(\bsl k_2)U^*_{n_3',n_3}(\bsl k_3)U^*_{n_4',n_4}(\bsl k_4).\nonumber
\eea
The Hamiltonian $\hat H$ in Eq.~\eqref{hnewbasis} is mathematically equivalent to Eq.~\eqref{Hgene}, just expressed with the new operators $\{\tilde d^\dagger_{\bsl k ,n}\}$ and updated coefficients $\tilde h$ and $\tilde V$. But once projected onto a truncated Hilbert space, different bases lead to different spectrum and eigenstates. We thus iteratively repeat the operation by computing the spectrum of Eq.~\eqref{hnewbasis}, updating the density matrix $\tilde{\rho}(\bsl k)$, and computing a new optimal basis out of it. This ED iteration procedure generates a sequence of new bases defined by the creation operators $\{\tilde d^{\dagger (0)}_{\bsl k ,n}\}$,...,$\{\tilde d^{\dagger ({\cal N}_{\text{it}})}_{\bsl k ,n}\}$, where ${\cal N}_{\text{it}}$ is the number of iterations and $\{\tilde d^{\dagger (0)}_{\bsl k ,n}\}\equiv\{d^\dagger_{\bsl k ,n}\}$. To decide when we should stop iterating, we propose the following convergence criteria. Denoting $\rho^{(s)}(\bsl k)$ the density matrix at iteration $s$ ($s=0$ being the density matrix prior to any transformation), we define the quantity
\bea
{\cal R}^{(s)}&=&\frac{{\text{max}}_{\{\bsl k,m\}}\left|\Lambda^{(s)}_{m,m}(\bsl k) - \Lambda^{(s-1)}_{m,m}(\bsl k)\right|}{\text{max}_{\{\bsl k,m\}}\Lambda^{(s)}_{m,m}(\bsl k)}\label{eq:convergencedef}
\eea
where $\Lambda^{(s)}_{m,m}(\bsl k)$ are the eigenvalues of $\rho^{(s)}(\bsl k)$ as defined by Eq.~\eqref{eq:deflamdbarho}. We then choose an accuracy $\varepsilon$ and fix the number of iteration ${\cal N}_{\text{it}}$ to be the minimal number such that
\bea
{\cal R}^{({\cal N}_{\text{it}})}&<&\varepsilon. \label{eq:convcriteria}
\eea

If we consider a single GS consisting of a product state with either exactly 0 or 1 particle per momentum sector, this procedure would be exact. Indeed, let $|\psi\rangle$ be
\be
|\psi\rangle=\prod_{\bsl k}\left(\sum_{m=0}^2 A_m(\bsl k) d^\dagger_{\bsl k,m}\right)|0\rangle\label{eq:productstate}
\ee 
where the $A_m(\bsl k)$'s are some arbitrary coefficients and the product over $\bsl k$ is only over the occupied momenta. The one-body density matrix is then
\bea
\rho_{mn}(\bsl k)&=&\langle \psi| d^\dagger_{\bsl k,n} d_{\bsl k,m}|\psi\rangle  \\
&=&A_n^*(\bsl k)A_m(\bsl k)\\
\eea
We immediately see that $\rho_{mn}$ is a rank one matrix. It has a single non-zero eigenvalue 1 whose corresponding eigenvector can be used to define the new band 0, i.e., $U_{m0}(\bsl k)=A_m(\bsl k)$. From the iteration formula in Eq.~\eqref{tilded}, we thus get 
\be
|\psi\rangle=\prod_{\bsl k} \tilde d^\dagger_{\bsl k, 0}|0\rangle\label{eq:productstateband0}
\ee
showing that the ED iteration procedure would converge after one step and would lead to a GS solely occupying one band.

Note that we did not prove that this algorithm converges irrespective of the initial Hamiltonian. In fact, we can find counter-examples (see App.~\ref{app_notconverge}). Still, for most cases, as we will discuss in Sec.~\ref{Sec_afteriter}, convergence can be reached quickly, typically after a few tens iterations for $\varepsilon=10^{-3}$. Similarly, the algorithm does not guarantee that the largest band occupation after the procedure is higher than the largest band occupation before any iteration (see App.~\ref{app_notconverge}). There are also trivial cases where the convergence is reached after a single iteration. As discussed previously, it is trivially true if no truncation is applied. Also, if the truncation parameter is chosen to be $\{0,0\}$, i.e., a single-band calculation, the physical quantities including the energy spectrum, PES, and occupation numbers will remain unchanged after a single step. Indeed, the basis rotation is only a $U(1)$ transformation that does not mix different bands.

\section{Numerical results with ED iteration method}
\label{Sec_afteriter}

\begin{figure*}
\centering
\includegraphics[width=6.8 in]{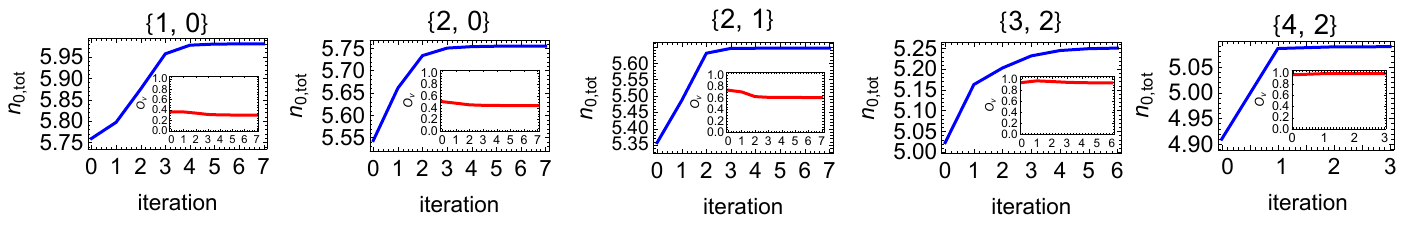}
\caption{  Total occupation number in band 0 summed over all momentum as a function of iteration for the $9\times 2$ system at 1/3 filling in the CN scheme in the HF basis. The inset shows the overlap between the ground state with Hilbert space truncation and the exact ground state without truncation as a function of the number of ED iterations.}
\label{fig_ovn0iter}
\end{figure*}

\begin{figure*}
\centering
\includegraphics[width=6.8 in]{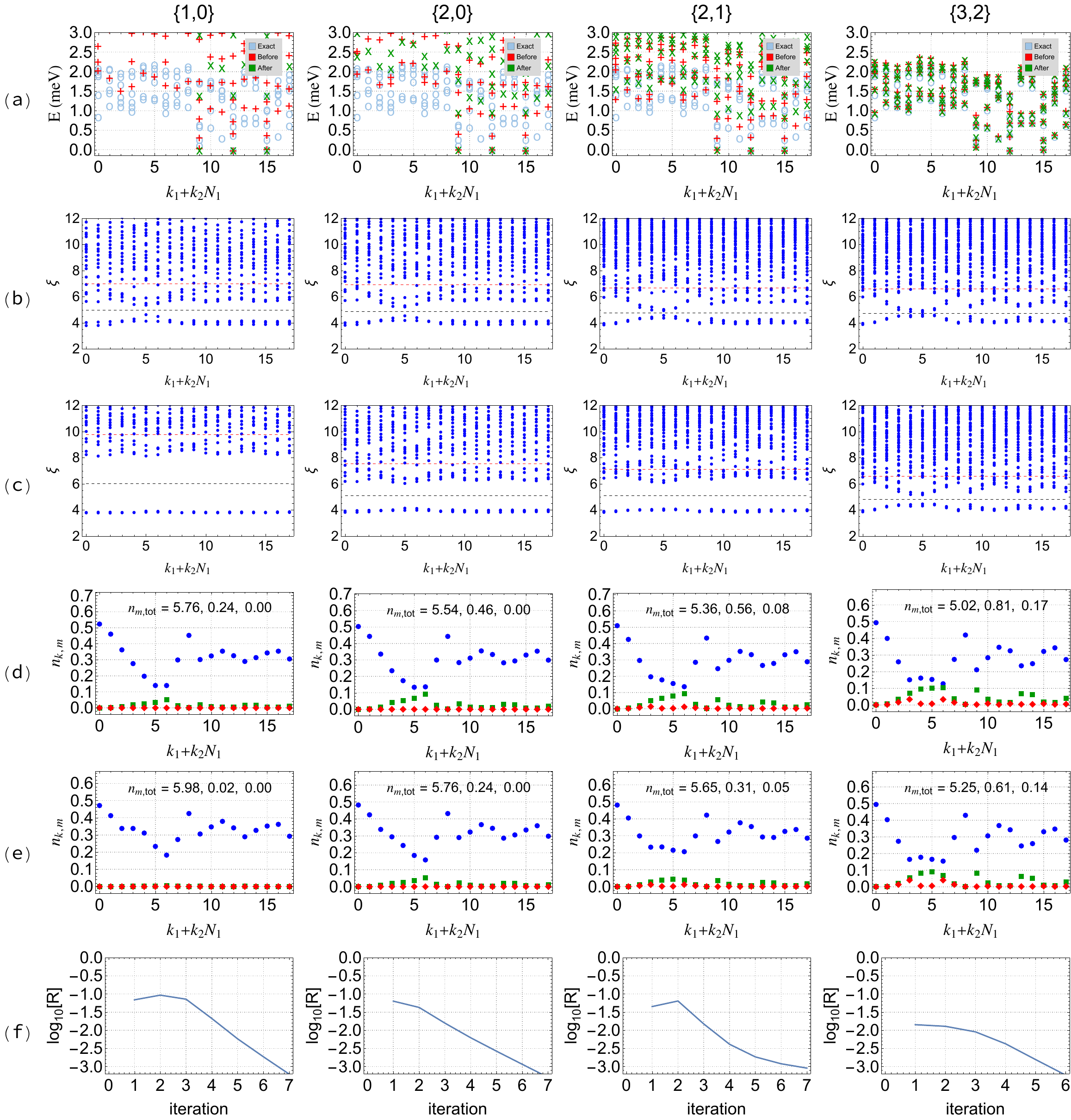}
\caption{ Energy spectrum, PES, and occupation number of the $9\times 2$ system at 1/3 filling in the CN scheme with $V=28$ meV. Different columns represent ED results computed at different truncation parameters. (a) Energy spectrum before and after iterations and the exact energy spectrum computed without truncation of the Hilbert space at $\{6,6\}$. Each energy spectrum is globally shifted by a constant such that the ground state energy is set to zero. (b,c) PES of the three lowest states at FCI momenta $9,12,15$ before (b) and after (c) iteration. There are 117 states below the red line, which marks the FCI counting, and 45 states below the black line, which marks the CDW counting. (d) Occupation number $n_{\bsl k,m}$ as defined in Eq.~\eqref{eq:bandoccupationperk}. Blue, green, and red colors correspond to $m=0,1,2$. The inset shows the total occupation number $n_{m,\text{tot}}$ for band $m=0,1,2$ respectively (see Eq.~\eqref{eq:totalbandoccupation}). (e) Occupation number computed in the basis after iteration. (f) Convergence criteria defined in Eq.~\eqref{eq:convergencedef} as a function of iteration step.}
\label{fig_CN9b2n6ch}
\end{figure*}

In this section, we apply the ED iteration method and investigate the results in the rotated basis. As for Sec.~\ref{Sec_beforeiter}, we focus on  the $9\times 2$ system, providing additional sizes in App.~\ref{app_allfig}. We start with the filling factor $\nu=1/3$ in the CN scheme. The $\{\text{GS}\}$ is chosen to contain the lowest state at each of the three FCI momenta. These momenta are also compatible with the CDW observed at intermediate band mixing (see Sec.~\ref{subsec:CNonethird}), \emph{no matter whether these sectors are the absolute lowest ones}. For all calculations, we always start from the HF basis with the lowest band having Ch=1. Their corresponding energy spectrum, momentum space occupation and the PES were already discussed in Sec.~\ref{subsec:CNonethird}. As shown in Fig.~\ref{fig_CN9b2n6ch}(f), the convergence is usually reached within less than 10 iterations irrespective of the band truncation parameters. As an indicator of the low energy manifold evolution under the procedure, we track the total occupation number $n_{0,\text{tot}}$ in band 0 as a function of the iteration step. In Fig.~\ref{fig_ovn0iter}, we show the evolution of $n_{0,\text{tot}}$ for different truncations of the Hilbert space $\{N_{\text{band}1},N_{\text{band}2}\}$. Note that for this system size and filling factor, we can actually perform the full calculation without truncation, corresponding here to $\{6,6\}$ which will be convenient for using as a comparison point or overlap calculations. For any truncation, we observe that the ED iteration indeed increases the occupation of the lowest band as expected from the algorithm, albeit moderately (at most around $6\%$). We show in { App.~\ref{app_nfluc}} that the fluctuations of the lowest band occupation are also damped.

\begin{figure*}
\centering
\includegraphics[width=6.0 in]{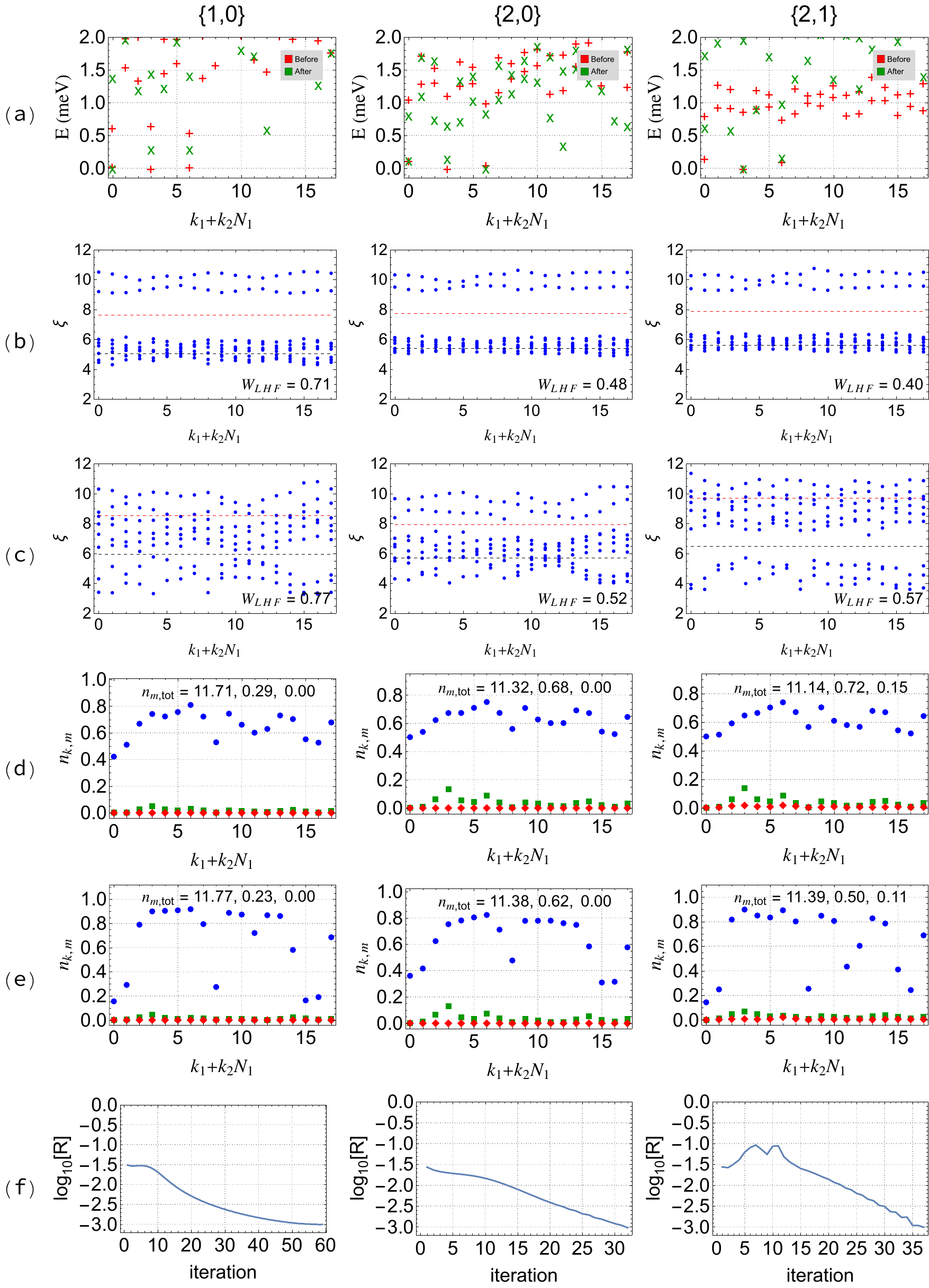}
\caption{ Energy spectrum, PES, and occupation number of the $9\times 2$ system at 2/3 filling in the AVE scheme with $V=22$ meV. Different columns represent ED results computed at different truncation parameters. (a) Energy spectrum before and after iterations. Each energy spectrum is globally shifted by a constant such that the ground state energy is set to zero. (b,c) PES computed from the three lowest states after single-band projection and PH transform at FCI momenta $0,3,6$ before (b) and after (c) iteration. There are 117 states below the red line, which marks the FCI counting, and 45 states below the black line, which marks the CDW counting. (d) Occupation number $n_{\bsl k,m}$ as defined in Eq.~\eqref{eq:bandoccupationperk}. Blue, green, and red colors correspond to $m=0,1,2$. The inset shows the total occupation number $n_{m,\text{tot}}$ for band $m=0,1,2$ respectively (see Eq.~\eqref{eq:totalbandoccupation}). (e) Occupation number computed in the basis after iteration. (f) Convergence criteria defined in Eq.~\eqref{eq:convergencedef} as a function of iteration step. }
\label{fig_AVE9b2n12ch}
\end{figure*}

A deeper analysis of the converged ground state manifold can be performed through the PES and the momentum resolved band occupation. For the former, Figs.~\ref{fig_CN9b2n6ch}(b) and (c) display the PES for the ground state manifold before and after the ED iteration procedure for different band truncation parameters. Interestingly, the ED iteration tends to reinforce the intermediate CDW phase discussed in Sec.~\ref{subsec:CNonethird}, increasing the entanglement gap between the CDW entanglement levels and the continuum. Similarly, the momentum resolved band occupation is provided in Figs.~\ref{fig_CN9b2n6ch}(d) and (e). Overall, the lowest band occupation $n_0(\bsl k)$ is enhanced while $n_1(\bsl k)$ and $n_2(\bsl k)$ are suppressed, in agreement with the above-mentioned results for $n_{0,\text{tot}}$. But we notice that the actual distribution with respect to the momentum $\bsl k$ is barely affected. 

To test if the ED iteration allows to alleviate the Hilbert space truncation, we further compute the overlap between the ground states obtained in the truncated Hilbert space and the exact ground states $|\Psi_a^{\text{ex}}\rangle$ without truncation, where $a=0,1,2$ labels the three exact ground states. Denote the projection operator to the manifold of exact ground states as $\mathcal{P}_{\text{GS}}^{\text{ex}}=\sum_a |\Psi_a^{\text{ex}}\rangle \langle \Psi_a^{\text{ex}}|$, and let $\mathcal{P}_{\text{GS}}$ be the projection to ground states computed with truncation, the overlap $O_v$ is defined by
\be
O_v= \frac{1}{N_{\text{GS}}}\text{Tr}\left[ \mathcal{P}_{\text{GS}}\mathcal{P}_{\text{GS}}^{\text{ex}}  \right].
\label{ovexact}
\ee
During the iteration process for a given Hilbert space truncation $\{N_{\text{band}1},N_{\text{band}2}\}$, we obtain a new basis $\tilde d_{\bsl k,\alpha}^{(i)}$ at each iteration step $i$. The ground state $|\Psi_a\rangle$ is computed in the basis $\tilde d_{\bsl k,\alpha}^{(i)}$ under the same Hilbert space truncation $\{N_{\text{band}1},N_{\text{band}2}\}$, while the exact ground state is obtained without truncation at $\{N,N\}$, where $N$ is the total particle number. We remind that in absence of truncation, the unitary transformation acting on the single particle basis does not affect the many-body Hamiltonian. We use this feature to bypass the direct transformation of many-body eigenstates under the change of single particle basis that can increase the complexity factorially with the number of particles.

The evolution of $O_v$ at each iteration step is shown in the inset of Fig.~\ref{fig_ovn0iter}. Overall, the ED iteration has no major effect on the overlap with the exact states. It even has a slightly tendency to decrease the overlap during the procedure except when the truncation of the Hilbert space is weak, i.e., close to the untruncated case, leading to a slight increase of the overlap. We could wonder if the ED iteration could at least brings us closer to a lower truncation of the Hilbert space. As discussed in App.~\ref{app_wfoverlap}, we indeed observed such a feature, although the improvement is not substantial in most of the cases.

Finally, we now turn to the effect of the ED iteration method beyond the GS manifold. The energy spectra before and after the ED iteration procedure for different truncation of the Hilbert space are given in Fig.~\ref{fig_CN9b2n6ch}(a). For sake a comparison, we also provide the low energy spectrum without any truncation.  After iteration, at truncation parameter $\{1,0\}$ there is an enhance of the energy gap, in line with the larger entanglement gap on top of the CDW entanglement levels as discussed previously. When increasing the band mixing, i.e., for larger $N_{\text{band}1}$ and $N_{\text{band}2}$, we observe that overall the low energy excitations display a quite similar dispersion with respect to the spectrum before iteration up to global shift (due to a larger gap on top of the GS manifold) that decreases when moving toward the untruncated Hilbert space.

The system at filling $\nu=2/3$ in the CN scheme or at $\nu=1/3$ in the AVE scheme hosts neither an FCI nor a CDW, making the ED iteration method less relevant due to the non obvious choice of a GS manifold. Still, we provide a complete study in App.~\ref{app:additionalED}. We also investigate the effect of ED iteration in the AVE scheme at 2/3 filling, where an FCI emerges both in the single-band limit $\{0,0\}$ and under weak band mixing with small $N_{\text{band}1}$ and $N_{\text{band}2}$ (see Sec.~\ref{subsec:AveHFBasis}) The results for the $9\times 2$ system are shown in Fig.~\ref{fig_AVE9b2n12ch}, with the ground states $\{\text{GS}\}$ chosen to contain the three lowest states at the FCI momenta $k_1+k_2N_1=0,3,6$. Due to the large number of particles at 2/3 filling, the exact energy spectrum in the untruncated Hilbert space (namely $\{12,12\}$) is beyond our computational capabilities. The energy spectra, PES and momentum resolved band occupations before and after iteration are shown in Figs.~\ref{fig_AVE9b2n12ch}(a), (b), (c), (d) and (e). We also provide the evolution of the convergence criteria in Fig.~\ref{fig_AVE9b2n12ch}(f). We observe that the convergence is slower than $\nu=1/3$ in the CN scheme, with around 30 iteration steps before reaching convergence. Remarkably, every indicator of the FCI phase, including the spread of the ground state manifold, the energy gap, the PES entanglement gap or the occupation number per momentum sector, points toward a destabilization of the FCI phase. Actually, the lowest three states are no longer at the FCI momenta for $\{2,1\}$ after iteration. Therefore for the AVE $\nu=2/3$ case, the ED iteration process drives the ground states away from the FCI phase, in line with the its fragility to large band mixing pointed out in Ref.~\cite{MFCIIV}.

\section{Conclusion}
\label{Sec_conclusion}

We have investigated the stability of fractional Chern insulating phases in rhombohedral pentalayer graphene on aligned hBN substrate using multi-band exact diagonalization. Our calculations focused on filling factors $\nu=1/3$ and $\nu=2/3$, considered interactions in both the charge neutrality and average schemes and incorporated inter-band mixing among the three lowest conduction bands.

In the CN scheme at 1/3 filling, we found that the FCI phase predicted by single-band ED is not stable under the minimal band mixing. Allowing even a small number of electrons to occupy higher bands immediately destroys the FCI, giving way to a CDW, as identified through the particle entanglement spectrum and density correlation functions. This illustrates how band mixing—unavoidable in a nearly gapless \text{moir\'e} system—can substantially alter the ground state. In contrast, the FCI at 2/3 filling in the AVE scheme exhibits a greater resilience to band mixing; small mixing with higher bands does not immediately eliminate the FCI. However, when band mixing is strong, the FCI phase still vanishes as indicated by the collapse of FCI energy gap~\cite{MFCIIV} and PES gap.

We also introduced the ED iteration method to tentatively optimize the single-particle basis such that the ground state manifold maximizes its weight on the lowest band. Although this strategy does not completely eliminate Hilbert space truncation error, it provides a way to test how different basis choices affect our results. Even after convergence of the ED iteration method, however, the FCI phases that are observed in single HF band models are washed out when higher-band states are partially occupied. Taken altogether, our comprehensive research shows the shortcomings of  \emph{all} current models to explain the experimental observations of FCIs in rhombohedral graphene.

\acknowledgments

We thank Erez Berg, Jiabin Yu, Jonah Herzog-Arbeitman and Yves Kwan for fruitful discussions. H.L. was supported by the European Research Council (ERC) under the European Unions Horizon 2020 research and innovation program (Grant Agreement No. 101020833). B.A.B. was supported by the Gordon and Betty Moore Foundation through Grant No.~GBMF8685 towards the Princeton theory program, the Gordon and Betty Moore Foundation’s EPiQS Initiative (Grant No.~GBMF11070), the Office of Naval Research (ONR Grant No.~N00014-20-1-2303), the Global Collaborative Network Grant at Princeton University, the Simons Investigator Grant No.~404513, the BSF Israel US foundation No.~2018226, the NSF-MERSEC (Grant No.~MERSEC DMR 2011750), the Simons Collaboration on New Frontiers in Superconductivity (SFI-MPS-NFS-00006741-01), and the Schmidt Foundation at the Princeton University. The Flatiron Institute is a division of the Simons Foundation.

\input{reference.bbl}

\appendix


\onecolumngrid

\tableofcontents

\clearpage
\newpage

\section{Review of the single particle model}
\label{Sec_h0}

Following Refs.~\cite{MFCI3,MFCIIV}, the single particle Hamiltonian $h_{0,\eta}(\bsl r)$ in valley $\eta=\pm K$ contains the rhombohedral pentalayer graphene Hamiltonian $h_{R5G}^\eta$ and the \text{moir\'e} potential term $h_{\text{moir\'e}}$ induced by the hBN near the bottom layer graphene:
\bea
h_{0,\eta}(\bsl r)&=&h_{\text{R5G}}^\eta(-i\nabla)+h_{\text{moir\'e}}(\bsl r).\ \ 
\eea
Here the Hamiltonians with lower case $h$ refers to the first-quantized form, and the second-quantized Hamiltonian is
\bea
H_{0,\eta}&=&\int d\bsl r [h_{0,\eta}(\bsl r)]_{l\sigma,l'\sigma'} c^\dagger_{\bsl{r}, l\sigma \eta s} c_{\bsl{r}, l'\sigma' \eta s} , 
\eea
where $\bsl{r}=(x,y)$ is the continuum 2D position, $l=0,1,...,4$ is the layer index, $\sigma=A,B$ represents the sublattice, $\eta=\pm \K$ labels the valley, and $s=\uparrow,\downarrow$ is the spin index. The hBN is near the bottom layer with $l=0$. The matrix Hamiltonian in the $\K$ valley for R5G is
\eqa{ 
\label{eq_main:H_K}
h_{\text{R5G}}^K(\bsl{p}) &= \bpm
v_F\mbf{p} \cdot \pmb{\sigma}  & t^\dag(\mbf{p}) & t'^\dagger &   &\\
t(\mbf{p}) & \ddots & \ddots & t'^\dagger \\
t' & \ddots & v_F\mbf{p} \cdot \pmb{\sigma} & t^\dagger(\mbf{p})\\
& t' & t(\mbf{p})  & v_F\mbf{p} \cdot \pmb{\sigma}
\epm+h_{ISP}+h_D, 
}
where $\bsl{p}=-\ii \nabla$, $\bsl{\sigma}=(\sigma_x,\sigma_y)$ are Pauli matrices in sublattice space, $t(\mathbf{p})$ and $t'$ are $2\times 2$ matrices given by:
\eq{
t(\mbf{p}) = -\bpm v_4 p_+ & -t_1 \\ v_3 p_- &  v_4 p_+ \epm, \qquad  \qquad t' = \bpm 0 & 0 \\ t_2 & 0 \epm\ .
}
Here $p_\pm = p_x \pm \ii p_y$,  $v_F$ is the Fermi velocity of graphene and $t_1,t_2,v_3,v_4$ are inter-layer hopping parameters. In this work we use the parameters obtained in Ref.~\cite{MFCI2} which incorporates the effect of moir\'e relaxation in the presence of hBN: $v_F =542.1\ \text{meV nm}, v_3 = v_4 = 34\ \text{meV nm},t_1 =  355.16\ \text{meV},  t_2 = -7\ \text{meV}$. The term $h_{ISP}$ is a polarization term that is inversion-symmetric with respect to the layer indices. It characterizes the chemical potential of each graphene layer, which has the form:
\eq{
[h_{ISP}]_{l l'} = V_{ISP} \delta_{ll'} \left| l - 2\right| \sigma_0\ .
}
Here the factor $l-2$ appears because the middle layer has $l=2$, and $V_{ISP} = 16.65$ meV is determined in Ref.~\cite{MFCI2} by fitting with the DFT band structure. The term $h_D$ induced by the displacement field reads
\eq{
\label{eq_main:H_D}
\null [h_{D}]_{l \sigma,l' \sigma'} = V \left(l - 2\right) \delta_{ll'} \delta_{\sigma \sigma'}.
}
When $V>0$, the electrons in the conduction are polarized away from the hBN layer where the moir\'e potential is strong. The moir\'e potential term $h_{\text{moir\'e}}$ induced by hBN has the form
\eq{
\label{eq_main:H_V}
\null [h_{\text{moir\'e}}(\bsl{r})]_{l \sigma,l' \sigma'} = \left[ V(\mbf{r}) \right]_{\sigma\sigma} \delta_{l0}\delta_{ll'}\ ,
}
\eqa{
\label{eq_main:Vxifinal}
V(\mbf{r}) &= V_0 + \left[V_1 e^{i\psi}\sum_{j=1}^3 e^{i \mbf{g}_j\cdot\mbf{r}}\bpm 1& \omega^{-j} \\ \omega^{j+1} &\omega \epm + h.c.\right]\ ,
}
which only acts on the bottom layer of R5G with $l=0$. Here $\omega=e^{i\frac{2\pi}{3}}$, $\mbf{g}_j = R(\frac{2\pi}{3}(j-1)) (\mbf{q}_2-\mbf{q}_3)$, where $R(\phi)$ is the rotation matrix by $\phi$ in counterclockwise direction. The $\mbf{q}$ vectors are defined as
\eq{
\label{eq_main:qvecmain}
\mbf{q}_1 = \mbf{K}_G - \mbf{K}_{\text{hBN}} = \frac{4\pi}{3 a_G}\left(1 - \frac{R(-\th)}{1+\epsilon_h} \right)\hat{x},
}
where $\th$ is the twist angle, $\mbf{K}_G$ and $\mbf{K}_{\text{hBN}}$ are the K vector of graphene and hBN respectively, $a_G = 2.46\AA$ is the graphene lattice constant, and $(1+\epsilon_h)a_G$ is the hBN lattice constant with $\epsilon_h=0.01673$. We also define $\bsl q_{2}=R(\frac{2\pi}{3})\bsl q_1$ and $\bsl q_{3}=R(\frac{2\pi}{3})\bsl q_2$ as $C_3$ partners of $\bsl q_1$.

The \text{moir\'e} potential depends on the stacking configuration of hBN, i.e., whether the carbon \(A\) site aligns with nitrogen ($\xi=1$) or boron ($\xi=0$) as shown in Fig.~\ref{fig_bandbare}(a) in the main text. In this work, we focus on $\xi=1$ with moir\'e parameters $V_0 = 1.50$\,meV, $V_1 = 7.37$\,meV and $\psi=16.55^\circ$. We also choose a twist angle $\theta=0.77^\circ$ in accordance with experiments~\cite{Lu2024PGexp,LuLong2025,PhysRevX.15.011045}. 

Combining all the terms into $h_{0,\eta}$, the single particle Hamiltonian in second-quantized form reads
\bea
H_{0,\eta}&=&\int d\bsl r [h_{0,\eta}(\bsl r)]_{l\sigma,l'\sigma'}\ c^\dagger_{\bsl{r}, l\sigma \eta s} c_{\bsl{r}, l'\sigma' \eta s}  \\
&=&\sum_{\bsl k} [h_{0,\eta}(\bsl k)]_{\bsl G l\sigma,\bsl G'l'\sigma'} \ c^\dagger_{\eta,\bsl{k},\bsl{G},l,\sigma , s} c_{\eta ,\bsl{k},\bsl{G'},l',\sigma' ,s}
\eea
The Fourier transformation is defined by
\eq{
\label{eq:plane_wave_basis}
c^\dagger_{\eta,\bsl{k},\bsl{G},l,\sigma , s} = \frac{1}{\sqrt{\V}} \int d^2 r \, e^{\ii (\bsl{k}+\bsl{G})\cdot\bsl{r}}
c^\dagger_{\bsl{r},l\sigma\eta s} \ ,
}
where $\V$ is the system area, $\bsl k$ is inside the MBZ and $\bsl G$ is a moir\'e reciprocal lattice vector. Diagonalizing the single particle Hamiltonian gives both the eigenvalue $\epsilon_{n}^\eta (\bsl k)$ and eigenvector $U^{\eta }_{n}(\bsl k)$ for the $n$-th band in valley $\eta$:
\bea
[h_{0,\eta}(\bsl k)]_{\bsl G l\sigma,\bsl G'l'\sigma'}&=&\sum_n [U^{\eta }_{n}(\bsl k)]_{\bsl G l\sigma}  \epsilon_{n}^\eta (\bsl k)  [U^{\eta }_{n}(\bsl k)]^*_{\bsl G' l'\sigma'} \nonumber\\
\eea
The creation operator in the band basis can be obtained by
\bea
c^\dagger_{\eta,\bsl{k},n ,s} =  \sum_{\bsl{G}l\sigma}c^\dagger_{\eta,\bsl{k},\bsl{G},l,\sigma , s} \left[ U_{n}^{\eta}(\bsl k) \right]_{\bsl{G}l\sigma}.
\label{cbaretransf}
\eea

\section{ Interaction schemes}
\label{Sec_appscheme}

In this appendix, we provide more details about the interaction Hamiltonian and introduce the different interaction schemes. We follow the notations in Refs.~\cite{MFCI3,MFCIIV}. In the AVE scheme, the interaction is written as
\bea
\label{eq:H_2D_int}
H_{\text{int}} = \int d^2 r d^2 r' V(\bsl{r}-\bsl{r}) \delta \hat{\rho}_{\bsl{r}} \delta \hat{\rho}_{\bsl{r}'}\ , \quad \delta \hat{\rho}_{\bsl{r}} = \sum_{\eta,l,\sigma,s} \left[ c^\dagger_{\bsl{r},l\sigma\eta s} c_{\bsl{r},l\sigma\eta s} - \frac{1}{2}\delta(\bsl{r})\right].
\eea
The Fourier-transformed interaction reads:
\eq{
V(\bsl{r}) = \frac{1}{\V} \sum_{\mbf{q}} \sum_{\mbf{G}} e^{\ii (\bsl{q}+\bsl{G})\cdot\bsl{r}} V(\bsl{q}+\bsl{G}).
}
We assume a layer-independent dual-gate screened Coulomb interaction given by
\eq{
V(\mbf{q}) = \frac{e^2}{2 \eps} \frac{\tanh |\mbf{q}| d_{sc}}{|\mbf{q}|},
}
where the dielectric constant is $\eps = 5\eps_0$ and the sample-to-gate distance is $d_{sc} = 10$\,nm. We can express $H_\text{int}$ in momentum space and decompose it into a normal-ordered two-body interaction term and a residual one-body term:
\eq{\label{eq:Hint_Hintnormord}
H_{\text{int}} = :H_{\text{int}}: + H_{b}^{\text{full}}.
}
Here, the normal-ordering operation $:\hat{O}:$ places all annihilation operators in the conduction band and all creation operators in the valence band to the right while keeping track of fermionic minus signs. Note that this differs from the conventional normal ordering with respect to the vacuum, which places all annihilation operators to the right. The term ``full'' in $H_{b}^{\text{full}}$ indicates that it includes contributions from all conduction and valence bands. The interaction term, written in the plane wave basis, is given by
\eqa{
:H_{\text{int}}: & = \frac{1}{2\V} \sum_{\bsl{q}\bsl{G}} \sum_{\bsl{k}_1 \bsl{G}_1 \eta_1 l_1 \sigma_1 s_1} \sum_{\bsl{k}_2 \bsl{G}_2 \eta_2 l_2 \sigma_2 s_2} V(\bsl{q}+\bsl{G}) \\
& \qquad \times :c^\dagger_{\eta_1,\bsl{k}_1+\bsl{q},\bsl{G}_1+\bsl{G},l_1\sigma_1 s_1} c^\dagger_{\eta_2,\bsl{k}_2-\bsl{q},\bsl{G}_2-\bsl{G},l_2\sigma_2s_2} c_{\eta_2,\bsl{k}_2,\bsl{G}_2,l_2\sigma_2 s_2} c_{\eta_1,\bsl{k}_1,\bsl{G}_1,l_1\sigma_1  s_1}:
}
In the band basis it can be written as
\eqa{
:H_{\text{int}}: & = \frac{1}{2\V} \sum_{\bsl{k}_1 \bsl{k}_2 \bsl{q} } \sum_{\eta_1 \eta_2} \sum_{s_1 s_2} \sum_{n_1 n_2 n_3 n_4}  V_{n_1 n_2 n_3 n_4}^{\eta_1 \eta_2}(\bsl{k}_1,\bsl{k}_2,\bsl{q}):c^\dagger_{\eta_1,\bsl{k}_1+\bsl{q},n_1 ,s_1} c^\dagger_{\eta_2,\bsl{k}_2-\bsl{q},n_2 ,s_2}  c_{\eta_2,\bsl{k}_2,n_3 ,s_2}  c_{\eta_1,\bsl{k}_1,n_4 ,s_1}:  \ ,
}
where
\eqa{\label{appeq:Vetaeta}
V_{n_1 n_2 n_3 n_4}^{\eta_1 \eta_2}(\bsl{k}_1,\bsl{k}_2,\bsl{q})  & = \sum_{\bsl{G}_1 l_1 \sigma_1} \sum_{\bsl{G}_2 l_2 \sigma_2}  \sum_{\bsl{G}}
V(\bsl{q}+\bsl{G}) \left[ U_{n_1}^{\eta_1}(\bsl{k}_1+\bsl{q})\right]^*_{(\bsl{G}_1 + \bsl{G}) l_1 \sigma_1} \left[ U_{n_2}^{\eta_2}(\bsl{k}_2-\bsl{q})\right]^*_{(\bsl{G}_2 - \bsl{G}) l_2 \sigma_2} \\
& \qquad \times \left[ U_{n_3}^{\eta_2}(\bsl{k}_2)\right]_{ \bsl{G}_2 l_2 \sigma_2}  \left[ U_{n_4}^{\eta_1}(\bsl{k}_1)\right]_{ \bsl{G}_1 l_1 \sigma_1}\\
& = \sum_{\bsl{G}} V(\bsl{q}+\bsl{G}) M_{n_1 n_4}^{ \eta_1}(\bsl{k}_1,\bsl{q}+\bsl{G})  M_{n_2 n_3}^{ \eta_2}(\bsl{k}_2,-\bsl{q}-\bsl{G}) 
}
and the form factor is
\eq{
M^{\eta}_{mn}(\bsl{k},\bsl{q}+\bsl{G}) = \sum_{\bsl{G}'l \sigma} \left[U^{\eta}_{m}(\bsl{k}+\bsl{q}+\bsl{G})\right]_{\bsl{G}'l\sigma}^* \left[U^{\eta}_{n}(\bsl{k})\right]_{\bsl{G}'l\sigma}\ .
}

So far, the interaction and single-particle terms have included operators in all conduction and valence bands. When performing ED, we focus on the active bands, which consist of the lowest three HF bands. Consequently, we consider a Hilbert space composed of many-body states where all valence bands are fully occupied, the three active bands are partially filled, and all higher bands remain empty. 

Thus, we restrict the electron operators in the interaction term and the background term $H_b^{\text{full}}$ to act only within the three active bands. Note that this does not imply that the valence bands are neglected. Instead, their effects manifest in the form of the background term $H_b$, which is restricted to the active bands:
\eq{
H_{b} = \sum_{\bsl{k} n_1 n_2 \eta s} c^\dagger_{\eta,\bsl{k},n_1, s} c_{\eta,\bsl{k},n_2, s}  h^{c,\eta}_{b,n_1 n_2}(\bsl{k})\ ,
}
where
\bea
h^{c,\eta}_{b,n_1 n_2}(\bsl{k})  &=& h^{H,\eta}_{b,n_1  n_2 }(\bsl{k}) + h^{F,\eta}_{b,n_1  n_2 }(\bsl{k}) \label{hbglong}
\nonumber\\
h^{H,\eta}_{b,n_1  n_2 }(\bsl{k})&=&\ \ \  - \frac{1}{\V} \sum_{ \bsl{G}} V(\bsl{G})  \sum_{\eta_1 \bsl{k}_1 \bsl{G}_1 l_1 \sigma_1} \left[ \tilde{P}_c^{\eta_1}(\bsl{k}_1) -  \tilde{P}_v^{\eta_1}(\bsl{k}_1) \right]_{\bsl{G}_1 l_1 \sigma_1,(\bsl{G}_1+\bsl{G})l_1 \sigma_1}^*  M^{ \eta}_{n_1 n_2}(\bsl{k},\bsl{G}) \nonumber\\
 h^{F,\eta}_{b,n_1  n_2 }(\bsl{k})&=& \frac{1}{2\V} \sum_{\bsl{q}\bsl{G}} V(\bsl{q}+\bsl{G}) \sum_{\bsl{G}_1'\sigma_1'\bsl{G}_2'\sigma_2' l_1 l_2}\left[ U^{\eta}_{n_1}(\bsl{k}) \right]_{\bsl{G}_1'-\bsl{G}l_2 \sigma_1'}^* \nonumber\\
&&\ \  \times \left[ \tilde{P}_c^{\eta}(\bsl{k} + \bsl{q})- \tilde{P}_v^{\eta}(\bsl{k} + \bsl{q}) \right]_{\bsl{G}_1' l_2 \sigma_1', \bsl{G}_2' l_1 \sigma_2'} \left[ U^{\eta}_{n_2 }(\bsl{k}) \right]_{\bsl{G}_2'-\bsl{G}l_1 \sigma_2'}.
\eea
Here $n_1$ and $n_2 $ are restricted to the three active bands, and we interpret the two terms as the Hartree and Fock terms from the normal ordering procedure. $\tilde{P}_c^\eta(\mbf{k})$ and $\tilde{P}_v^\eta(\mbf{k})$ are projections to the conduction bands and valence bands respectively, which are written as
\bea
\tilde{P}_c^\eta(\mbf{k}) &=& \sum_{n_c} U^\eta_{n_c}(\mbf{k})U^{\eta \dag}_{n_c}(\mbf{k}), \\
\tilde{P}_v^\eta(\mbf{k}) &=& \sum_{n_v} U^\eta_{n_v}(\mbf{k})U^{\eta \dag}_{n_v}(\mbf{k}).
\eea
Here $n_c$ summed over all the conduction bands (not limited to the low-energy active bands) and $n_v$ is summed over all the valence bands.

We make a further assumption by only considering the case where the electrons in the active bands are fully polarized in valley $\K$ and spin $\uparrow$ motivated by the HF computation at integer filling $\nu=1$~\cite{MFCI3,MFCIIV}. Thus we finally arrive at the following effective total many-body Hamiltonian in the AVE scheme:
\eqa{
H_{\text{AVE}} & = \sum_{\bsl{k} n_1  n_2  } c^\dagger_{\K,\bsl{k},n_1 , \uparrow} c_{\K,\bsl{k},n_2 , \uparrow}  \left[h_{0,K}(\bsl{k})+ h^{H,K}_b(\bsl{k})  + h^{F,K}_b(\bsl{k}) \right]_{n_1  n_2 } \\
& \quad + \frac{1}{2\V} \sum_{\bsl{k}_1 \bsl{k}_2 \bsl{q} }  \sum_{n_1 n_2 n_3 n_4}  V_{n_1 n_2 n_3 n_4}^{\K \K}(\bsl{k}_1,\bsl{k}_2,\bsl{q})c^\dagger_{K,\bsl{k}_1+\bsl{q},n_1 ,\uparrow} c^\dagger_{\K,\bsl{k}_2-\bsl{q},n_2 ,\uparrow}  c_{\K,\bsl{k}_2,n_3 ,\uparrow}  c_{\K,\bsl{k}_1,n_4 ,\uparrow}
}
The total Hamiltonian in the CN scheme is obtained by dropping the background terms $h^{H,K}_b(\bsl{k})$ and $h^{F,K}_b(\bsl{k})$:
\eqa{
H_{\text{CN}} & = \sum_{\bsl{k} n_1  n_2  } c^\dagger_{\K,\bsl{k},n_1 , \uparrow} c_{\K,\bsl{k},n_2 , \uparrow}  \left[h_{0,K}(\bsl{k}) \right]_{n_1  n_2 } \\
& \quad + \frac{1}{2\V} \sum_{\bsl{k}_1 \bsl{k}_2 \bsl{q} }  \sum_{n_1 n_2 n_3 n_4}  V_{n_1 n_2 n_3 n_4}^{\K \K}(\bsl{k}_1,\bsl{k}_2,\bsl{q})c^\dagger_{K,\bsl{k}_1+\bsl{q},n_1 ,\uparrow} c^\dagger_{\K,\bsl{k}_2-\bsl{q},n_2 ,\uparrow}  c_{\K,\bsl{k}_2,n_3 ,\uparrow}  c_{\K,\bsl{k}_1,n_4 ,\uparrow}
}

\section{Properties of CDW}
\label{app_CDWktot}

In this appendix, we derive the number of ground states and the corresponding total momentum in ED spectrum for CDW states at $\nu=1/3$ in different system sizes. The CDW order observed in the main text is $K$-CDW, and here we will also include the discussion of $M$-CDW for completeness.

\subsection{ Total momentum and ground state degeneracy of $K$-CDW}

\begin{figure}
\centering
\includegraphics[width=6.8 in]{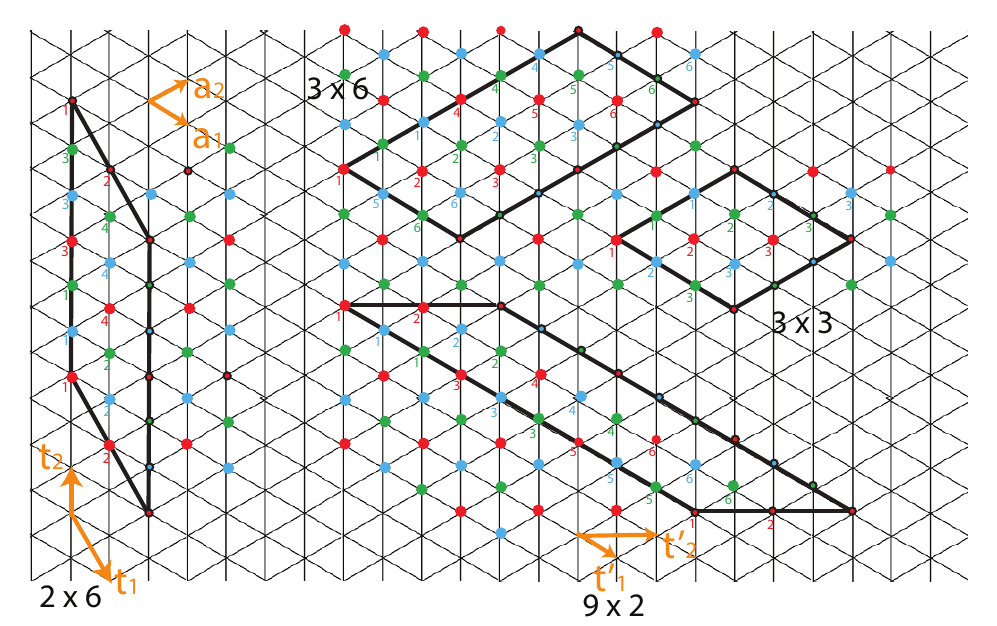}
\caption{ K-CDW order in different system sizes. Red, blue and green dots represent three distinct CDW states. }
\label{fig_Kcdwall}
\end{figure}

The real space pattern of $K$-CDW can be visualized in Fig.~\ref{fig_Kcdwall}, where red, green and blue dots represent charge concentration coming from the three different CDW states. Each CDW state is made of a product of creation operators at sites with the same color. Let $\bsl a_1$ and $\bsl a_2$ be the lattice vectors as in Fig.~\ref{fig_Kcdwall}. The minimal translation vector that keeps the CDW state invariant is $\bsl a_1+\bsl a_2$ with a magnitude of $\sqrt{3}$ times the original lattice vector, which realizes a $\sqrt{3}\times\sqrt{3}$ order. The thick parallelograms represent different system sizes that we are going to study, including $3\times 3,\ 3\times 6,\ 2\times 6$ and $9\times 2$. For each system size we have labeled the sites with the same color by different numbers. Denote red, green, blue sites by $r,g,b$ and each creation operator in real space can be labeled as $c^\dagger_{\mu,n}$ with $\mu=r,g,b$ and $n$ is a positive integer labeling the location inside the extended unit cell. Note that $\mu,n$ is just a different way to label the original unit cell coordinate. We also define $\mu_{i_1...i_n}\equiv c^\dagger_{\mu,i_1}...c^\dagger_{\mu,i_n}|0\rangle$ with $\mu=r,g,b$. For example, $r_{123}=c^\dagger_{r,1}c^\dagger_{r,2}c^\dagger_{r,3}|0\rangle$.

\begin{figure}
\centering
\includegraphics[width=2.5 in]{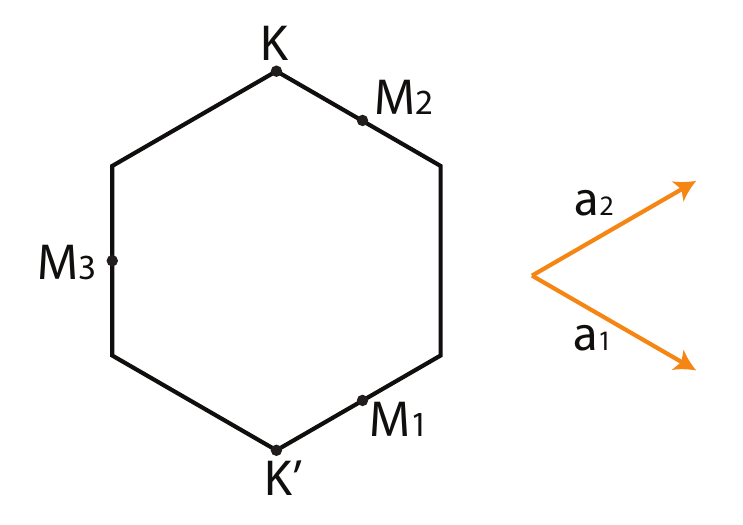}
\caption{ Momentum points in the Brillouin zone.}
\label{fig_BZ}
\end{figure}

\subsubsection{ $K$-CDW in $3\times 3$ system}

As shown in Fig.~\ref{fig_Kcdwall}, the three CDW states in $3\times 3$ lattice are $r_{123},g_{123}$ and $b_{123}$. Under translations by $\bsl a_1,\bsl a_2$, these state transform as:
\bea
&&\bsl a_1:\ r_{123}\rightarrow b_{231}=b_{123},\ b_{123}\rightarrow g_{312}=g_{123},\ g_{123}\rightarrow r_{231}=r_{123} \\
&&\bsl a_2:\ r_{123}\rightarrow g_{123},\ b_{123}\rightarrow r_{123},\ g_{123}\rightarrow b_{123}
\label{trans3b3}
\eea
To construct eigenstates of translation that can be observed in momentum space ED, we define
\be
\psi_j=(r_{123}+e^{i\frac{2\pi j}{3}}g_{123}+e^{i\frac{4\pi j}{3}}b_{123})/\sqrt{3},
\ee
with $j=0,1,2$. Then $\psi_j$ is an eigenstate of translation operator $\hat T(\bsl a_i)$ which shifts all electron creation operators by $\bsl a_i$. The total momentum $\bsl k^{tot}_{j}$ is obtained by:
\bea
\hat T(\bsl a_i)\psi_j=\lambda_j(\bsl a_i)\psi_j,\ \lambda_j(\bsl a_i)=e^{-i\bsl k^{tot}_j\cdot \bsl a_i}
\eea
From Eq.\eqref{trans3b3} we get
\be
\lambda_j(\bsl a_1)=e^{i\frac{2\pi j}{3}},\ \ \lambda_j(\bsl a_2)=e^{-i\frac{2\pi j}{3}}
\ee
We plot the BZ in Fig.~\ref{fig_BZ}. Note that $e^{i\bsl K\cdot \bsl a_1}=e^{-i\frac{2\pi j}{3}},\ e^{i\bsl K\cdot \bsl a_2}=e^{i\frac{2\pi j}{3}},\ e^{i\bsl M_1\cdot \bsl a_1}=-1,\ e^{i\bsl M_1\cdot \bsl a_2}=1$. Therefore we get the total momentum as:
\be
\bsl k^{tot}_0=\Gamma,\ \bsl k^{tot}_1=\bsl K,\ \bsl k^{tot}_2=\bsl K'
\ee

\subsubsection{$K$-CDW in $3\times 6$ system}

As shown in Fig.~\ref{fig_Kcdwall}, the three CDW states in $3\times 6$ lattice are $r_{123456},\ g_{123456}$ and $b_{123456}$. Under translation by $\bsl a_1+\bsl a_2$ and $\bsl a_2$, they transform as:
\bea
&&\bsl a_1+\bsl a_2:\ r_{123456}\rightarrow r_{234561}=-r_{123456},\ g_{123456}\rightarrow g_{234561}=-g_{123456},\ b_{123456}\rightarrow b_{234561}=-b_{123456} \\
&&\bsl a_2:\ r_{123456}\rightarrow g_{123456},\ g_{123456}\rightarrow b_{123456},\ b_{123456}\rightarrow r_{456123}=-r_{123456}
\eea
Note the appearance of minus signs from anti-commutation of fermion operators. The eigenstates of translation can be constructed by:
\be
\psi_j=(r_{123456}+e^{i\frac{2\pi j}{3}+i\frac{\pi}{3}}g_{123456}+e^{i\frac{4\pi j}{3}+i\frac{2\pi}{3}}b_{123456})/\sqrt{3}
\ee
The translation eigenvalues are:
\be
\lambda_j(\bsl a_1+\bsl a_2)=-1,\ \lambda_j(\bsl a_2)=e^{-i\frac{2\pi j}{3}-i\frac{\pi}{3}}
\ee
The total momentum of these states are:
\be
\bsl k^{tot}_0=\bsl K'+\bsl M_2,\ \bsl k^{tot}_1=\bsl M_2,\ \bsl k^{tot}_2=\bsl K+\bsl M_2
\ee
The fermion parity contributes an overall shift of momentum $\bsl M_2$ to the total momentum.

\subsubsection{$K$-CDW in tilted $2\times 6$ system}

The tilted $2\times 6$ lattice in Fig.~\ref{fig_Kcdwall} is generated by translations $\bsl t_1=2\bsl a_1 -\bsl a_2$ and $\bsl t_2=\bsl a_2-\bsl a_1$. Under these translations, the CDW states transform as:
\bea
&&\bsl t_1=2\bsl a_1 -\bsl a_2:\ r_{1234}\rightarrow r_{2143}=r_{1234},\ b_{1234}\rightarrow b_{2143}=b_{1234},\ g_{1234}\rightarrow g_{2143}=g_{1234} \\
&&\bsl t_2=\bsl a_2-\bsl a_1:\ r_{1234}\rightarrow b_{1234},\ b_{1234}\rightarrow g_{1234},\ g_{1234}\rightarrow r_{1234}
\eea
The translation eigenstates can be constructed by:
\be
\psi_j=(r_{1234}+e^{i\frac{2\pi j}{3}}g_{1234}+e^{i\frac{4\pi j}{3}}b_{1234})/\sqrt{3}.
\ee
The translation eigenvalues are:
\be
\lambda_j(2\bsl a_1-\bsl a_2)=1,\ \lambda_j(\bsl a_2-\bsl a_1)=e^{i\frac{2\pi j}{3}}.
\ee
These are equivalent to:
\be
\lambda_j(\bsl a_1)=e^{i\frac{2\pi j}{3}},\ \ \lambda_j(\bsl a_2)=e^{-i\frac{2\pi j}{3}}.
\ee
Therefore the total momentum of these states are
\be
\bsl k^{tot}_0=\Gamma,\ \bsl k^{tot}_1=\bsl K,\ \bsl k^{tot}_2=\bsl K'.
\ee

\subsubsection{$K$-CDW in tilted $9\times 2$ system}

The tilted $9\times 2$ lattice discussed in the main text is shown in Fig.~\ref{fig_Kcdwall}, which is generated by translations $\bsl t'_1=\bsl a_1$ and $\bsl t'_2=\bsl a_1+\bsl a_2$. Under these translations, the CDW states transform as:
\bea
&&\bsl a_1:\ r_{123456}\rightarrow b_{123456},\ b_{123456}\rightarrow g_{123456},\ g_{123456}\rightarrow a_{345612}=a_{123456} \\
&&\bsl a_1+\bsl a_2:\ r_{123456}\rightarrow r_{214365}=-r_{123456},\ b_{123456}\rightarrow b_{214365}=-b_{123456},\ g_{123456}\rightarrow g_{214365}=-g_{123456}
\eea
The translation eigenstates can be constructed by:
\be
\psi_j=(r_{123456}+e^{i\frac{2\pi j}{3}}g_{123456}+e^{i\frac{4\pi j}{3}}b_{123456})/\sqrt{3}.
\ee
The translation eigenvalues are:
\be
\lambda_j(\bsl a_1)=e^{i\frac{2\pi j}{3}},\ \lambda_j(\bsl a_1+\bsl a_2)=-1.
\ee
Notice that $e^{i\bsl M_2\cdot \bsl a_2}=-1$, therefore the total momentum of these states are
\be
\bsl k^{tot}_0=\bsl M_2,\bsl K+\bsl M_2,\bsl K'+\bsl M_2.
\ee
The fermion parity contributes an overall shift of momentum $\bsl M_2$ to the total momentum. This is consistent with the CDW observed in $9\times 2$ system in Sec.~\ref{subsec:CNonethird} in the main text. 

\subsection{Momentum space occupation and PES of $K$-CDW}\label{app_CDWPES}

The formula of $\psi_j$ given above for different system sizes provides the explicit form of CDW wave function with a given total momentum, from which the PES and distribution of occupation number can be computed. 

The PES of this ideal CDW state can give a guideline for identifying CDW states in realistic ED computations, because the counting of PES should be the same for both cases. The form of $\psi_j$ explicitly shows that it is a superposition of three product states which are made of distinct single-particle states. For any of the examples discussed above, the density matrix is  given by
\be
\rho =\frac{1}{3}\sum_{j=0}^{2} \psi_j \psi_j^\dagger =\frac{1}{3}  \sum_{\mu=\{r,g,b\}}\mu_{12...n} \mu_{12...n}^\dagger
\ee
where $n$ is the number of CDW unit cell and $3n$ is the system size. We immediately see that $\rho$ is a sum of the 3 density matrices for product states. Since the PES does not depend on the underlying single-particle basis (as opposed to the orbital entanglement spectrum), the counting of PES is equivalent to that in the thin-torus limit discussed in Ref.~\cite{bernevig2012thintoruslimitfractionaltopological}, which is Eq.\eqref{eq:countingCDW} in the main text.

We can also compute momentum space occupation number $n_{\bsl k}$ of the CDW states, which is given by $n_{\bsl k}=\text{Tr}[\rho \hat n_{\bsl k} ]$ with $\hat n_{\bsl k}=d^\dagger_{\bsl k}d_{\bsl k}$. In the following, we show that the distribution of $n_{\bsl k}$ is uniform and independent of $\bsl k$.

The CDW states can be rewritten with operators in momentum space as
\bea
&&\mu_{12...n}=\frac{1}{(3n)^{n/2}}\sum_{\bsl k_1,...,\bsl k_n}e^{-i\bsl k_1\cdot \bsl \mu_1}e^{-i\bsl k_2\cdot \bsl \mu_2}...e^{-i\bsl k_n\cdot \bsl\mu_n}d^\dagger_{\bsl k_1}...d^\dagger_{\bsl k_n}|0\rangle \\
&&\ \ \ \ \ \ \ \ \ \ \ \ =\ \frac{1}{(3n)^{n/2}}\sum_{\bsl k_1<\bsl k_2<...<\bsl k_n}\text{det}[M^\mu(\bsl k_1,\bsl k_2,...,\bsl k_n)]d^\dagger_{\bsl k_1}...d^\dagger_{\bsl k_n}|0\rangle. \\
&&M^\mu(\bsl k_1,\bsl k_2,...,\bsl k_n)_{\alpha\beta}=e^{-i\bsl k_{\alpha}\cdot \bsl \mu_{\beta}},\ 1\le \alpha,\beta\le n.
\eea
Here $\bsl \mu_i$ is the real space location of site $i$ with color $\mu$ in Fig.~\ref{fig_Kcdwall}, and we have defined an ordering for the 2D momentun space such that $\bsl k<\bsl k'$ means $k_x+k_yN_x<k'_x+k'_yN_x$. The occupation number is then given by
\bea
n_{\bsl k}=\text{Tr}[\rho \hat n_{\bsl k} ]=\frac{1}{3}\sum_{\mu}\sum_{\bsl k_2<...<\bsl k_n}\frac{1}{3n}|\text{det}[M^\mu(\bsl k,\bsl k_2,...,\bsl k_n)]|^2
\eea
Notice that the determinant in the above formula has a property that if all momenta are shifted by a constant $\bsl \delta$, the determinant only changes by a phase, hence its magnitude is unchanged:
\be
\text{det}[M^\mu(\bsl k_1+\bsl \delta,\bsl k_2+\bsl \delta,...,\bsl k_n+\bsl \delta)]=\text{det}[M^\mu(\bsl k_1,\bsl k_2,...,\bsl k_n)]e^{-i\bsl \delta\cdot(\bsl \mu_1+\bsl \mu_2+...+\bsl \mu_n)}.
\label{Mphaseproperty}
\ee
This property can be used to show that $n_{\bsl k}$ is independent of $\bsl k$. Indeed, for arbitrary $\bsl \delta$ in the momentum space, we have
\bea
n_{\bsl k+\bsl \delta}&=&\frac{1}{3}\sum_{\mu}\sum_{\bsl k_2<...<\bsl k_n}\frac{1}{3n}|\text{det}[M^\mu(\bsl k+\bsl \delta,\bsl k_2,...,\bsl k_n)]|^2 \nonumber\\
&=&\frac{1}{3}\sum_{\mu}\sum_{\bsl k_2<...<\bsl k_n}\frac{1}{3n}|\text{det}[M^\mu(\bsl k,\bsl k_2-\bsl \delta,...,\bsl k_n-\bsl \delta)]|^2 \nonumber\\
&=&\frac{1}{3}\sum_{\mu}\sum_{\bsl k'_2<...<\bsl k'_n}\frac{1}{3n}|\text{det}[M^\mu(\bsl k,\bsl k'_2,...,\bsl k'_n)]|^2 \nonumber\\
&=&n_{\bsl k}.
\eea
Here in the first step we used Eq.\eqref{Mphaseproperty} with a $-\bsl \delta$ shift and then a re-labeling of $\bsl k_2...\bsl k_n$ inside the summation plus the fact that reordering of $k'_2,...,k'_n$ only leads to a potential minus sign that is irrelevant for the norm. 

Therefore, the distribution of $n_{\bsl k}$ is uniform in the momentum space, which is the same as the thin-torus limit~\cite{bernevig2012thintoruslimitfractionaltopological}. Note that the uniform distribution of $n_{\bsl k}$ is not required for a realistic CDW state.

\subsection{$M$-CDW with $2\times 1$ periodicity }

We move on to discuss other types of CDW with ordering vector at $M$. The $M$-CDW with $2\times 1$ periodicity has ordering vector at either $M_1,M_2$ or $M_3$, which is shown in Fig.~\ref{fig_MCDW1M}(a). For each ordering vector $M_i$, there are two CDW states represented by the red and the blue dots. The CDW with different $M$ are related by $C_3$ symmetry. In ED, if the system has $C_3$ symmetry and the system prefers to develop a CDW with one ordering vector at $M$, all these six CDW states will show up as ground states. Now we find the total momentum of these six states.

\subsubsection{$2\times 2$ system}

\begin{figure}
\centering
\includegraphics[width=5.0 in]{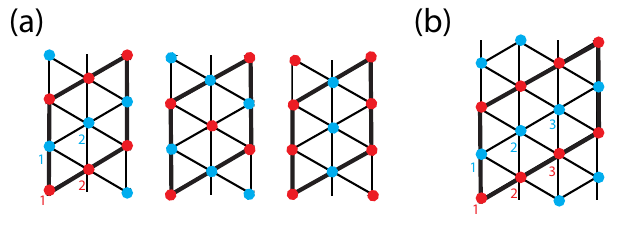}
\caption{(a) Three configurations of $M$-CDW with $2\times 1$ periodicity in $2\times 2$ system. (b) $M$-CDW state with ordering vector $\bsl M_1$ in $2\times 3$ system without $C_3$ symmetry. }
\label{fig_MCDW1M}
\end{figure}

Take the first configuration in Fig.~\ref{fig_MCDW1M}(a), which has ordering vector at $\bsl M_1$. There are two CDW states with this ordering vector, which are $r_{12}$ and $b_{12}$. Under translation they transform as:
\bea
&&\bsl a_2: \ r_{12}\rightarrow r_{21}=-r_{12},\ b_{12}\rightarrow b_{21}=-b_{12} \\
&&\bsl a_2 -\bsl a_1:\ r_{12}\rightarrow b_{12},\ b_{12}\rightarrow r_{12}
\eea
The translation eigenstates can be constructed by:
\be
\psi_{\pm}=(r_{12}\pm b_{12})/\sqrt{2}
\ee
The translation eigenvalues are:
\be
\lambda_\pm (\bsl a_2)=-1,\ \lambda_\pm (\bsl a_2-\bsl a_1)=\pm 1.
\ee
The total momentum of these states are
\be
\bsl k^{tot}_+=\bsl M_3,\ \bsl k^{tot}_-=\bsl M_2
\ee
The other two configures in the second and third figures in Fig.~\ref{fig_MCDW1M}(a) will also appear in the ED solution. Those states are obtained by $C_3$ rotation, hence the total momentum of the two states in the second configuration are at $\bsl M_1$ and $\bsl M_3$, and the last configuration has total momentum $\bsl M_1$ and $\bsl M_2$. Therefore, the ED spectrum has six ground states in total, with each $\bsl M_i$ momentum sector having two states.

\subsubsection{$2\times 3$ system}

This system size does not have $C_3$ symmetry, hence there is only one $M$-CDW at $\bsl M_1$, as in Fig.~\ref{fig_MCDW1M}(b). There are two CDW states denoted by the red and blue dots. Under translation they transform as:
\bea
&&\bsl a_2: \ r_{123}\rightarrow r_{231}=r_{123},\ b_{123}\rightarrow b_{231}=b_{123} \\
&&\bsl a_2 -\bsl a_1:\ r_{123}\rightarrow b_{123},\ b_{123}\rightarrow r_{123}
\eea
The translation eigenstates can be constructed by:
\be
\psi_{\pm}=(r_{123}\pm b_{123})/\sqrt{2}
\ee
The translation eigenvalues are:
\be
\lambda_\pm (\bsl a_2)=1,\ \lambda_\pm (\bsl a_2-\bsl a_1)=\pm 1.
\ee
The total momentum of these states are
\be
\bsl k^{tot}_+=\Gamma,\ \bsl k^{tot}_-=\bsl M_1
\ee
Therefore, there are two ground states with total momentum $\Gamma$ and $\bsl M_1$.

\subsection{$M$-CDW with $2\times 2$ periodicity}

\begin{figure}
\centering
\includegraphics[width=2.0 in]{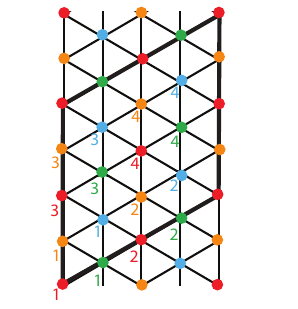}
\caption{ $M$-CDW state with $2\times 2$ periodicity. }
\label{fig_MCDW4}
\end{figure}

The $M$-CDW can also have $2\times 2$ periodicity, as shown in Fig.~\ref{fig_MCDW4}. There are four CDW states represented by the red, blue, green and yellow dots. For the $4\times 4$ lattice, these states transform under translation as:
\bea
&&\bsl a_2:\ r_{1234}\rightarrow g_{1234},\ g_{1234}\rightarrow r_{2143}=r_{1234},\ y_{1234}\rightarrow b_{1234},\ b_{1234}\rightarrow y_{2143}=y_{1234} \\
&& \bsl a_2-\bsl a_1:\ r_{1234}
\rightarrow y_{1234},\ g_{1234}\rightarrow b_{1234},
\ y_{1234}\rightarrow r_{3412}=r_{1234},\ b_{1234}\rightarrow g_{3412}=g_{1234}
\eea
The translation eigenstates can be constructed by:
\be
\psi_{mn}=(r_{1234}+ e^{i\pi m} g_{1234}+e^{i\pi n}y_{1234}+e^{i\pi(m+n)}b_{1234})/2,\ \ m,n=0,1
\ee
The total momentum of these four states are:
\be
\bsl k^{\text{tot}}_{00}=\Gamma,\ \bsl k^{\text{tot}}_{01}=\bsl M_1,\ \bsl k^{\text{tot}}_{10}=\bsl M_3,\ \bsl k^{\text{tot}}_{11}=\bsl M_2
\ee
Therefore, there are four ground states with total momentum $\Gamma,\bsl M_1,\bsl M_2$ and $\bsl M_3$.

\section{Derivation of density correlation function and structure factor }
\label{Sec_cor}

The structure factor $S(\bsl q)$ is the Fourier transformation of the density-density correlation function, which can be used to identify the existence of CDW. The structure factor is given by~\cite{Wilhelm2021FCITBG}:
\bea
S(\bsl q)&=&\frac{1}{N_k}(\langle\hat\rho_{\bsl q}\hat{\rho}_{-\bsl q})\rangle - \langle\hat\rho_{\bsl q}\rangle\langle\hat\rho_{-\bsl q}\rangle )
\eea
We focus on the spin and valley polarized scenario, and $N_k$ is the number of momentum points in the MBZ. The density operator is written as
\bea
\hat\rho_{\bsl q}&=&\sum_{\bsl{k},\bsl G,l,\sigma} c^\dagger_{K,\bsl k+\bsl q,\bsl G,l,\sigma,\uparrow} c_{K,\bsl k,\bsl G,l,\sigma,\uparrow}\\
&=& \sum_{\mbf{k} mn } M_{mn}^{K}(\mbf{k},\mbf{q}+\mbf{G}) c^\dag_{K,\mbf{k}+\mbf{q}, m,\uparrow} c_{K,\mbf{k}, n,\uparrow}.
\eea
The ED computation is performed in the HF basis defined by $\gamma^\dagger_{\mbf{k},n}$. To compute the structure factor, we need to switch to the HF basis. Using the basis transformations in Eq.~\eqref{eq:HF_basis} and Eq.~\eqref{cbaretransf}
\bea
&&\gamma^\dag_{\mbf{k},n} = \sum_{n'} c^\dag_{K,\bsl{k}, n',\uparrow}  \tilde{U}_{ n'n}(\mbf{k}),\\
&&c^\dagger_{K,\bsl{k},n' ,\uparrow} =  \sum_{\bsl{G}l\sigma}c^\dagger_{K,\bsl{k},\bsl{G},l,\sigma , \uparrow} \left[ U_{n'}^{K}(\bsl k) \right]_{\bsl{G}l\sigma} 
\eea
we obtain
\bea
&&\gamma^\dag_{\mbf{k},n} = \sum_{\bsl{G}l\sigma} c^\dag_{K,\bsl{k},\bsl{G},l,\sigma , \uparrow}  \left[\bar{U}_{n}(\mbf{k})\right]_{\bsl G l\sigma},\\ &&\left[\bar{U}_{n}(\mbf{k})\right]_{\bsl G l\sigma}=\sum_{n'}\tilde{U}_{ n'n}(\mbf{k})\left[ U_{n'}^{K}(\bsl k) \right]_{\bsl{G}l\sigma} 
\eea

The density operator becomes
\bea
\hat\rho_{\bsl q}&=&\sum_{\bsl{k},\bsl G,l,\sigma,n,n'} \gamma^\dagger_{\bsl{k}+\bsl q,n } \gamma_{\bsl{k},n' } \left[ \bar U^*_{n}(\bsl{k}+\bsl q) \right]_{\bsl{G}l\sigma} \left[ \bar U_{n'}(\bsl{k}) \right]_{\bsl{G}l\sigma} \nonumber\\
&=&\sum_{\bsl{k},n,n'} \gamma^\dagger_{\bsl{k}+\bsl q,n } \gamma_{\bsl{k},n' } \bar M_{nn'}(\bsl k,\bsl q),
\eea
where the form factor is given by
\be
\bar M_{mn}(\bsl{k},\bsl{q}) = \sum_{\bsl{G}'l \sigma} \left[\bar U_{m}(\bsl{k}+\bsl{q})\right]_{\bsl{G}'l\sigma}^* \left[\bar U_{n}(\bsl{k})\right]_{\bsl{G}'l\sigma}\ .
\ee
The $\langle \rho_{\bsl q}\rangle$ term is nonzero only when $\bsl q$ is a moir\'e reciprocal lattice vector. We focus on $\bsl q$ inside the MBZ, then $\langle \rho_{\bsl q}\rangle=N\delta_{\bsl q,0}$ with $N$ being the number of particles. The structure factor becomes
\bea
S(\bsl q)&=&\frac{1}{N_k}(\langle\hat\rho_{\bsl q}\hat\rho_{-\bsl q}\rangle - N^2\delta_{\bsl q,0} ) \nonumber\\
&=&\frac{1}{N_k}\left(\langle \sum_{\bsl k,m,n}\gamma^\dagger_{\bsl{k}+\bsl q,m}\gamma_{\bsl{k},n } \bar M_{mn}(\bsl{k},\bsl{q}) \sum_{\bsl k',m',n'}\gamma^\dagger_{\bsl{k'}-\bsl q,m' }\gamma_{\bsl{k'}n' } \bar M_{m'n'}(\bsl{k'},\bsl{-q}) \rangle - N^2\delta_{\bsl q,0} \right) \nonumber\\
&=&\frac{1}{N_k}\left(\sum_{\bsl k,\bsl k',m,n,m',n'}\bar M_{mn}(\bsl{k},\bsl{q})\bar M_{m'n'}(\bsl{k'},\bsl{-q})\langle \gamma^\dagger_{\bsl{k}+\bsl q,m}\gamma_{\bsl{k},n }  \gamma^\dagger_{\bsl{k'}-\bsl q,m' }\gamma_{\bsl{k'}n' }  \rangle - N^2\delta_{\bsl q,0} \right) \nonumber\\
&=&\frac{1}{N_k}\left(\sum_{\bsl k,\bsl k',m,n,m',n'}\bar M_{mn}(\bsl{k},\bsl{q})\bar M_{m'n'}(\bsl{k'},\bsl{-q}) (\delta_{ \bsl k,\bsl k'-\bsl q}\delta_{n,m'}\langle \gamma^\dagger_{\bsl{k}+\bsl q,m} \gamma_{\bsl{k'}n' }  \rangle+\langle \gamma^\dagger_{\bsl{k}+\bsl q,m} \gamma^\dagger_{\bsl{k'}-\bsl q,m' }\gamma_{\bsl{k'}n' }\gamma_{\bsl{k},n }   \rangle) - N^2\delta_{\bsl q,0}\right)  \nonumber\\
&=&\frac{1}{N_k}\left( \sum_{\bsl k,m,n}|\bar M_{mn}(\bsl{k},-\bsl{q})|^2 \langle \gamma^\dagger_{\bsl{k},m} \gamma_{\bsl{k},n }  \rangle+   \sum_{\bsl k,\bsl k',m,n,m',n'}\bar M_{mn}(\bsl{k},\bsl{q})\bar M_{m'n'}(\bsl{k'},\bsl{-q}) \langle \gamma^\dagger_{\bsl{k}+\bsl q,m} \gamma^\dagger_{\bsl{k'}-\bsl q,m' }\gamma_{\bsl{k'}n' }\gamma_{\bsl{k},n }   \rangle - N^2\delta_{\bsl q,0}\right)  \nonumber\\
\eea

\section{ Additional numerical results}
\label{app:additionalED}

\subsection{Fluctuations of total occupation number}
\label{app_nfluc}

\begin{figure}
\centering
\includegraphics[width=\textwidth]{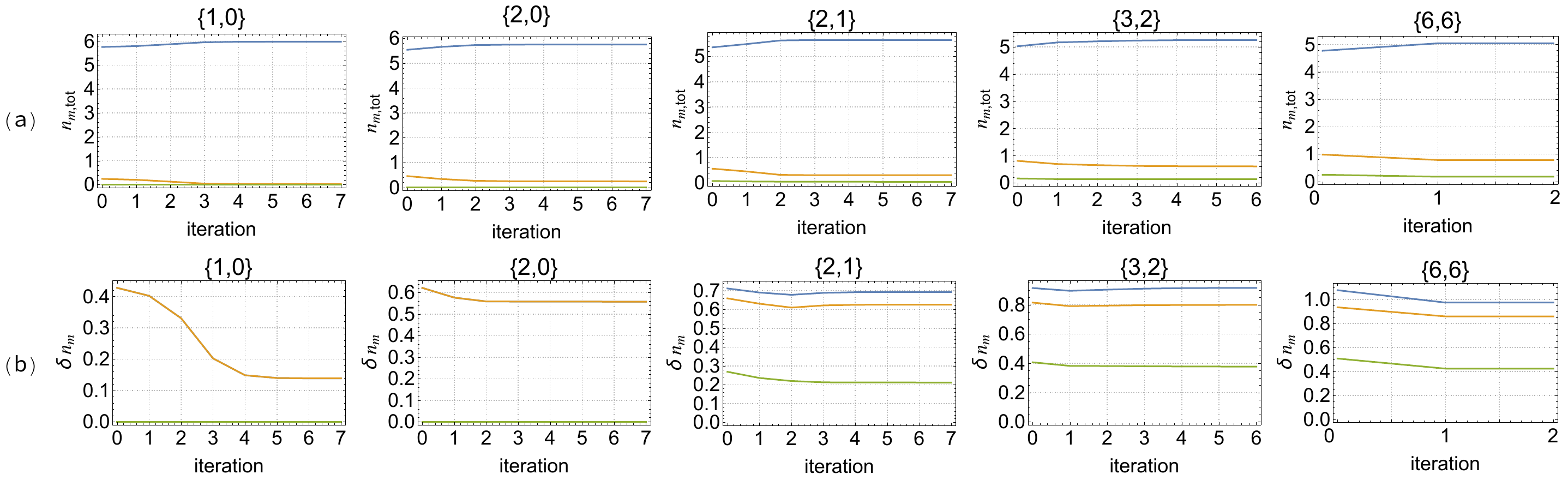}
\caption{
Total occupation number per band $n_{m,\text{tot}}$ (a) and its fluctuation $\delta n_m$ (b) as a function of iteration step for $9\times 2$ system at 1/3 filling in CN scheme. The number of iteration is the minimal number to make the convergence quantity $\mathcal{R}$ in Eq.~\eqref{eq:convergencedef} smaller than $10^{-3}$. Blue, orange and green curves correspond to $m=0,1,2$ respectively. The blue and orange curves in the first two figures in (b) are on top of each other,  since the sum of occupation number in band 0 and band 1 is fixed when $N_{\text{band}2}=0$, hence their fluctuations are identical.
}
\label{fig_ntfluc186}
\end{figure}

\begin{figure}
\centering
\includegraphics[width=5.8in]{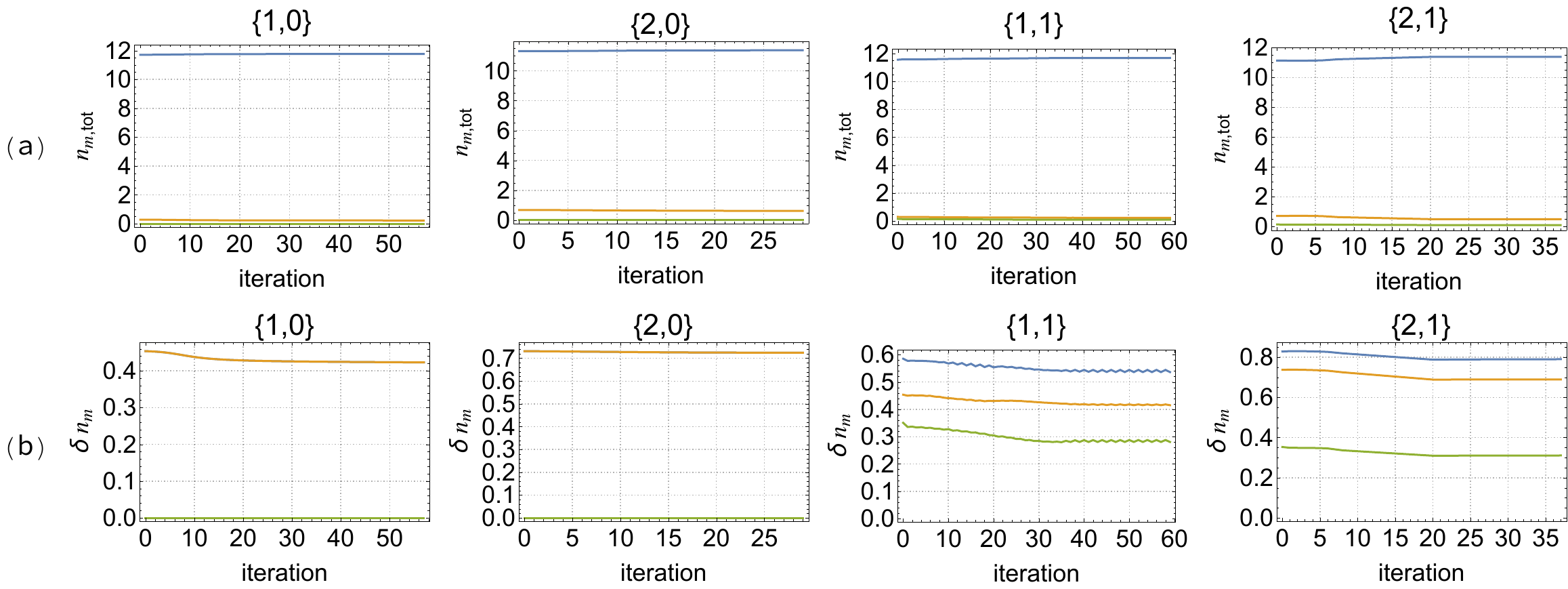}
\caption{
Total occupation number per band $n_{m,\text{tot}}$ and its fluctuation $\delta n_m$ (b) for $9\times 2$ system at 2/3 filling in AVE scheme. The blue and orange curves in (b) are on top of each other when $N_{\text{band}2}=0$. This is because the sum of occupation number in band 0 and band 1 is fixed when $N_{\text{band}2}=0$, hence their fluctuations are identical.
}
\label{fig_ntfluc1812}
\end{figure}

\begin{figure}
\centering
\includegraphics[width=6.8in]{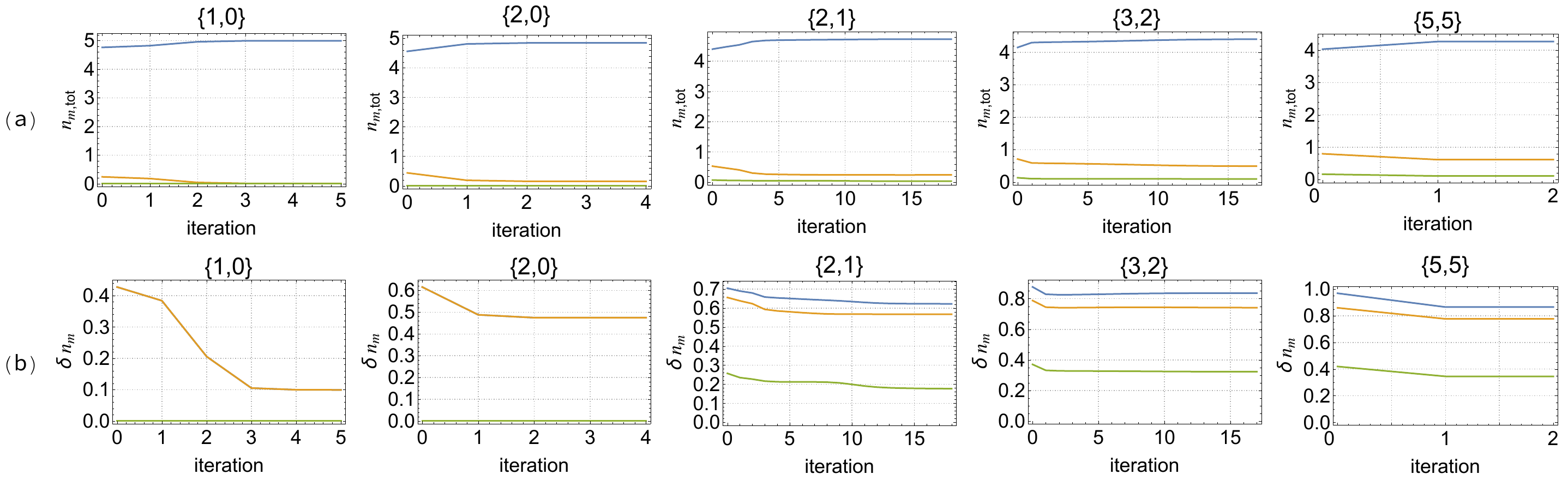}
\caption{
Total occupation number per band $n_{m,\text{tot}}$ and its fluctuation $\delta n_m$ (b) for $15\times 1$ system at 1/3 filling in CN scheme. The blue and orange curves in (b) are on top of each other when $N_{\text{band}2}=0$. This is because the sum of occupation number in band 0 and band 1 is fixed when $N_{\text{band}2}=0$, hence their fluctuations are identical.
}
\label{fig_ntfluc155}
\end{figure}

\begin{figure}
\centering
\includegraphics[width=5.4in]{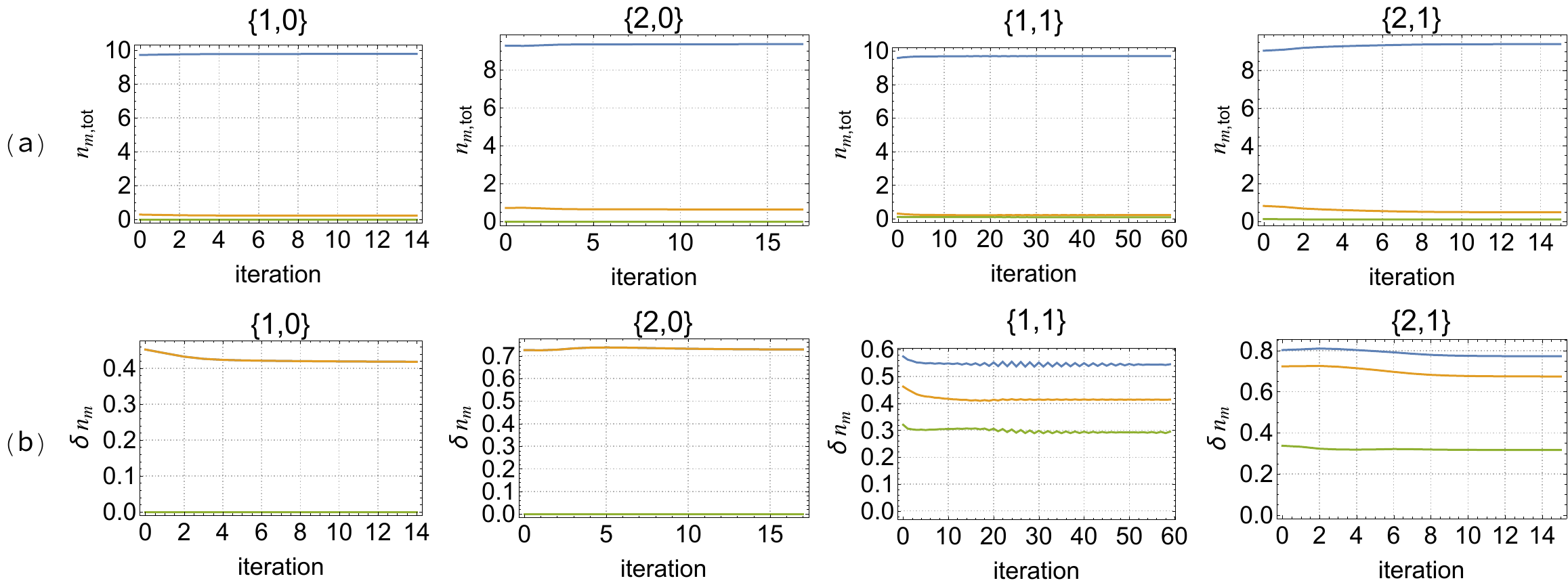}
\caption{
Total occupation number per band $n_{m,\text{tot}}$ and its fluctuation $\delta n_m$ (b) for $15\times 1$ system at 2/3 filling in AVE scheme. The blue and orange curves in (b) are on top of each other when $N_{\text{band}2}=0$. This is because the sum of occupation number in band 0 and band 1 is fixed when $N_{\text{band}2}=0$, hence their fluctuations are identical. }
\label{fig_ntfluc1510}
\end{figure}

The ED iteration procedure presented in Sec.~\ref{Sec_itermethod} is built to maximize the average occupation of the band 0 at each momentum sector. Solely focusing on the average might lead to a biased perspective, since fluctuations around the average could still be important. In this appendix, we check whether the iteration process might also reduce the fluctuation of the total occupation number in the lowest band. The total occupation number per band $n_{m,\text{tot}}$ is defined in Eq.~\eqref{eq:totalbandoccupation}, and its fluctuation is given by:
\be
\delta n_m=\sqrt{\sum_{\bsl k,\bsl k'}\frac{1}{N_{\text{GS}}}{\text Tr} \left[\gamma^\dagger_{\bsl k,m}\gamma_{\bsl k,m}\gamma^\dagger_{\bsl k',m} \gamma_{\bsl k',m} \mathcal{P}_{\text{GS}} \right] - \left( n_{m,\text{tot}} \right)^2  }
\ee
From Fig.~\ref{fig_ntfluc186} to Fig.~\ref{fig_ntfluc1510} we plot the occupation number $n_{m,\text{tot}}$ and its fluctuation $\delta n_m$ as a function of iteration step for CN scheme at 1/3 filling and AVE scheme at 2/3 filling in $9\times 2$ and $15\times 1$ systems respectively. In these cases the three lowest energy states are at the FCI momenta before iteration. We first focus on the untruncated case, namely $\{6,6\}$ for 1/3 filling in $9\times 2$ system and remind that iteration 0 corresponds to the initial state and the ED iteration procedure lets the system unchanged after one step if applied to the untruncated Hamiltonian. We observe that after just one iteration, the average band occupation of band 1 and 2 or any of the fluctuations are non-zero (although slightly lower than the original state), and are still substantial (for example $\frac{\delta n_0}{n_{0,\text{tot}}}\simeq 0.2$). The fact that we cannot set the fluctuations to zero is just the signature that the ground state is not a simple product state. In presence of truncation, the picture is slightly different. For weak mixing such as $\{1,0\}$, since the original system is mostly hosted in one band and close to a CDW (thus a product state), the ED iteration allows for a larger damping of the band occupation fluctuations. The closer we move toward the untruncated case, the smaller is the reduction of these fluctuations under the ED iteration procedure.

\subsection{Evolution of wave function overlap with iteration }
\label{app_wfoverlap}

In Sec.~\ref{Sec_afteriter}, we have shown how the overlap between the exact ground state (in absence of Hilbert space truncation) and the ground state at a given iteration step evolves with the number of ED iterations. In particular, we observed that this overlap when applying the ED iteration might be lower than the overlap without any iteration, especially for low band mixing. Here, we discuss the evolution of wave function overlap between different truncation parameters as a function of iteration steps. Denote the ground state wave functions at truncation parameters  $\{N_{\text{band}1},N_{\text{band}2}\}=\{a,b\}$ as $|\Psi_i^{a,b}\rangle$ with $i=1,...,N_{\text{GS}}$, and define the projector to this ground state manifold as
\be
\mathcal{P}_{\text{GS}}^{a,b}=\sum_{i} | \Psi_i^{a,b} \rangle \langle \Psi_i^{a,b}|
\ee
The overlap between ground state wave functions at different truncations $\{a,b\}$ and $\{a',b'\}$ is defined as:
\be
O_v(a,b,a',b')=\frac{1}{N_{\text{GS}}}\text{Tr}\left[ \mathcal{P}_{\text{GS}}^{a,b}\mathcal{P}_{\text{GS}}^{a',b'}  \right]
\label{ovabab}
\ee
Here the projectors $\mathcal{P}_{\text{GS}}^{a,b}$ and $\mathcal{P}_{\text{GS}}^{a',b'}$ are both computed after iteration. One can also define the overlap similar to Eq.~\eqref{ovabab} but with $\mathcal{P}_{\text{GS}}^{a,b}$ computed in the basis before iteration and $\mathcal{P}_{\text{GS}}^{a',b'}$ computed in the basis after iteration. In that case, $\mathcal{P}_{\text{GS}}^{a,b}$ and $\mathcal{P}_{\text{GS}}^{a',b'}$ involve different one-body basis. Explicitly, the basis before iteration is built from Slater determinants made of direct products of single particle states in the HF basis $\gamma^\dagger_{\bsl k,m}|0\rangle$, whereas the basis after iteration is made of the rotated basis $\tilde \gamma^\dagger_{\bsl k,m}|0\rangle$ which is related to $\gamma^\dagger_{\bsl k,m}|0\rangle$ by a unitary transformation determined by the iteration process. This difference in one-body basis dramatically increases the complexity of the overlap computation (scaling as $N!$) and puts this approach out of reach. Therefore, we keep both projectors in Eq.~\eqref{ovabab} to be after iteration.

\begin{figure}
\centering
\includegraphics[width=\textwidth]{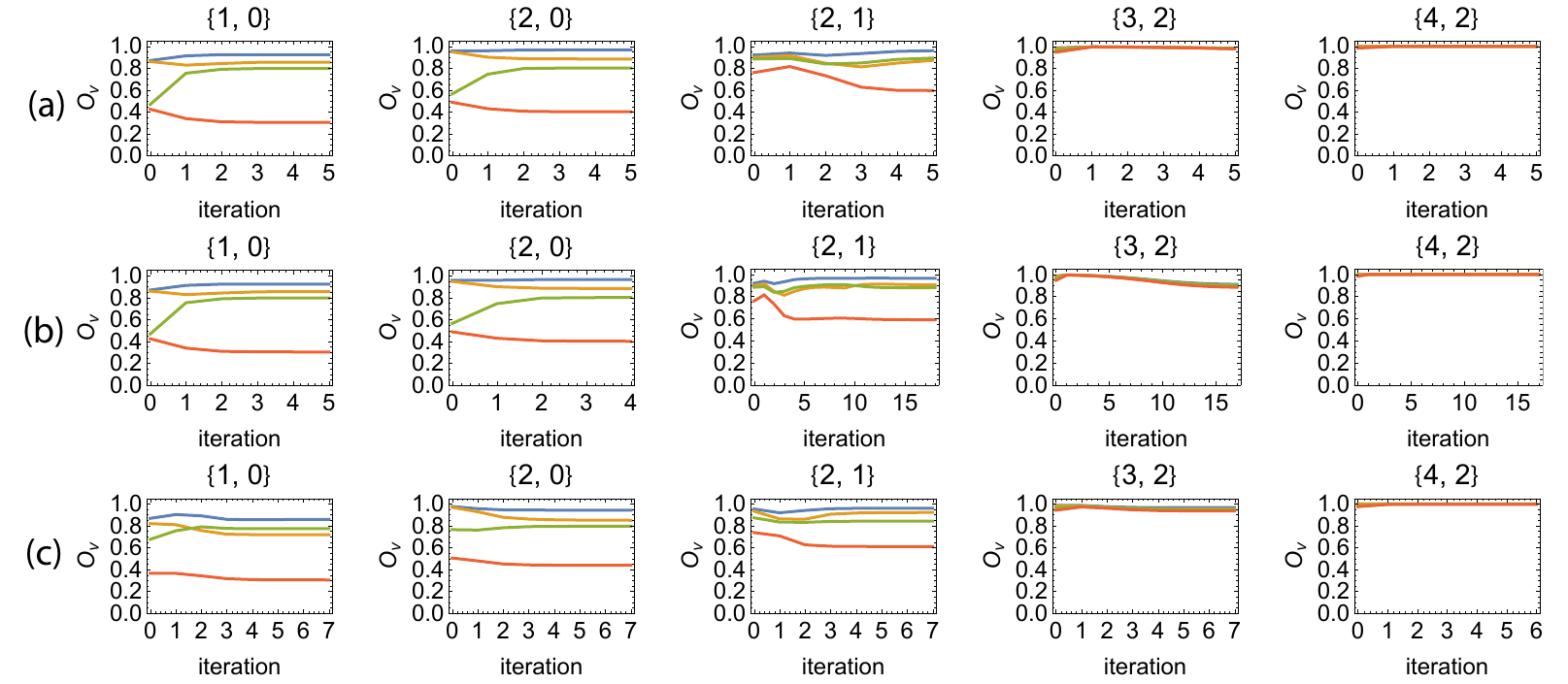}
\caption{
Overlap between ground state wave function at different truncation parameters in CN scheme at 1/3 filling for (a) $2\times 6$, (b) $15\times 1$ and (c) $9\times 2$ systems. The truncation parameter $\{a,b\}$ shown in the title of each plot. The blue, orange, green and red curves represent $O_v(a,b,a+1,b),O_v(a,b,a+2,b),O_v(a,b,a+1,b+1)$ and $O_v(a,b,N,N)$ respectively.
}
\label{fig_oveiterall}
\end{figure}

We will once again consider the system at $\nu=1/3$ in the CN scheme for three different system sizes: $2 \times 6$, $15 \times 1$ and $9 \times 2$. We remind that the overlap with the exact GS in absence of truncation on $9 \times 2$ system is provided as insets of Fig.~\ref{fig_ovn0iter} of the main text. As shown in Fig.~\ref{fig_oveiterall}, for each truncation parameter $\{a,b\}$ and each system size we compute the overlap $O_v(a,b,a+1,b),O_v(a,b,a+2,b),O_v(a,b,a+1,b+1)$ and $O_v(a,b,N,N)$ at each iteration, where $N$ is the total particle number. It shows the overlap is larger when the truncation is weaker (larger $a,b$) and when $a,b$ are closer to $a',b'$, consistent with the anticipation that at weaker truncation the wave functions will be closer to the non-truncated (exact) limit. There are cases for which the ED iteration brings the ground states at a given truncation closer to the ground states computed in a less-truncated Hilbert space, as shown by the increase of the overlaps $O_v(a,b,a+1,b)$ and $O_v(a,b,a+1,b+1)$ in $\{1,0\}$ and $\{2,0\}$. However, these overlaps might behave non-monotonically with the iteration step, as shown in the cases with $a=2,b=1$ and $a=3,b=2$.

\subsection{ED iterations for $15\times 1,\ 9\times 2,\ 21\times 1$ systems}
\label{app_allfig}

\begin{figure}
\centering
\includegraphics[width=6.8 in]{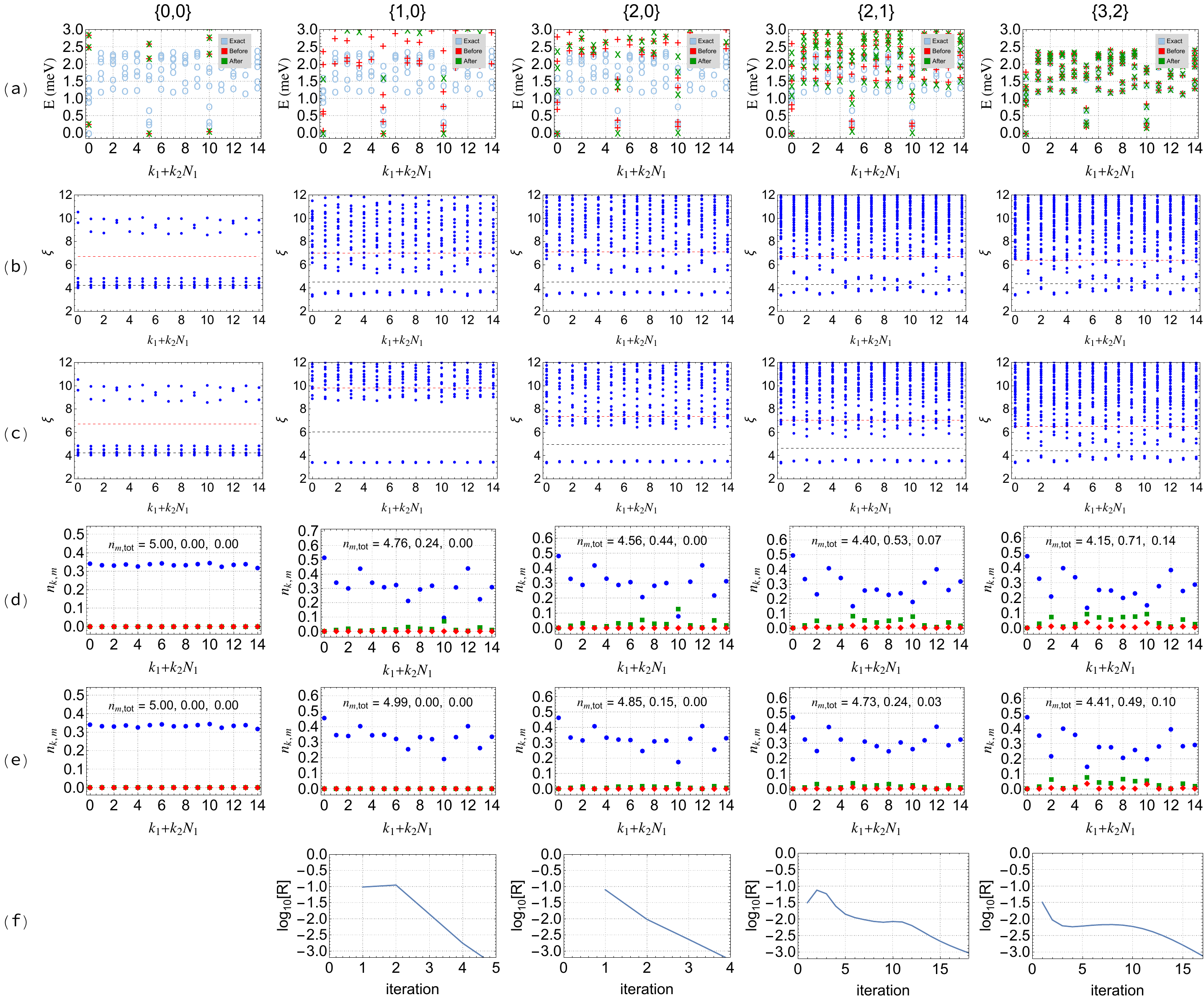}
\caption{
Energy spectrum, PES and occupation number of $15\times 1$ system at 1/3 filling with $V=28$ meV in HF basis in CN scheme. Different columns represent ED results computed at different truncation parameters. (a) Energy spectrum before and after iterations. (b,c) PES of the three lowest states at FCI momenta $0,5,10$ before (b) and after (c) iteration. There are 75 states below the red line which marks the FCI counting, and 30 states below the black line which marks the CDW counting. (d) Occupation number $n_{\bsl k,m}$ as defined in Eq.~\eqref{eq:bandoccupationperk}. Blue, green, and red colors correspond to $m=0,1,2$. The inset shows the total occupation number $n_{m,\text{tot}}$ for band $m=0,1,2$ respectively (see Eq.~\eqref{eq:totalbandoccupation}). (e) Occupation number computed in the basis after iteration. (f) Convergence quantity defined in Eq.~\eqref{eq:convergencedef} as a function of iteration number. The plot at $\{0,0\}$ is empty because the density matrix does not change with iteration in the one-band limit and $\mathcal R$ remains zero.
}
\label{fig_CN15b1n5}
\end{figure}

\begin{figure}
\centering
\includegraphics[width=6.8 in]{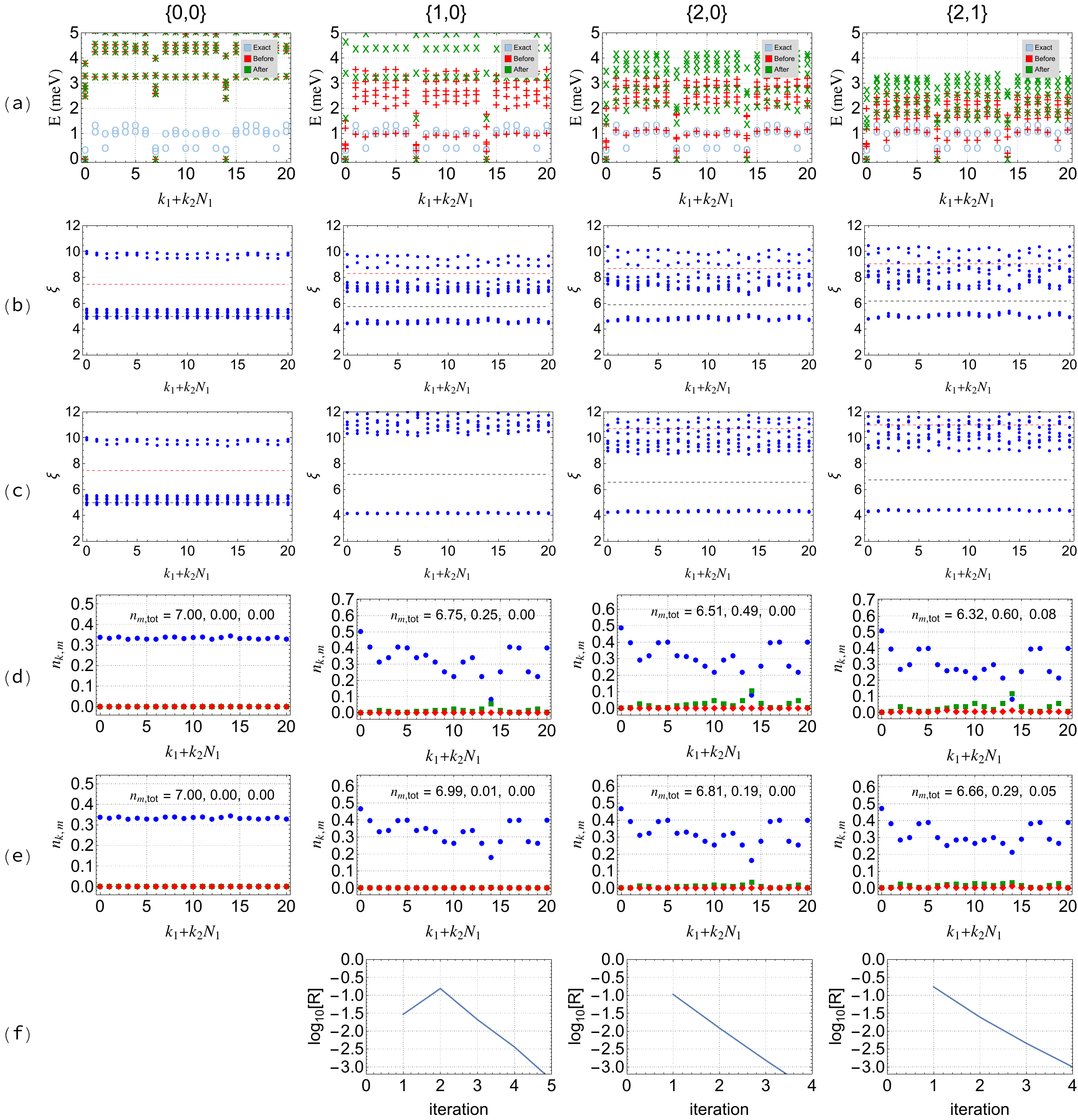}
\caption{
Energy spectrum, PES and occupation number of $21\times 1$ system at 1/3 filling with $V=28$ meV in HF basis in CN scheme. Different columns represent ED results computed at different truncation parameters. (a) Energy spectrum before and after iterations. (b,c) PES of the three lowest states at FCI momenta $0,7,14$ before (b) and after (c) iteration. There are 168 states below the red line which marks the FCI counting, and 63 states below the black line which marks the CDW counting. (d) Occupation number $n_{\bsl k,m}$ as defined in Eq.~\eqref{eq:bandoccupationperk}. Blue, green, and red colors correspond to $m=0,1,2$. The inset shows the total occupation number $n_{m,\text{tot}}$ for band $m=0,1,2$ respectively (see Eq.~\eqref{eq:totalbandoccupation}). (e) Occupation number computed in the basis after iteration. (f) Convergence quantity defined in Eq.~\eqref{eq:convergencedef} as a function of iteration number. The plot at $\{0,0\}$ is empty because the density matrix does not change with iteration in the one-band limit and $\mathcal R$ remains zero.
}
\label{fig_CN21b1n7}
\end{figure}

\begin{figure}
\centering
\includegraphics[width=6.0 in]{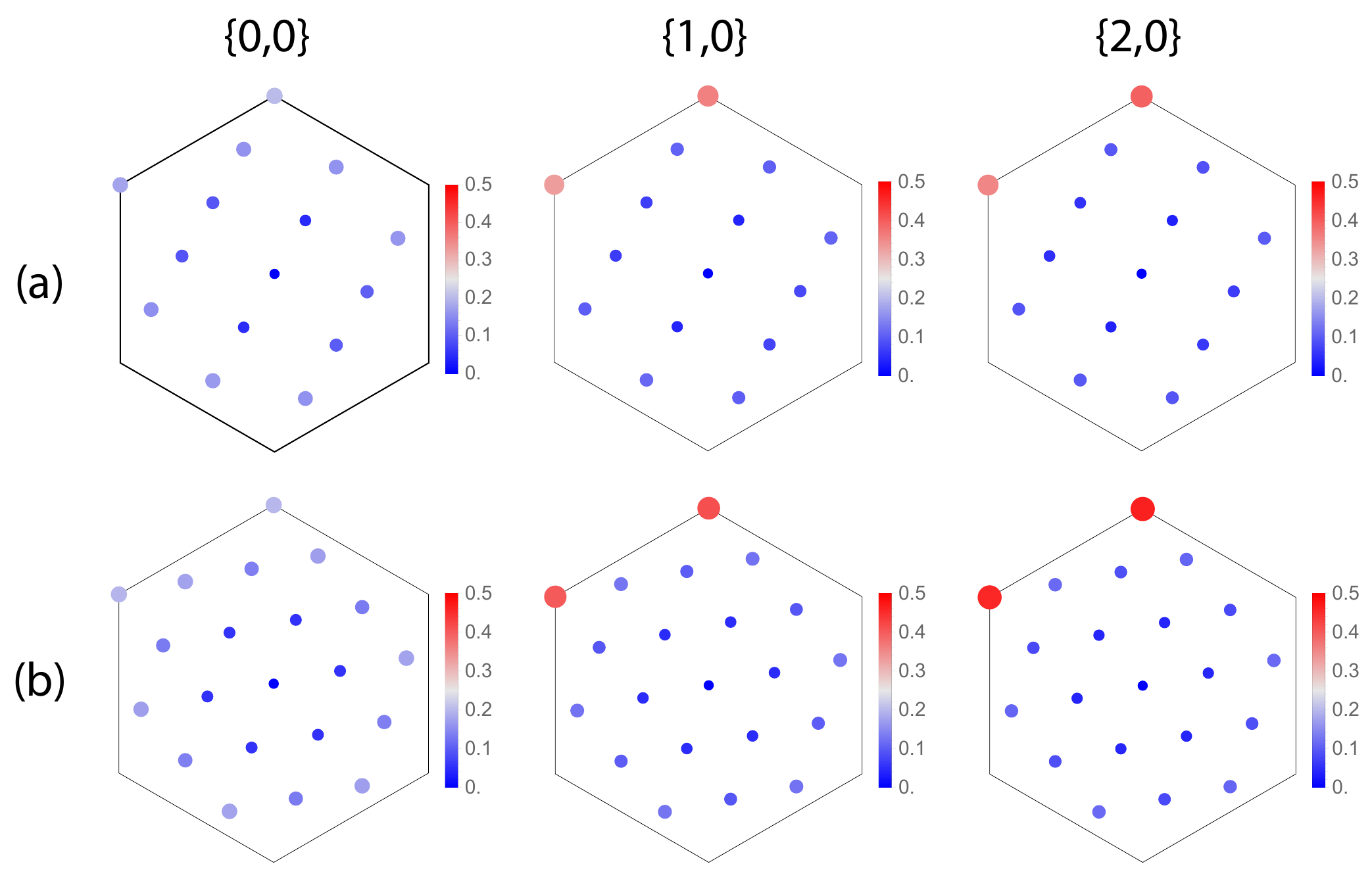}
\caption{
Structure factor $S(\bsl q)$ in the $15\times 1$ system (a) and $21\times 1$ system (b) at 1/3 filling in the CN scheme in the HF basis at different truncation parameters $\{0,0\}$, $\{1,0\}$, $\{2,0\}$. The hexagon represents the MBZ.
}
\label{fig_cor1521}
\end{figure}

\begin{figure}
\centering
\includegraphics[width=6.8 in]{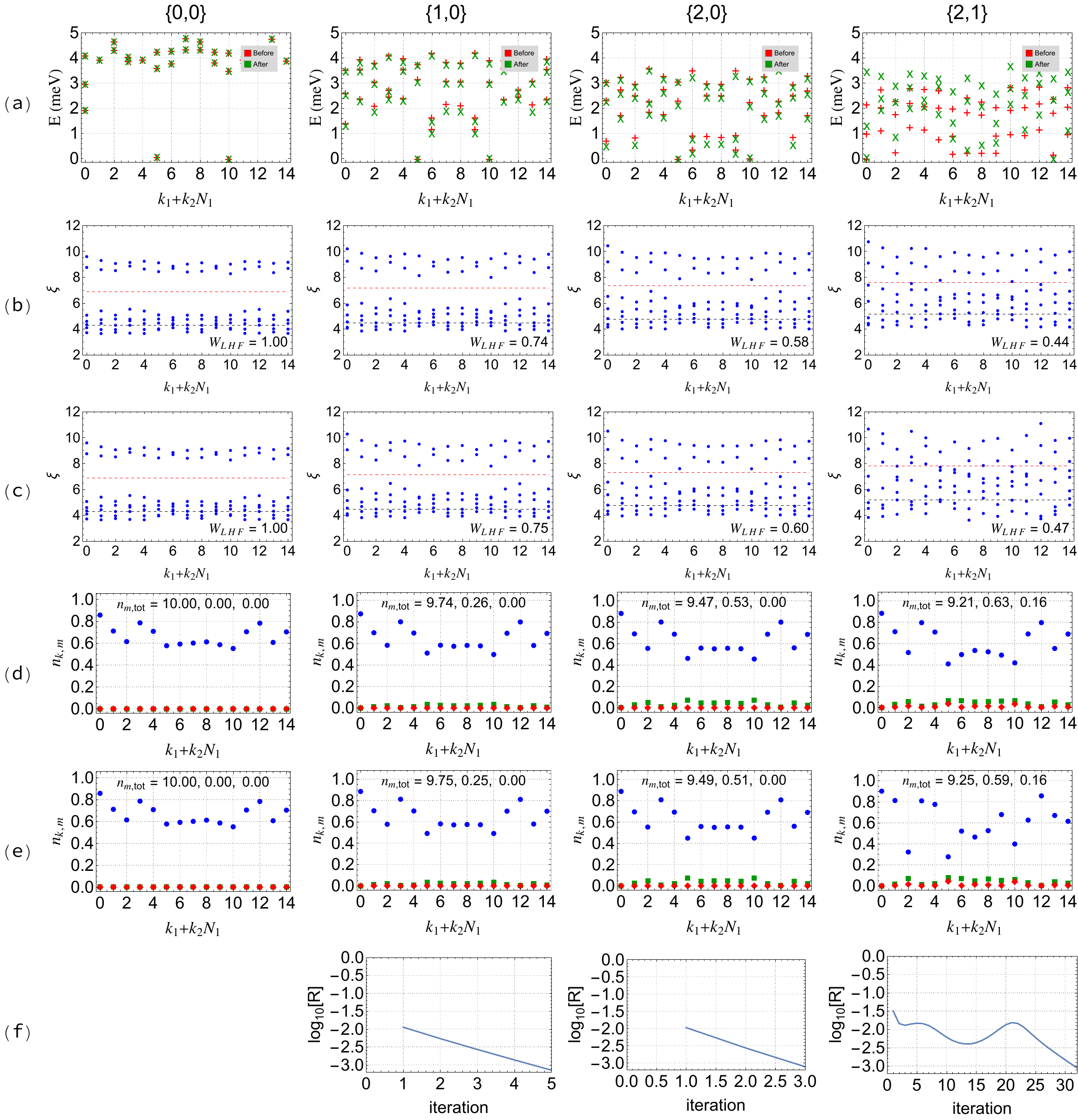}
\caption{
Energy spectrum, PES and occupation number of $15\times 1$ system at 2/3 filling with $V=28$ meV in HF basis in CN scheme. Different columns represent ED results computed at different truncation parameters. (a) Energy spectrum before and after iterations. (b,c) PES of the three lowest states after single-band projection and PH transform at FCI momenta $0,5,10$ before (b) and after (c) iteration. There are 75 states below the red line which marks the FCI counting, and 30 states below the black line which marks the CDW counting. (d) Occupation number $n_{\bsl k,m}$ as defined in Eq.~\eqref{eq:bandoccupationperk}. Blue, green, and red colors correspond to $m=0,1,2$. The inset shows the total occupation number $n_{m,\text{tot}}$ for band $m=0,1,2$ respectively (see Eq.~\eqref{eq:totalbandoccupation}). (e) Occupation number computed in the basis after iteration. (f) Convergence quantity defined in Eq.~\eqref{eq:convergencedef} as a function of iteration number. The plot at $\{0,0\}$ is empty because the density matrix does not change with iteration in the one-band limit and $\mathcal R$ remains zero.
}
\label{fig_CN15b1n10}
\end{figure}

\begin{figure}
\centering
\includegraphics[width=6.8 in]{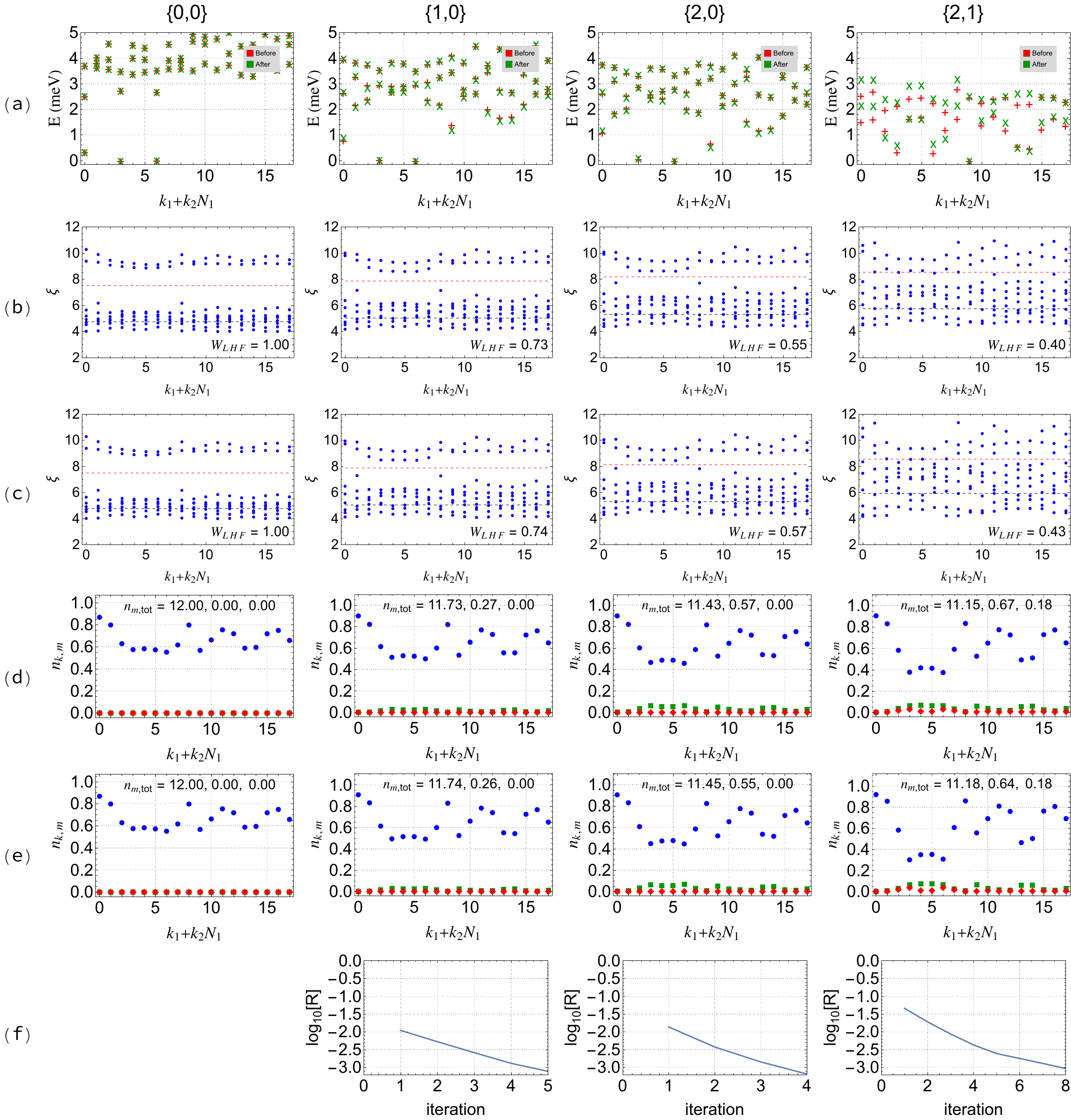}
\caption{ Energy spectrum, PES and occupation number of $9\times 2$ system at 2/3 filling in CN scheme with $V=28$ meV. Different columns represent ED results computed at different truncation parameters. (a) Energy spectrum before and after iterations. (b,c) PES computed from the three lowest states after single-band projection and PH transform at FCI momenta $0,3,6$ before (b) and after (c) iteration. There are 117 states below the red line which marks the FCI counting, and 45 states below the black line which marks the CDW counting. (d) Occupation number $n_{\bsl k,m}$ as defined in Eq.~\eqref{eq:bandoccupationperk}. Blue, green, and red colors correspond to $m=0,1,2$. The inset shows the total occupation number $n_{m,\text{tot}}$ for band $m=0,1,2$ respectively (see Eq.~\eqref{eq:totalbandoccupation}). (e) Occupation number computed in the basis after iteration. (f) Convergence quantity defined in Eq.~\eqref{eq:convergencedef} as a function of iteration number. The plot at $\{0,0\}$ is empty because the density matrix does not change with iteration in the one-band limit and $\mathcal R$ remains zero.  }
\label{fig_CN9b2n12ch}
\end{figure}

\begin{figure}
\centering
\includegraphics[width=6.8 in]{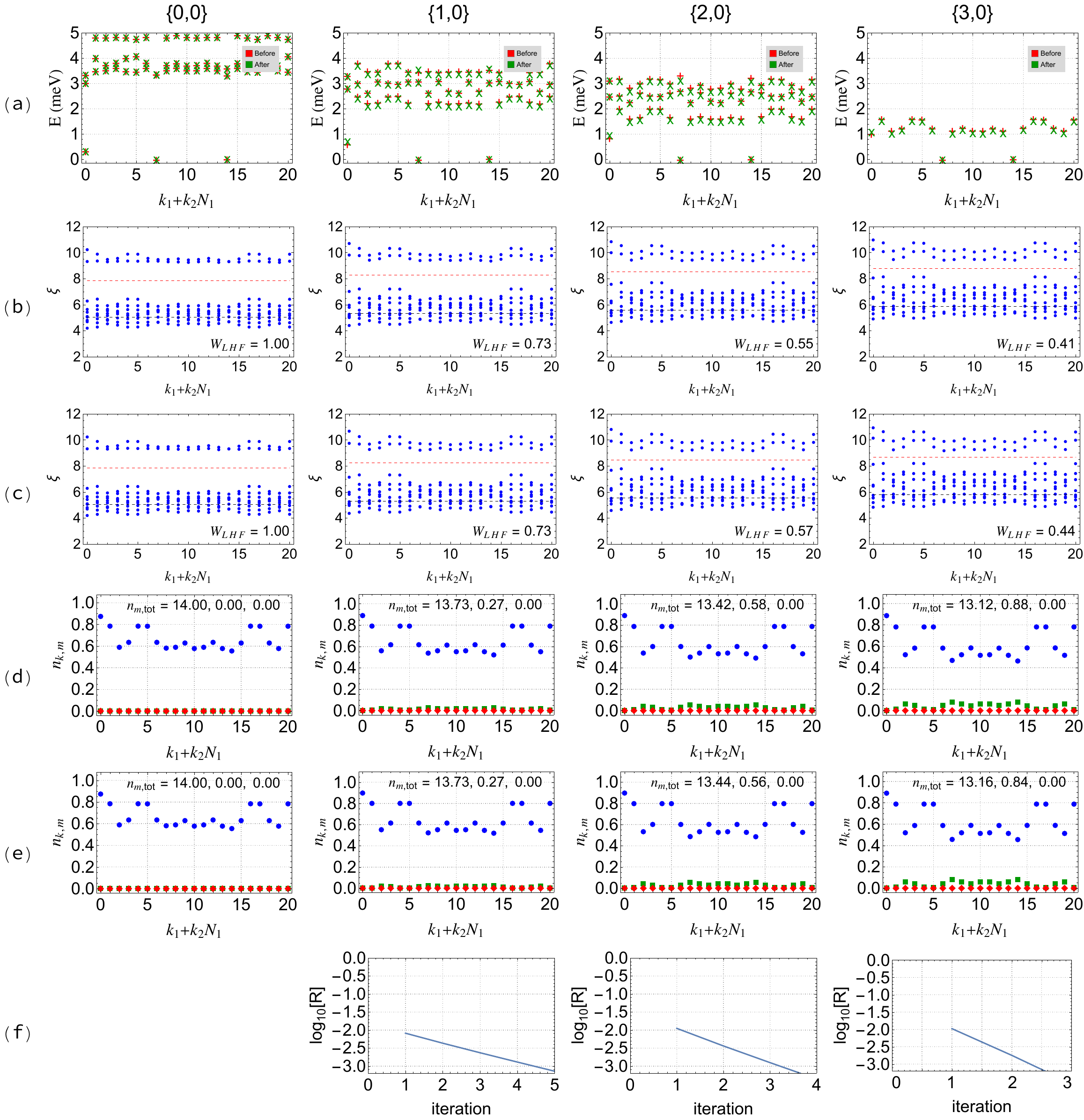}
\caption{
Energy spectrum, PES and occupation number of $21\times 1$ system at 2/3 filling with $V=28$ meV in HF basis in CN scheme. Different columns represent ED results computed at different truncation parameters. (a) Energy spectrum before and after iterations. (b,c) PES of the three lowest states after single-band projection and PH transform at FCI momenta $0,7,14$ before (b) and after (c) iteration. There are 168 states below the red line which marks the FCI counting, and 63 states below the black line which marks the CDW counting. (d) Occupation number $n_{\bsl k,m}$ as defined in Eq.~\eqref{eq:bandoccupationperk}. Blue, green, and red colors correspond to $m=0,1,2$. The inset shows the total occupation number $n_{m,\text{tot}}$ for band $m=0,1,2$ respectively (see Eq.~\eqref{eq:totalbandoccupation}). (e) Occupation number computed in the basis after iteration. (f) Convergence quantity defined in Eq.~\eqref{eq:convergencedef} as a function of iteration number. The plot at $\{0,0\}$ is empty because the density matrix does not change with iteration in the one-band limit and $\mathcal R$ remains zero.}
\label{fig_CN21b1n14}
\end{figure}

\begin{figure}
\centering
\includegraphics[width=6.8 in]{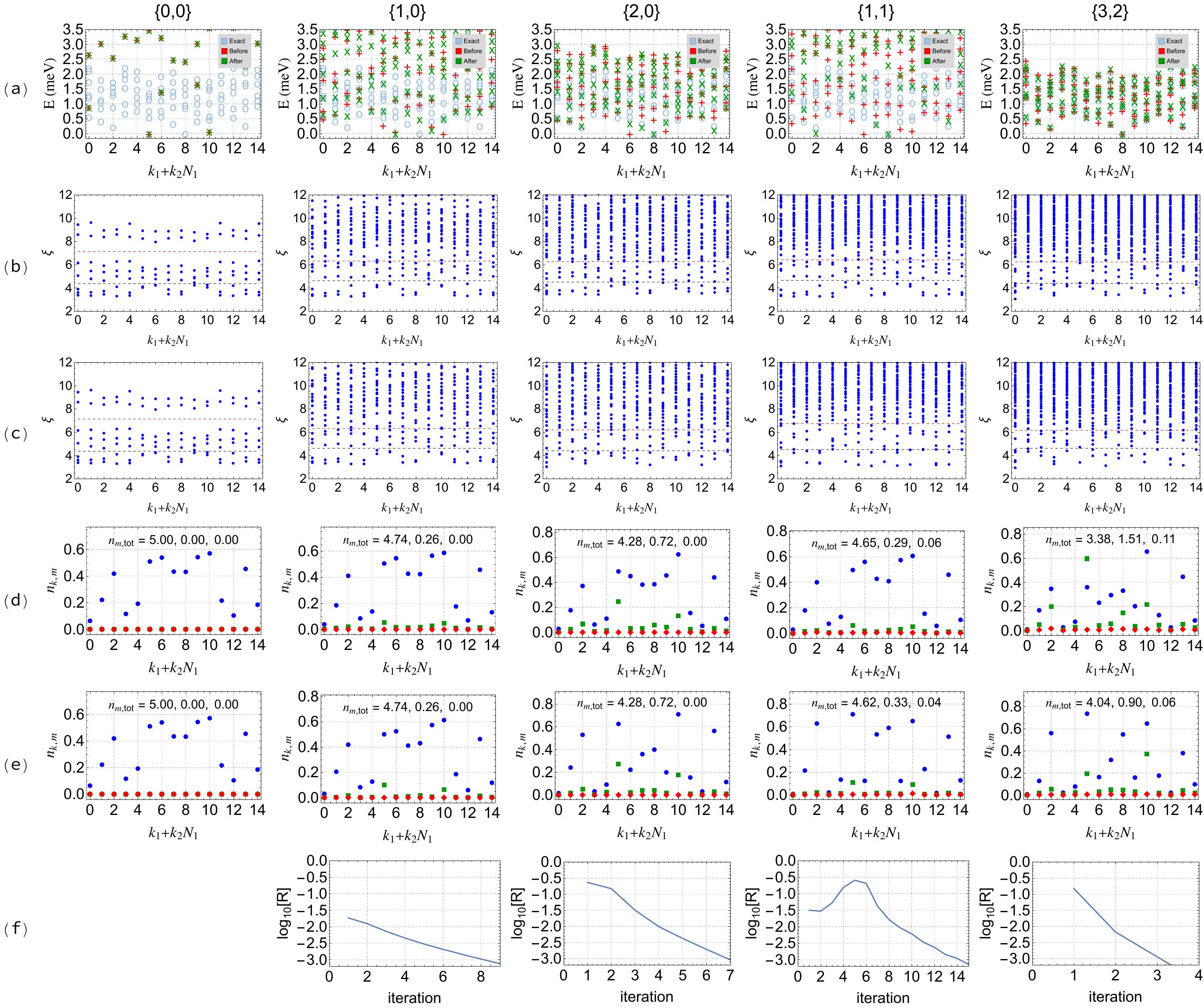}
\caption{
Energy spectrum, PES and occupation number of $15\times 1$ system at 1/3 filling with $V=22$ meV in HF basis in AVE scheme. Different columns represent ED results computed at different truncation parameters. (a) Energy spectrum before and after iterations. (b,c) PES of the three lowest states at FCI momenta $0,5,10$ before (b) and after (c) iteration. There are 75 states below the red line which marks the FCI counting, and 30 states below the black line which marks the CDW counting. (d) Occupation number $n_{\bsl k,m}$ as defined in Eq.~\eqref{eq:bandoccupationperk}. Blue, green, and red colors correspond to $m=0,1,2$. The inset shows the total occupation number $n_{m,\text{tot}}$ for band $m=0,1,2$ respectively (see Eq.~\eqref{eq:totalbandoccupation}). (e) Occupation number computed in the basis after iteration. (f) Convergence quantity defined in Eq.~\eqref{eq:convergencedef} as a function of iteration number. The plot at $\{0,0\}$ is empty because the density matrix does not change with iteration in the one-band limit and $\mathcal R$ remains zero.
}
\label{fig_AVE15b1n5}
\end{figure}

\begin{figure}
\centering
\includegraphics[width=6.8 in]{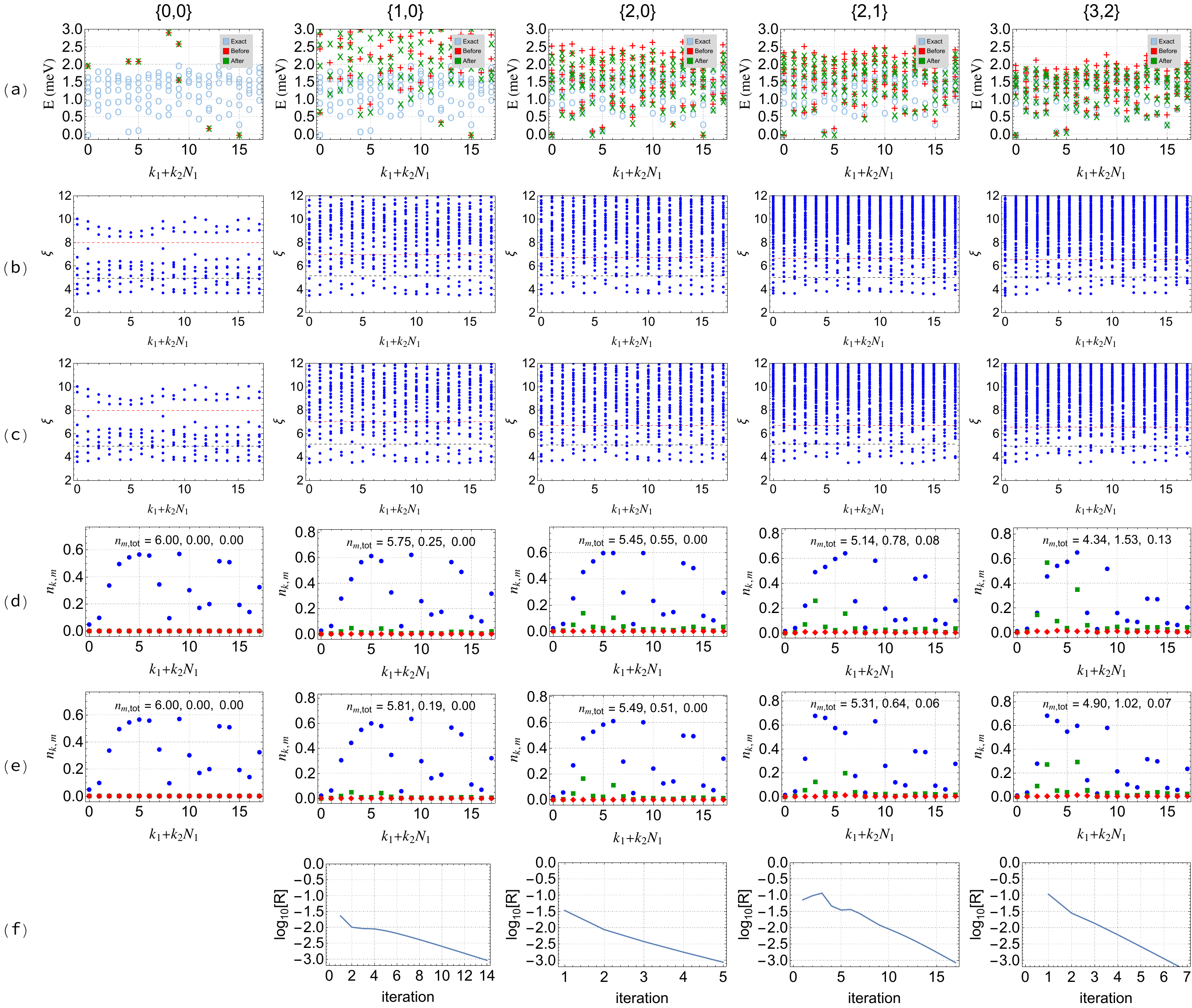}
\caption{ Energy spectrum, PES and occupation number of $9\times 2$ system at 1/3 filling in AVE scheme with $V=22$ meV. Different columns represent ED results computed at different truncation parameters. (a) Energy spectrum before and after iterations. (b,c) PES of the three lowest states at FCI momenta $9,12,15$ before (b) and after (c) iteration. There are 117 states below the red line which marks the FCI counting, and 45 states below the black line which marks the CDW counting. (d) Occupation number $n_{\bsl k,m}$ as defined in Eq.~\eqref{eq:bandoccupationperk}. Blue, green, and red colors correspond to $m=0,1,2$. The inset shows the total occupation number $n_{m,\text{tot}}$ for band $m=0,1,2$ respectively (see Eq.~\eqref{eq:totalbandoccupation}). (e) Occupation number computed in the basis after iteration. (f) Convergence quantity defined in Eq.~\eqref{eq:convergencedef} as a function of iteration number. The plot at $\{0,0\}$ is empty because the density matrix does not change with iteration in the one-band limit and $\mathcal R$ remains zero. }
\label{fig_AVE9b2n6ch}
\end{figure}

\begin{figure}
\centering
\includegraphics[width=6.8 in]{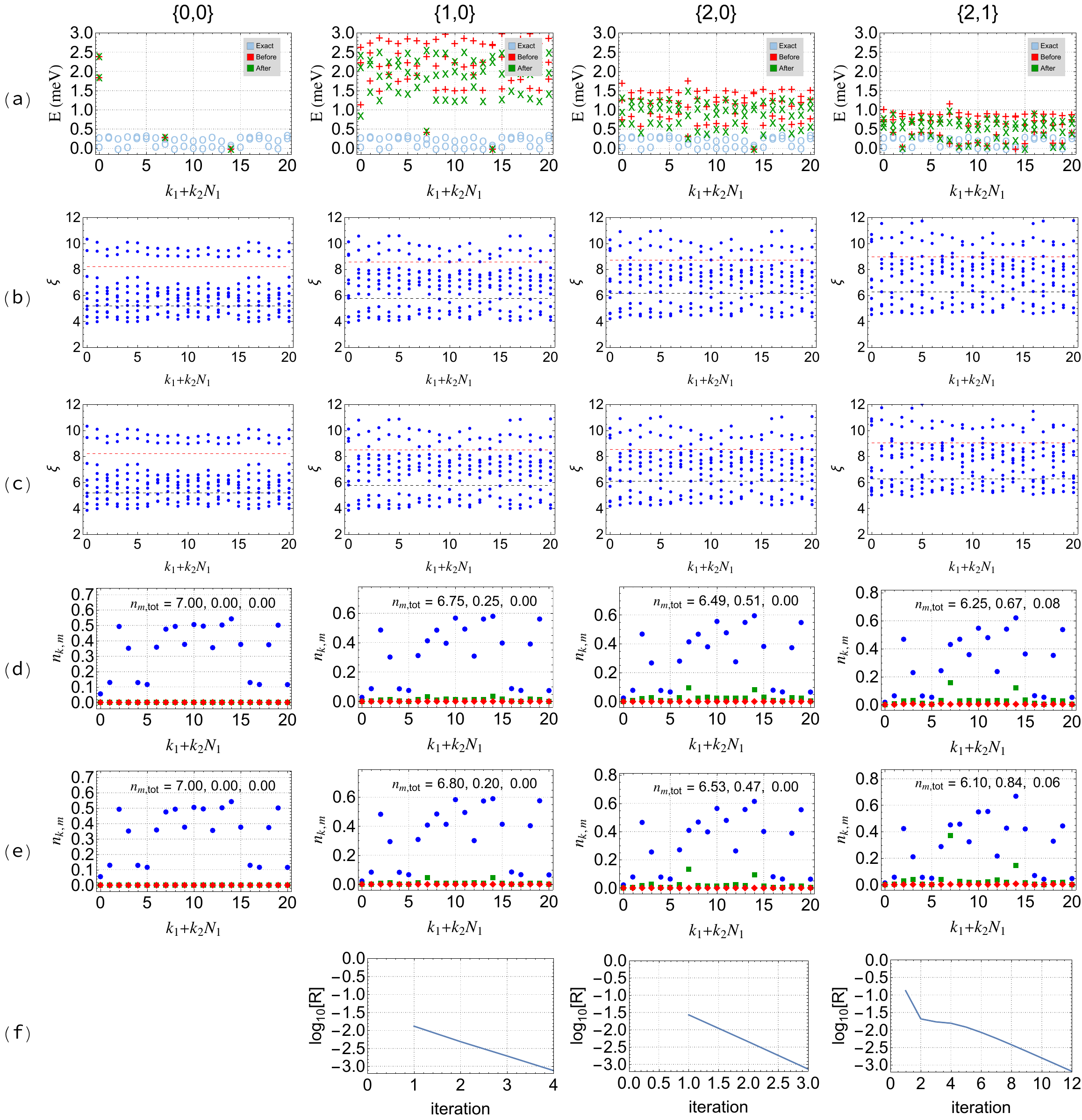}
\caption{Energy spectrum, PES and occupation number of $21\times 1$ system at 1/3 filling with $V=22$ meV in HF basis in AVE scheme. Different columns represent ED results computed at different truncation parameters. (a) Energy spectrum before and after iterations. (b,c) PES of the three lowest states at FCI momenta $0,7,14$ before (b) and after (c) iteration. There are 168 states below the red line which marks the FCI counting, and 63 states below the black line which marks the CDW counting. (d) Occupation number $n_{\bsl k,m}$ as defined in Eq.~\eqref{eq:bandoccupationperk}. Blue, green, and red colors correspond to $m=0,1,2$. The inset shows the total occupation number $n_{m,\text{tot}}$ for band $m=0,1,2$ respectively (see Eq.~\eqref{eq:totalbandoccupation}). (e) Occupation number computed in the basis after iteration. (f) Convergence quantity defined in Eq.~\eqref{eq:convergencedef} as a function of iteration number. The plot at $\{0,0\}$ is empty because the density matrix does not change with iteration in the one-band limit and $\mathcal R$ remains zero.  }
\label{fig_AVE21b1n7}
\end{figure}

\begin{figure}
\centering
\includegraphics[width=6.8 in]{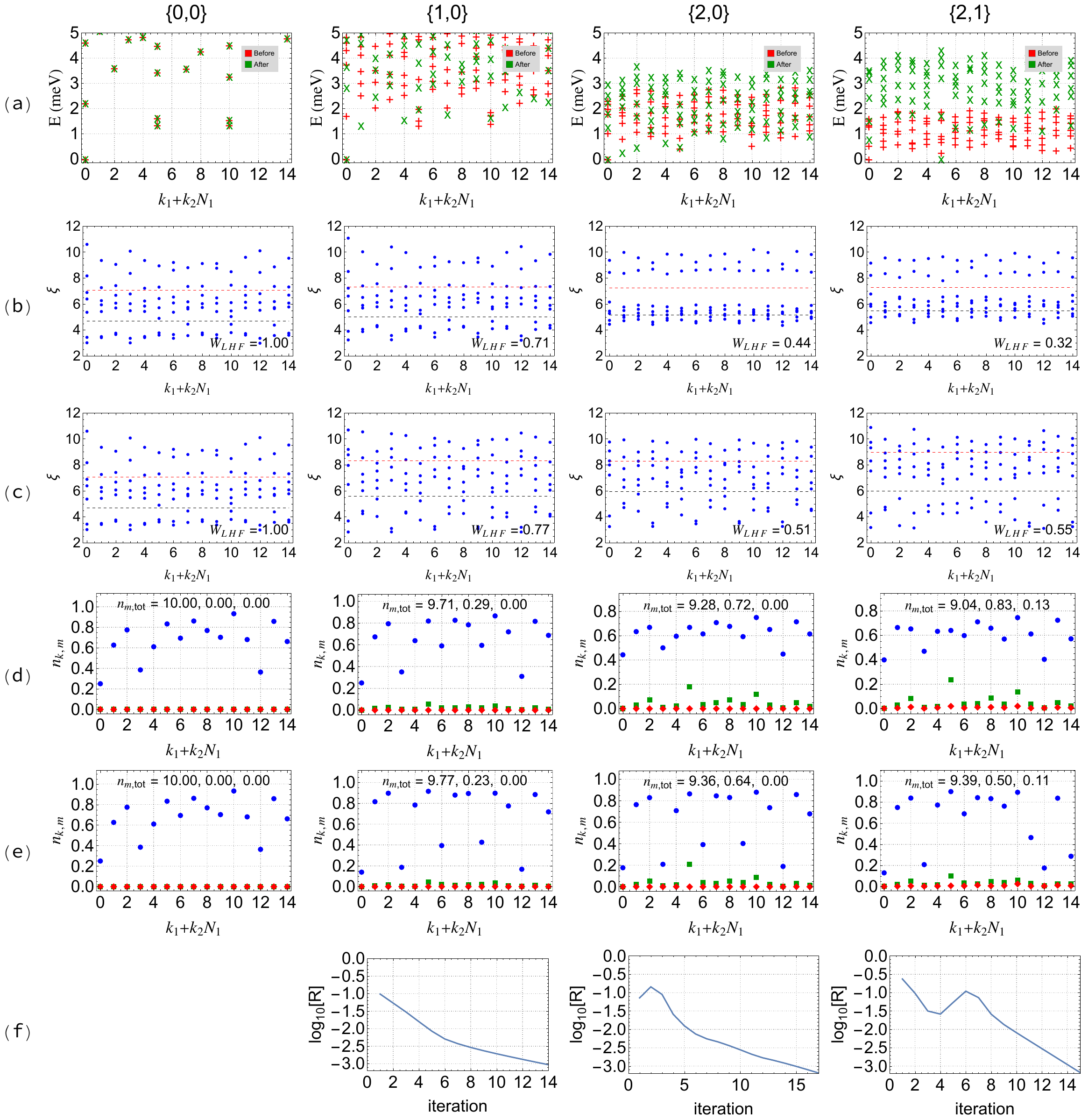}
\caption{
Energy spectrum, PES and occupation number of $15\times 1$ system at 2/3 filling with $V=22$ meV in HF basis in AVE scheme. Different columns represent ED results computed at different truncation parameters. (a) Energy spectrum before and after iterations. (b,c) PES of the three lowest states after single-band projection and PH transform at FCI momenta $0,5,10$ before (b) and after (c) iteration. There are 75 states below the red line which marks the FCI counting, and 30 states below the black line which marks the CDW counting. (d) Occupation number $n_{\bsl k,m}$ as defined in Eq.~\eqref{eq:bandoccupationperk}. Blue, green, and red colors correspond to $m=0,1,2$. The inset shows the total occupation number $n_{m,\text{tot}}$ for band $m=0,1,2$ respectively (see Eq.~\eqref{eq:totalbandoccupation}). (e) Occupation number computed in the basis after iteration. (f) Convergence quantity defined in Eq.~\eqref{eq:convergencedef} as a function of iteration number. The plot at $\{0,0\}$ is empty because the density matrix does not change with iteration in the one-band limit and $\mathcal R$ remains zero.
}
\label{fig_AVE15b1n10}
\end{figure}

\begin{figure}
\centering
\includegraphics[width=\textwidth]{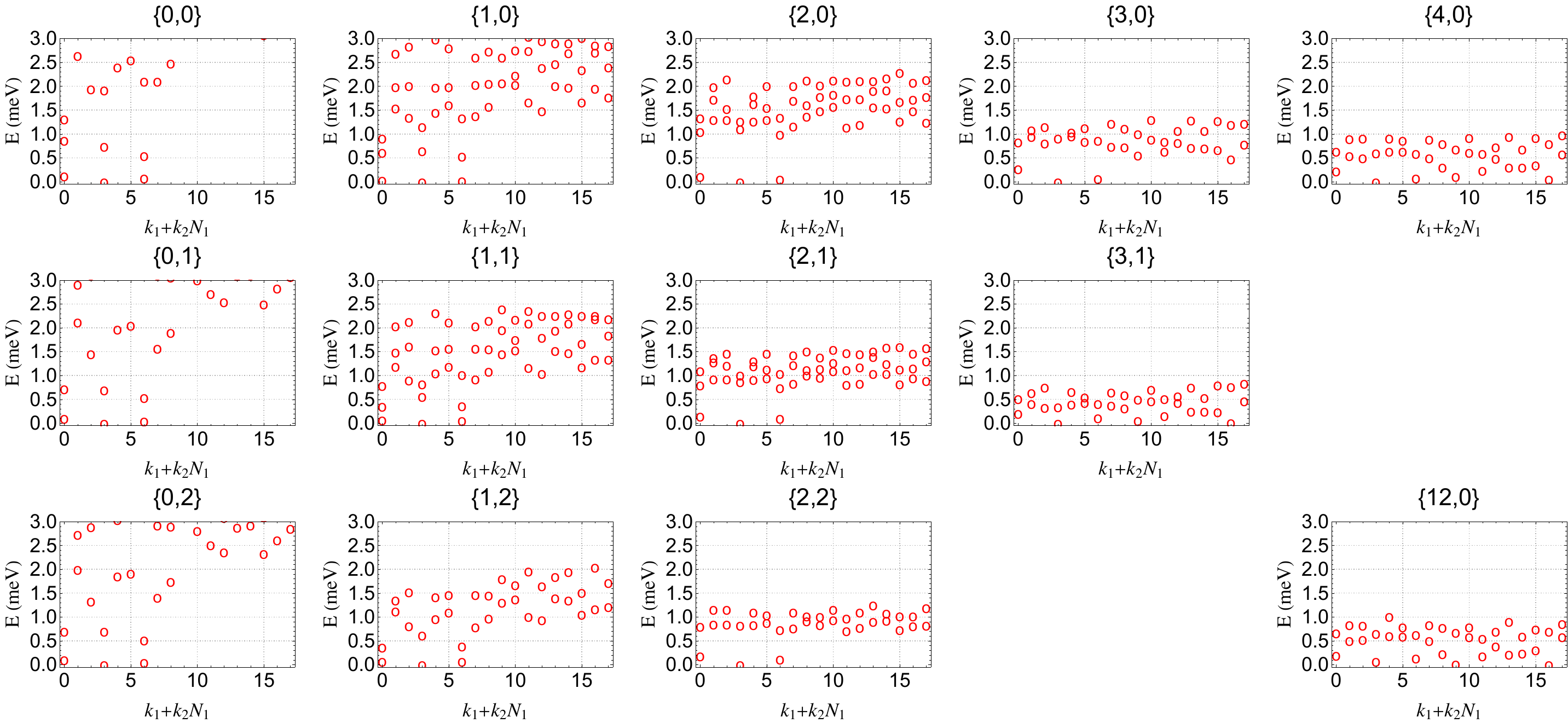}
\caption{
Energy spectrum of $9\times 2$ system at 2/3 filling in AVE scheme with $V=22$ meV at different truncation parameters. When band-mixing is strong, the low energy states are no longer at the FCI momenta. This plot has been computed in Ref.~\cite{MFCIIV}. The energy spectrum at $\{12,0\}$ is shown in the corner for convenience, and the similarity between $\{12,0\}$ and $\{4,0\}$ indicates the convergence of energy spectrum with respect to $N_{\text{band1}}$.
}
\label{fig_EgAVE9b2n12}
\end{figure}

\begin{figure}
\centering
\includegraphics[width=\textwidth]{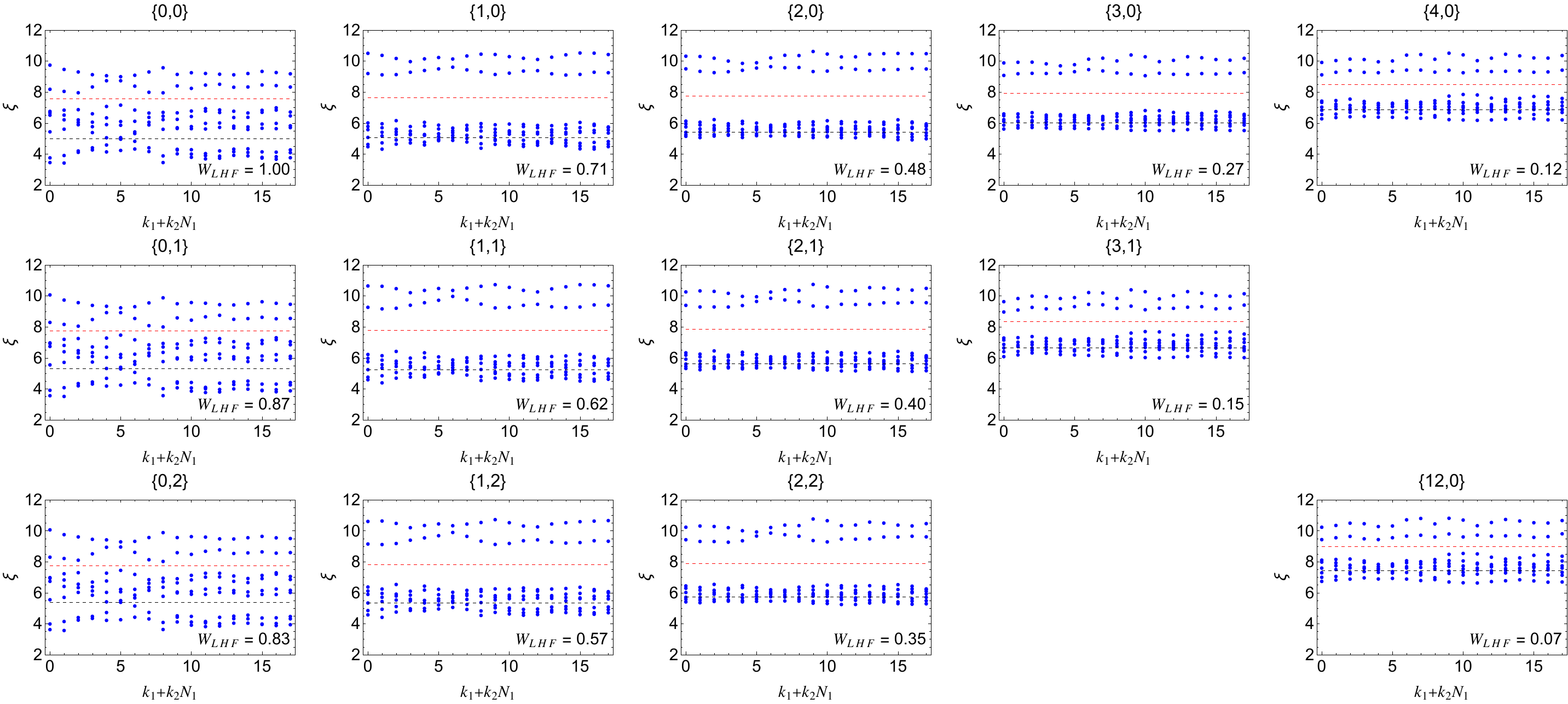}
\caption{
PES of $9\times 2$ system at 2/3 filling in AVE scheme with $V=22$ meV at different truncation parameters. The PES is computed using the lowest states after single-band projection and PH transform at FCI momenta 0,3,6, no matter whether those states are the global lowest energy states. The red line labels the FCI counting and the black line labels the CDW counting. The PES at $\{12,0\}$ is still gapped at FCI counting, and it is shown in the corner for convenience. 
}
\label{fig_PESAVE9b2n12}
\end{figure}

\begin{figure}
\centering
\includegraphics[width=4.8 in]{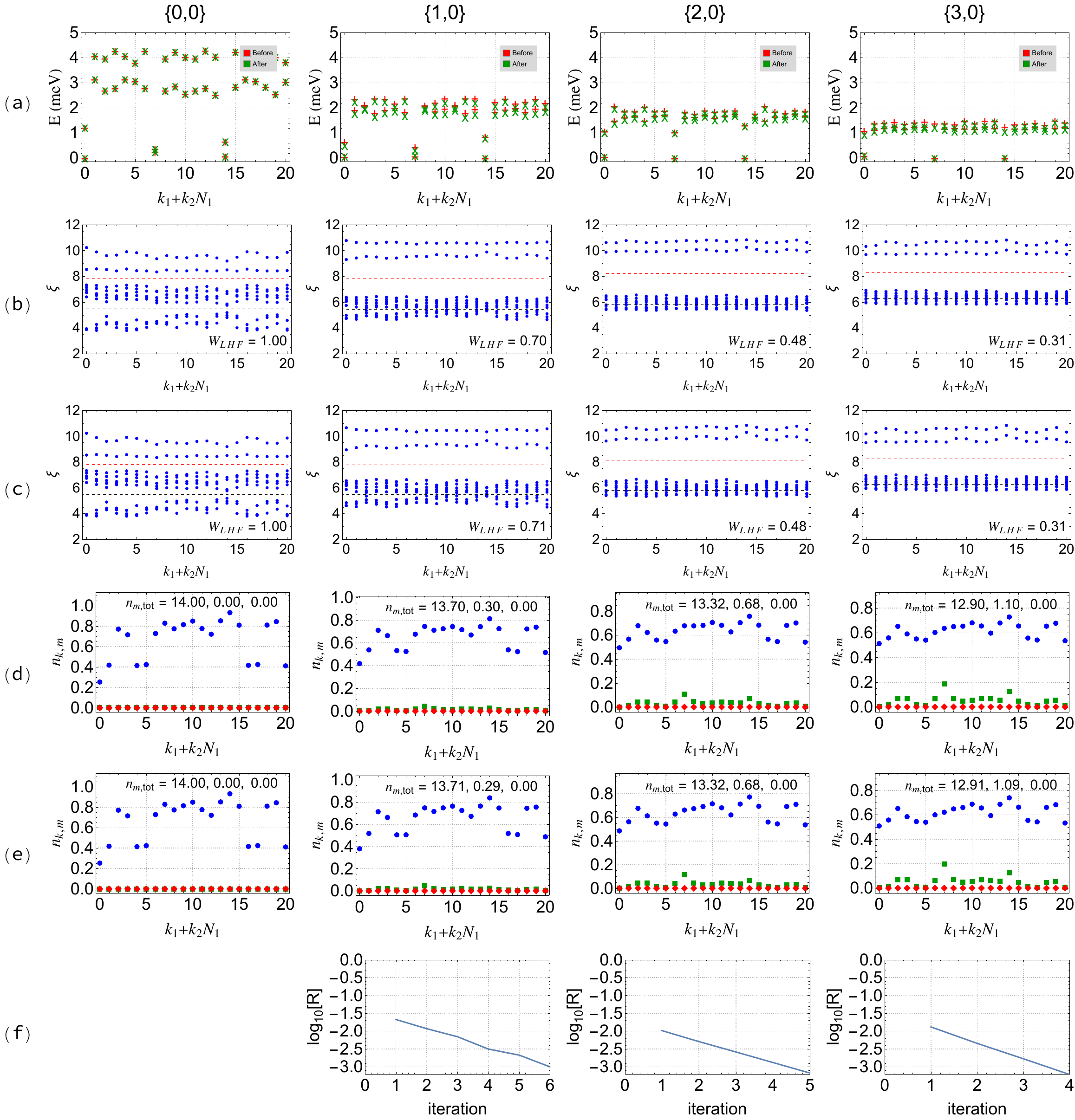}
\caption{Energy spectrum, PES and occupation number of $21\times 1$ system at 2/3 filling with $V=22$ meV in HF basis in AVE scheme. Different columns represent ED results computed at different truncation parameters. (a) Energy spectrum before and after iterations. (b,c) PES of the three lowest states after single-band projection and PH transform at FCI momenta $0,7,14$ before (b) and after (c) iteration. There are 168 states below the red line which marks the FCI counting, and 63 states below the black line which marks the CDW counting. (d) Occupation number $n_{\bsl k,m}$ as defined in Eq.~\eqref{eq:bandoccupationperk}. Blue, green, and red colors correspond to $m=0,1,2$. The inset shows the total occupation number $n_{m,\text{tot}}$ for band $m=0,1,2$ respectively (see Eq.~\eqref{eq:totalbandoccupation}). (e) Occupation number computed in the basis after iteration. (f) Convergence quantity defined in Eq.~\eqref{eq:convergencedef} as a function of iteration number. The plot at $\{0,0\}$ is empty because the density matrix does not change with iteration in the one-band limit and $\mathcal R$ remains zero. }
\label{fig_AVE21b1n14}
\end{figure}

In this appendix, we provide the energy spectrum, PES and occupation number obtained from the ED iteration method for different system sizes and both the CN and AVE scheme, complementing the data provided in Sec.~\ref{Sec_afteriter}. The results are shown from Figs.~\ref{fig_CN15b1n5} to~\ref{fig_AVE21b1n14}. Among these results, the energy spectra, in absence of ED iterations, have also been computed in Ref.~\cite{MFCIIV} and are given for convenience. In most cases the iteration procedure converges with $\mathcal{R}<10^{-3}$ after sufficiently large iterations, but there are exceptions in which the iteration does not converge, which are discussed in App.~\ref{app_notconverge}.

\paragraph*{Filling 1/3 in the CN scheme.} The plots for system sizes $15\times 1$ and $21\times 1$ are shown in Figs.~\ref{fig_CN15b1n5} and~\ref{fig_CN21b1n7} respectively. These two systems exhibit the same features as those discussed in the main text for $9\times 2$: (i) In the single-band limit $\{0,0\}$ the ground state is FCI, from the energy spectrum with three lowest energy states at FCI momenta, the PES with an entanglement gap at FCI counting and the approximate uniform distribution of occupation number. (ii) Under the smallest band-mixing (such as $\{1,0\}$), the ground state is no longer an FCI, but instead it becomes a CDW with ordering vector $K$ as indicated by the structure factor calculations shown in Fig.~\ref{fig_cor1521} and also the PES gap above the CDW counting or the non-uniform occupation number in momentum space. (iii) At large band-mixing, the energy gap and PES gap of CDW collapses, and the ground state in the untruncated Hilbert space is neither CDW nor FCI. (iv) The ED iteration procedure enhances the CDW energy gap and PES gap at small band-mixing without bringing the energy spectrum closer to the full Hilbert space despite the increased occupation number for band 0.

\paragraph*{Filling 2/3 in the CN scheme and filling 1/3 in the AVE scheme.} The plots for $\nu=2/3$ in the CN scheme in system sizes $15\times 1,\ 9\times 2$ and $21\times 1$ are available in Figs.~\ref{fig_CN15b1n10},~\ref{fig_CN9b2n12ch} and~\ref{fig_CN21b1n14} respectively, and for filling 1/3 in the AVE scheme in system sizes $15\times 1,\ 9\times 2$ and $21\times 1$ are shown in Figs.~\ref{fig_AVE15b1n5},~\ref{fig_AVE9b2n6ch} and~\ref{fig_AVE21b1n7} respectively. Note that the data before any ED iteration for $9\times 2$ in the CN scheme at $\nu=2/3$ are identical to those of Sec.~\ref{subsec:CNtwothird} and are provided here for convenience. Moreover, in the single-band limit the ground state is an FCI, but with finite band-mixing the lowest three states for $15\times 1$ and $9\times 2$ systems are no longer at FCI momenta. At $\nu=2/3$ in CN scheme, the FCI in $21\times 1$ system seems to survive larger band-mixing than those in the smaller systems sizes, but the gap still decreases monotonously. Looking at the ED iteration procedures, in most cases the occupation number in band 0 increases after iteration, but there are exceptions. For example, $n_{0,\text{tot}}$ decreases after iteration in $21\times 1$ system at 1/3 filling in the AVE scheme with truncation parameter $\{2,1\}$ (see  App.~\ref{app_notconverge} for more details). Also, the ED iteration procedure neither enhances any PES gaps above the FCI counting nor induces any intermediate CDW phase that would be detected by the PES.

\paragraph*{Filling 2/3 in the AVE scheme.} The plots for system sizes $15\times 1$ and $21\times 1$ are shown in Figs.~\ref{fig_AVE15b1n10} and~\ref{fig_AVE21b1n14} respectively. The results for these two sizes are in line with those  of Fig.~\ref{fig_AVE9b2n12ch} for $9\times 2$ system discussed in the main text. In particular we notice that when the band mixing is increased but remains small (such as \{2,0\}), the FCI gap in the PES is enlarged compared to $\{0,0\}$, indicating an enhancement of FCI in line with the increase of the FCI energy gap observed in Ref.~\cite{MFCIIV}. Note that under the ED iteration procedure, while the total occupation number in band 0 increases, both the FCI gaps in the energy spectrum and PES decrease after convergence. This is in sharp contrast with $\nu=1/3$ in the CN scheme where the ED iteration was able to enlarge both the energy and PES gap of the CDW.

As previously mentioned and contrary to the CN scheme at 1/3 filling where the ground state is an FCI at \{0,0\} and CDW at \{1,0\}, the FCI ground state  in the AVE scheme at $\nu=2/3$ persists at small band-mixing. For sake of completeness, we look at the fate of FCI states in the AVE scheme at 2/3 filling under stronger band-mixing from the PES perspective, backing up the results of Ref.~\cite{MFCIIV} that was focusing on the energy signature. We present the energy spectrum~\cite{MFCIIV} in Fig.~\ref{fig_EgAVE9b2n12} and PES in Fig.~\ref{fig_PESAVE9b2n12} of the $9\times 2$ system across a wide range of truncation parameters. When the band-mixing is weak, the ground states are FCIs at momenta $k_1+k_2N_1=0,3,6$ while the FCI gap collapses and the lowest three states are no longer at FCI momenta when the band mixing becomes larger. The PES gap above the FCI counting does not decrease monotonously when increasing band mixing, i.e., it first increases when moving away from the single band approximation, thus following the same trend as the energy gap. Note that if we focus on the FCI momentum sectors (which correspond now to excited states at large band mixing), the PES still exhibits a clear gap above the FCI counting indicated by the red line in Fig.~\ref{fig_PESAVE9b2n12}, although the weight of projection to the lowest HF band becomes very small.

\subsection{Examples of ED iteration procedure failure}
\label{app_notconverge}

As discussed in Sec.~\ref{Sec_afteriter} and in App.~\ref{app_allfig}, in most cases spanning from different interaction schemes, system sizes or Hilbert space truncation, the total occupation number in band 0 increases after the ED iteration process. Moreover the convergence as defined by $\mathcal{R}^{(\mathcal {N}_{\text{it}})}<10^{-3}$ (see Eq.~\eqref{eq:convergencedef} and Sec.~\ref{Sec_itermethod}) is usually reached after a few tenth of iterations. However as pointed out in the main text, there are some cases where the ED iteration does not converge even with a large number of iterations or where the occupation of band 0 is lower after convergence than its initial value. 

\begin{figure}
\centering
\includegraphics[width=6.0 in]{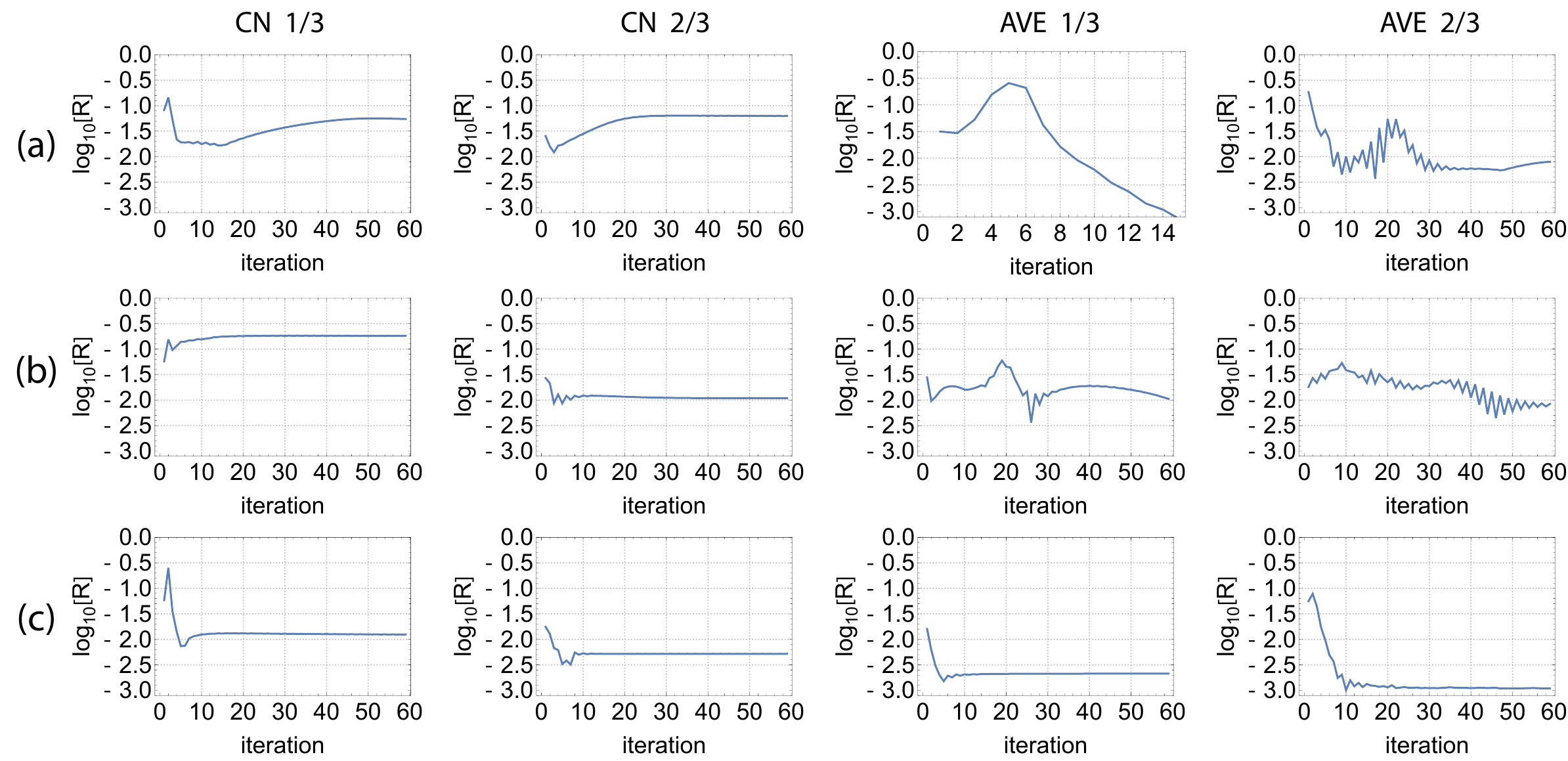}
\caption{
Convergence quantity $\mathcal{R}$ as a function of iteration at truncation parameter $\{1,1\}$ in $15\times 1$ (a), $9\times 2$ (b) and $21\times 1$ (c) systems at 1/3 and 2/3 filling in the CN and AVE schemes.
}
\label{fig_itnotconv}
\end{figure}

We first address the case of the absence of convergence. The non-converged cases occur at truncation parameter $\{1,1\}$, as shown in Fig.~\ref{fig_itnotconv} that solely focuses on this Hilbert space truncation for different filling factors, interaction schemes and sizes. For example, the 9 × 2 system at $\nu=2/3$ in CN scheme has its convergence parameter $\mathcal{R}$ exhibiting  oscillatory behavior with iteration, and at large iteration steps it almost approaches a constant and shows no significant change. In all the systems that  we have studied, the non-convergence behavior at $\{1,1\}$ is observed for both CN and AVE scheme at filling 1/3 and 2/3, except for $15\times 1$ system at 1/3 filling in the AVE scheme. {Non-convergence also occurs for truncations such as $\{1,2\}$ and $\{1,3\}$. We present in App.~\ref{app:analyticaloscillations} an analytical model that captures the reason of these oscillations. In a nutshell, oscillations arise from the switching of ordering in the density matrix eigenvalues between band 1 and band 2.}

\begin{figure}
\centering
\includegraphics[width=5.0 in]{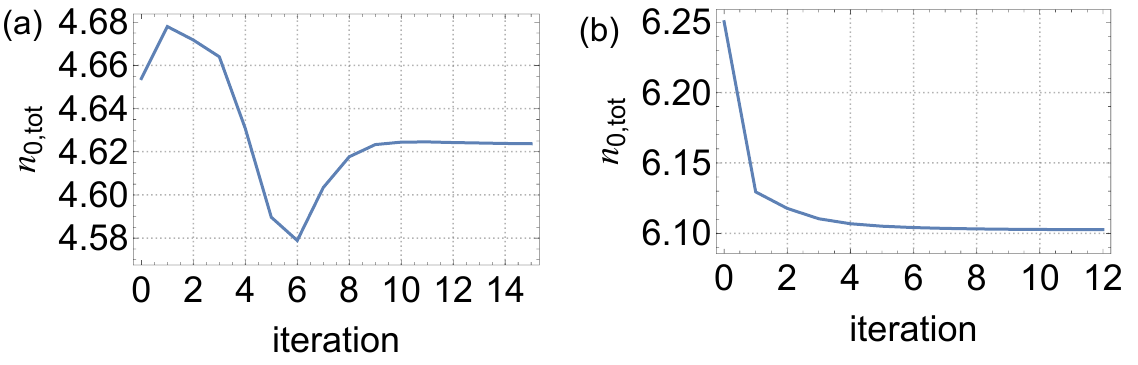}
\caption{
Total occupation number in band 0 as a function of iteration for (a) $15\times 1$ system at 1/3 filling in AVE scheme with truncation parameter $\{1,1\}$ and (b) $21\times 1$ system at 1/3 filling in AVE scheme with truncation parameter $\{2,1\}$. These cases exemplify that the value of $n_{0,\text{tot}}$ after convergence could be smaller than the initial value. Note that the evolution of the convergence parameter is provided in Figs~\ref{fig_AVE15b1n5} and \ref{fig_AVE21b1n7}, respectively. 
}
\label{fig_ntotdecrease}
\end{figure}

We now turn to cases where the ED iteration procedure converges but where the total occupation number $n_{0,\text{tot}}$ in band 0 after convergence is smaller than the initial value. In Table.~\ref{tablenchange} we show the relative change of $n_{0,\text{tot}}$ before iteration and after convergence in different systems. In most cases $n_{0,\text{tot}}$ has increased once convergence is reached, but there are exceptions, as shown in Fig.~\ref{fig_ntotdecrease}. But for all the cases where we have observed this decrease of $n_{0,\text{tot}}$, the lowest three many-body states at FCI momenta are not the ground states of the system, i.e., the three states of the FCI manifold are not the ground state. Note that the converse is not true. For example, if we consider $\nu=1/3$ in the AVE scheme on $9\times 2$ (shown in Fig.~\ref{fig_AVE9b2n6ch}), we have an increase of $n_{0,\text{tot}}$ despite not all of three lowest states being at FCI momenta. 

\begin{figure}
\centering
\includegraphics[width=6.0 in]{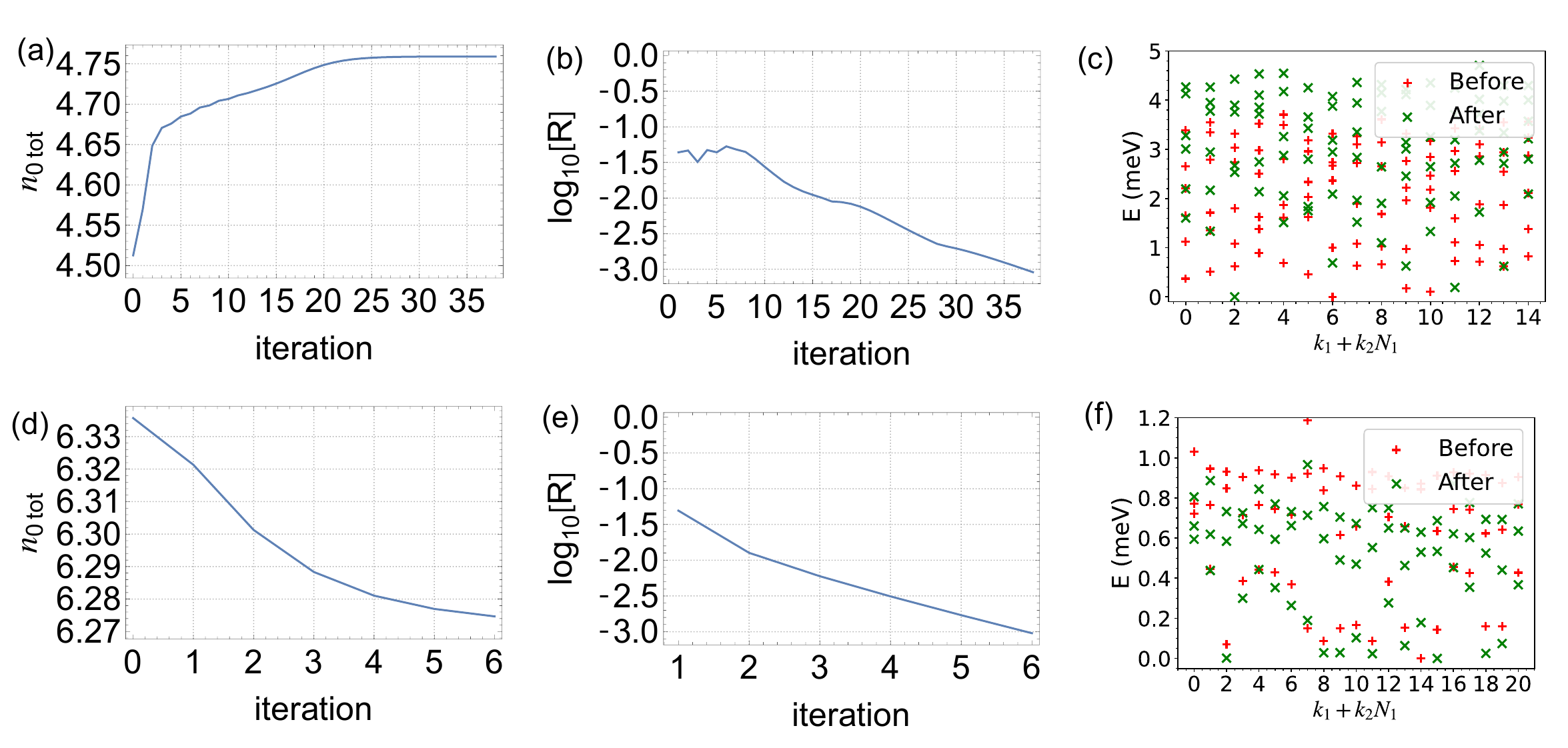}
\caption{ (a-c): ED iteration for $15\times 1$ system at 1/3 filling in AVE scheme  with $\{\text{GS}\}$ made of the lowest energy many-body state at $k_1+k_2N_1=6$ and truncation parameter $\{1,1\}$. (a) Evolution of the total occupation number in band 0. (b) Evolution of the convergence quantity. (c) Energy spectrum before and after the iteration. (d-f): Similar to (a-c) but for $21\times 1$ system at 1/3 filling in AVE scheme with $\{\text{GS}\}$ made of the lowest energy many-body state at $k_1+k_2N_1=14$ and truncation parameter $\{2,1\}$.
}
\label{fig_itern0gs1}
\end{figure}

We could wonder if $n_{0,\text{tot}}$ would increases if we would perform the ED iteration procedure relying on the absolute (and single) ground state rather than the 3 lowest states at the FCI momenta. To test this hypothesis, we denote the momentum sector of this many-body ground state as $k_0$ before any iteration, we fix $\{\text{GS}\}$ as the lowest state at the same momentum $k_0$, and we compute $n_{0,\text{tot}}$ for this state. The evolution with iteration is shown in Fig.~\ref{fig_itern0gs1} for the cases pointed out in Table.~\ref{tablenchange}, namely $15\times 1$ at $\nu=2/3$ and $21\times 1$ at $\nu=1/3$ both the AVE scheme. In these two cases, the momentum sector of the absolute ground state after convergence is no longer $k_0$. Moreover, while $n_{0,\text{tot}}$ for $15\times 1$ system at 1/3 filling now increases after convergence as opposed to the ED iteration using three states in $\{\text{GS}\}$, the $21\times 1$ system at 1/3 filling still exhibits a decrease of $n_{0,\text{tot}}$. These examples illustrate that the iteration procedure does not guaranteed to increase the occupation number in band 0, even though such cases are rare.

\begin{table}[htbp]
\centering
\begin{tabular}{lllccccc}
\hline
Size & Scheme & Filling  & \{1,0\} & \{2,0\}& \{1,1\} & \{2,1\} & \{3,2\} \\
\hline
$15\times 1$ & CN  & $1/3$ & 0.0485 & 0.0630 & \text{N.C.}  & 0.0747 & 0.0612 \\
$15\times 1$ & CN  & $2/3$ & 0.0005 & 0.0023 & \text{N.C.} & 0.0040 &        \\
$15\times 1$ & AVE & $1/3$ & 0.0002 & 0.0020 & {\bf -0.0065}    & 0.0569 & 0.1940 \\
$15\times 1$ & AVE & $2/3$ & 0.0063 & 0.0089 & \text{N.C.} & 0.0384 &        \\
$9\times 2$  & CN  & $1/3$ & 0.0384 & 0.0382 & \text{N.C.}  & 0.0538 & 0.0453 \\
$9\times 2$  & CN  & $2/3$ & 0.0004 & 0.0018 & \text{N.C.} & 0.0024 &        \\
$9\times 2$  & AVE & $1/3$ & 0.0106 & 0.0068 & \text{N.C.}  & 0.0333 & 0.1293 \\
$9\times 2$  & AVE & $2/3$ & 0.0048 & 0.0050 & \text{N.C.} & 0.0229 &        \\
$21\times 1$ & CN  & $1/3$ & 0.0352 & 0.0466 & \text{N.C.} & 0.0548 &        \\
$21\times 1$ & CN  & $2/3$ & 0.0004 & 0.0015 & \text{N.C.}  &        &        \\
$21\times 1$ & AVE & $1/3$ & 0.0064 & 0.0076 & \text{N.C.} & {\bf -0.0236} &       \\
$21\times 1$ & AVE & $2/3$ & 0.0004 & 0.0001 & \text{N.C.}   &        &        \\
\hline
\end{tabular}
\caption{ The relative change of total occupation number in band 0 after convergence $n_{0,\text{tot}}^{(\mathcal{N}_{\text{it}})}$ with respect to that before any iteration $n_{0,\text{tot}}^{(0)}$ for different systems and different truncation parameters. The relative change is defined as $\left(n_{0,\text{tot}}^{(\mathcal{N}_{\text{it}})}-n_{0,\text{tot}}^{(0)}\right)/n_{0,\text{tot}}^{(0)}$. N.C. indicates when convergence cannot be reached and entries are left empty when the calculations cannot be performed due to computational resources. In most of the cases, the occupation number in band 0 increases after convergence. But a few exceptions, corresponding to negative relative changes, can be found (in bold).}
\label{tablenchange}
\end{table}

\section{Analytical model for ED iteration procedures without convergence }\label{app:analyticaloscillations}

In this appendix, we provide a simple analytical model as an example to demonstrate that there exist certain cases where the ED iteration procedure cannot converge.

\subsection{A simple model for oscillations at truncation $\{1,1\}$}

\begin{figure}
\centering
\includegraphics[width=6.8 in]{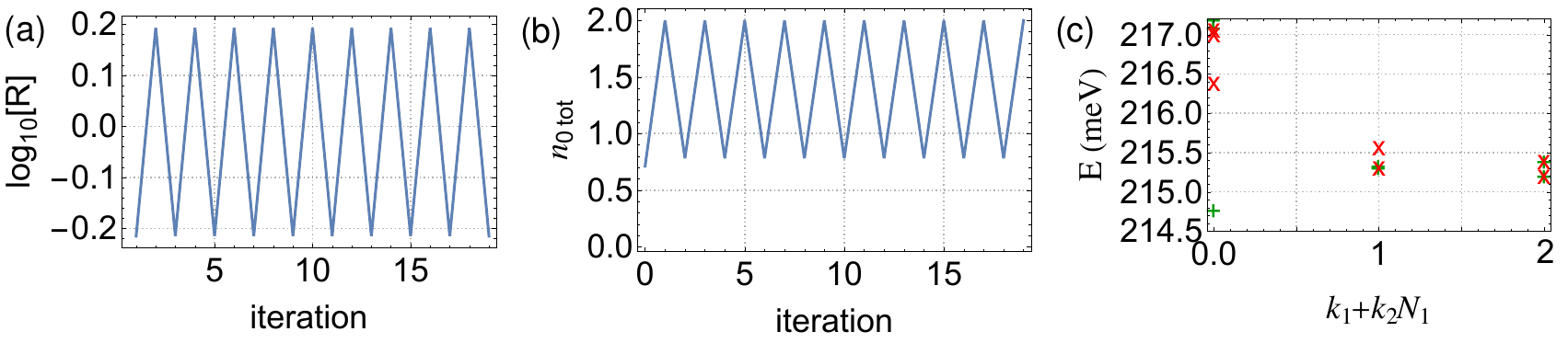}
\caption{
(a) Convergence criteria as a function of iteration step for $1\times 3$ system in CN scheme with 2 particles at truncation parameter $\{1,1\}$. The iteration procedure exhibits oscillatory behavior without convergence. (b) Total occupation number in band 0 as a function of iteration step. (c) The energy spectrum of the two oscillating solutions.
}
\label{fig_oscillate1b3}
\end{figure}

Consider a small $1\times 3$ system with only $\Gamma,K,K'$ points in the CN scheme with two particles, i.e., $\nu=2/3$. We find that the ED iteration result in oscillations between two distinct solutions without convergence at truncation parameter $\{1,1\}$ with $\{\text{GS}\}$ containing the lowest state at $\Gamma$. In Fig.~\ref{fig_oscillate1b3}, we show the evolution of convergence criteria and occupation number in band 0 as a function of iteration steps, both exhibiting clear oscillatory behavior. We plot the energy spectrum of the two oscillating solutions in Fig.~\ref{fig_oscillate1b3}(c). To find out the origin of this oscillation, we investigate the Hilbert space and the Hamiltonian. Before the Hilbert space truncation, there are 12 many-body states in the total momentum sector $\Gamma$, with three of them having both particles at momentum $\Gamma$ and 9 of them having one particle at $K$ and one at $K'$. The three states with double occupancy at $\Gamma$ have very high energy because the energies of the second- and third-lowest conduction band at $\Gamma$ are high, as seen in Fig.~\ref{fig_bandbare} in the main text. Hence those states have negligible contribution to the ground state. With truncation $\{1,1\}$, the two states with both particles in either band 1 or band 2 are discarded. Hence there are 7 states remaining that can have significant contributions to the ground state, which are listed as follows:
\bea
\{\Psi^{0}\}&=&\{d^\dagger_{\bsl k_a,0}d^\dagger_{\bsl k_b,0}|0\rangle,d^\dagger_{\bsl k_a,2}d^\dagger_{\bsl k_b,1}|0\rangle ,d^\dagger_{\bsl k_a,1}d^\dagger_{\bsl k_b,2}|0\rangle,\nonumber\\
&&\ \ d^\dagger_{\bsl k_a,0}d^\dagger_{\bsl k_b,2}|0\rangle,d^\dagger_{\bsl k_a,0}d^\dagger_{\bsl k_b,1}|0\rangle,d^\dagger_{\bsl k_a,2}d^\dagger_{\bsl k_b,0}|0\rangle ,d^\dagger_{\bsl k_a,1}d^\dagger_{\bsl k_b,0}|0\rangle\},
\label{psiall}
\eea
where $\bsl k_a=K$ and $\bsl k_b=K'$. In general, the oscillation does not have to occur at $K$ and $K'$, hence in the following we use the generic labels $\bsl k_a$ and $\bsl k_b$. With sufficient number of iteration steps (here around two steps), the results oscillate between two distinct solutions with different Hamiltonian matrices $H_a$ and $H_b$. We further find that the ground state of each Hamiltonian can be captured by at most the first three states in $\{\Psi^0\}$ due to the $C_3$ symmetry. As an example, we present the numerical value of these Hamiltonians as follows:
\bea
|H_a|=\left(
\begin{array}{ccccccc}
 0 & 0.97 & 0.94 & 0 & 0 & 0 & 0 \\
 0.97 & 0.26 & 0.91 & 0 & 0 & 0 & 0 \\
 0.94 & 0.91 & 0.40 & 0 & 0 & 0 & 0 \\
 0 & 0 & 0 & 1.70 & 0 & 0.51 & 0 \\
 0 & 0 & 0 & 0 & 7.92 & 0 & 7.09 \\
 0 & 0 & 0 & 0.51 & 0 & 1.82 & 0 \\
 0 & 0 & 0 & 0 & 7.09 & 0 & 8.20 \\
\end{array}
\right),\label{eq:ha}
\eea
\bea
|H_b|=\left(
\begin{array}{ccccccc}
 0 & 0 & 0 & 0 & 0 & 0 & 0 \\
 0 & 1.51 & 0 & 0.57 & 0 & 0 & 0.52 \\
 0 & 0 & 8.37 & 0 & 6.97 & 6.97 & 0 \\
 0 & 0.57 & 0 & 1.70 & 0 & 0 & 0.51 \\
 0 & 0 & 6.97 & 0 & 7.92 & 7.09 & 0 \\
 0 & 0 & 6.97 & 0 & 7.09 & 8.20 & 0 \\
 0 & 0.52 & 0 & 0.51 & 0 & 0 & 1.82 \\
\end{array}
\right).\label{eq:hb}
\eea
Here the notation $|H|$ means that we only give the magnitude of each matrix element, and the diagonal elements are shifted by an overall constant. The appearance of zero elements is due to $C_3$ symmetry. Since both $\bsl k_a$ and $\bsl k_b$ are invariant under $C_3$, the Hamiltonian can only couple states with the same $C_3$ eigenvalue. For $H_a$, the $C_3$ eigenvalues are $\omega^*,\omega,1$ for bands $0,1,2$ respectively at both $\bsl k_a$ and $\bsl k_b$ with $\omega=e^{\frac{2\pi i}{3}}$. For $H_b$, the $C_3$ eigenvalues at $\bsl k_a$ is the same as $H_a$, whereas at $\bsl k_b$ they become $\omega^*,1,\omega$ due to a rotation of basis by the iteration procedure. Then the $C_3$ eigenvalues of many-body states in Eq.\eqref{psiall} are $\omega,\omega,\omega,\omega^*,1,\omega^*,1$ for $H_a$ and $\omega,1,\omega^*,1,\omega^*,\omega^*,1$ for $H_b$. This explains the location of the zero matrix elements. While the  $C_3$ symmetry simplifies our Hamiltonians, it should be noted that this symmetry is not crucial since oscillations also occur for momentum meshes that are not $C_3$ invariant {such as $9 \times 2$}.

The ground state of $H_a$ has finite component only in the first three states, and the ground state of $H_b$ is fully in the first state. This motivates us to build a model made of the first three states in Eq.\eqref{psiall} and investigate its behavior under iteration. However, the first three states are not closed under iteration: Indeed, the rotation of basis during iteration will involve two addition states $d^\dagger_{\bsl k_a,1}d^\dagger_{\bsl k_b,1}|0\rangle$ and $d^\dagger_{\bsl k_a,2}d^\dagger_{\bsl k_b,2}|0\rangle$. Therefore we take these two additional states into account and consider a basis made of the following five states:
\be
\{\Psi\}=\{d^\dagger_{\bsl k_a,0}d^\dagger_{\bsl k_b,0}|0\rangle,d^\dagger_{\bsl k_a,2}d^\dagger_{\bsl k_b,1}|0\rangle ,d^\dagger_{\bsl k_a,1}d^\dagger_{\bsl k_b,2}|0\rangle,d^\dagger_{\bsl k_a,1}d^\dagger_{\bsl k_b,1}|0\rangle,d^\dagger_{\bsl k_a,2}d^\dagger_{\bsl k_b,2}|0\rangle   \},
\label{statessel}
\ee

Next we will show that for a generic Hamiltonian in the space spanned by the states in Eq.\eqref{statessel}, as long as certain condition is met (Eq.\eqref{assumecab} below), the ED iteration procedure with truncation $\{1,1\}$ will oscillate between two different ground states without convergence. We begin with the generic form of Hamiltonian in this space, which is written as:
\bea
H=\left(\begin{array}{ccccc}
h_{00} & h_{01} & h_{02} & h_{03}& h_{04}  \\
h_{10} & h_{11} & h_{12}& h_{13}& h_{14}  \\
h_{20} & h_{21} & h_{22}& h_{23}& h_{24} \\
h_{30} & h_{31} & h_{32}& h_{33}& h_{34}\\
h_{40} & h_{41} & h_{42}& h_{43}& h_{44}
\end{array}\right).
\label{Horig}
\eea
During the ED iteration procedure, some of the states in $\{\Psi\}$ can be removed due to the truncation of Hilbert space. Hence it would be useful to discuss the truncated Hamiltonian. If the last two states $d^\dagger_{\bsl k_a,1}d^\dagger_{\bsl k_b,1}|0\rangle$ and $d^\dagger_{\bsl k_a,2}d^\dagger_{\bsl k_b,2}|0\rangle$ are removed, the truncated Hamiltonian formed by the first three states is denoted as $H^{012}$. If the states $d^\dagger_{\bsl k_a,2}d^\dagger_{\bsl k_b,1}|0\rangle$ and $d^\dagger_{\bsl k_a,1}d^\dagger_{\bsl k_b,2}|0\rangle$ are removed, the truncated Hamiltonian denoted as $H^{034}$. They are given by:
\bea
H^{012}=\left(\begin{array}{ccc}
h_{00} & h_{01} & h_{02}  \\
h_{10} & h_{11} & h_{12}  \\
h_{20} & h_{21} & h_{22}
\end{array}\right),\ \ H^{034}=\left(\begin{array}{ccc}
h_{00} & h_{03} & h_{04}  \\
h_{30} & h_{33} & h_{34}  \\
h_{40} & h_{43} & h_{44}
\end{array}\right).
\label{H012034}
\eea
Denote the lowest eigenstate of $H^{012}$ and $H^{034}$ by $\psi^{012}$ and $\psi^{034}$ respectively, which are written as
\be
\psi^{012}=(c,a,b)^T,\ \ \psi^{034}=(c',a',b')^T
\ee
The iteration procedure tends to increase the occupation in band 0, hence it is expected that the largest weight is in the state $d^\dagger_{\bsl k_a,0}d^\dagger_{\bsl k_b,0}|0\rangle$, i.e., $c$ and $c'$ are the largest component in these eigenvectors:
\be
|c|>\text{max}(|a|,|b|),\ \ |c'|>\text{max}(|a'|,|b'|).
\label{assumecab}
\ee
Such a property is obvious for the $H_b$ given by Eq.\eqref{eq:hb}, and for $H_a$ we get $|c|=0.63$. In the following, we will show that the iteration procedure with truncation parameter $\{1,1\}$ will oscillate without convergence when the condition in Eq.\eqref{assumecab} is valid.

The truncation $\{1,1\}$ removes the states $d^\dagger_{\bsl k_a,1}d^\dagger_{\bsl k_b,1}|0\rangle$ and $d^\dagger_{\bsl k_a,2}d^\dagger_{\bsl k_b,2}|0\rangle$, hence the truncated Hamiltonian is $H^{012}$ with lowest eigenstate $\psi^{012}$, and the truncated basis is
\bea
\{\Psi^{\{1,1\}}\}&=&\{d^\dagger_{\bsl k_a,0}d^\dagger_{\bsl k_b,0}|0\rangle,d^\dagger_{\bsl k_a,2}d^\dagger_{\bsl k_b,1}|0\rangle ,d^\dagger_{\bsl k_a,1}d^\dagger_{\bsl k_b,2}|0\rangle   \}.
\label{psi012}
\eea
The density matrix can be computed to be:
\bea
\rho(\bsl k_a)=\left(\begin{array}{ccc}
|c|^2 & 0 & 0  \\
 0& |b|^2 & 0  \\
0 & 0 & |a|^2 
\end{array}\right),\ \ \rho(\bsl k_b)=\left(\begin{array}{ccc}
|c|^2 & 0 & 0  \\
 0& |a|^2 & 0  \\
0 & 0 & |b|^2 
\end{array}\right).
\eea
Notice that the density matrices are diagonal because the states in Eq.\eqref{psi012} cannot be transformed into each other by changing the band index at one momentum point. During the ED iteration, we use a unitary matrix to diagonalize the density matrix and bring the eigenvalues in descending order. Depending on the relative magnitude of $a$ and $b$, different unitary matrices are obtained. Without loss of generality, we focus on the case with $|a|>|b|$, and the other case with $|a|<|b|$ can be analyzed similarly by interchanging $\bsl k_a$ and $\bsl k_b$. There are situations where symmetry requires $|a|=|b|$, but numerically $|a|$ and $|b|$ cannot be exactly equal at machine precision, hence it still follows the discussion of either $|a|>|b|$ or $|a|<|b|$. When $|a|>|b|$, to bring the eigenvalues of density matrices in descending order $|c|^2>|a|^2>|b|^2$, the unitary matrices take the form:
\be
U(\bsl k_a)=\left(\begin{array}{ccc}
1 & 0 & 0  \\
 0& 0 & 1  \\
0 & 1 & 0 
\end{array}\right),\ \ U(\bsl k_b)=\left(\begin{array}{ccc}
1 & 0 & 0  \\
 0& 1 & 0  \\
0 & 0 & 1 
\end{array}\right).
\label{Uflip}
\ee
In the next iteration step, the one-body basis $d^\dagger_{\bsl k,n}$ will be rotated according to these unitary matrices, which flips $d^\dagger_{\bsl k_a,1}$ and $d^\dagger_{\bsl k_a,2}$. Explicitly, the rotated basis with $\tilde d^\dagger_{\bsl k,n}$ is given by
\bea
\tilde d^\dagger_{\bsl k_a,0}= d^\dagger_{\bsl k_a,0},\ \tilde d^\dagger_{\bsl k_a,1}= d^\dagger_{\bsl k_a,2},\ \tilde d^\dagger_{\bsl k_a,2}= d^\dagger_{\bsl k_a,1} ,
\eea
whereas $\tilde d^\dagger_{\bsl k_b,n}=d^\dagger_{\bsl k_b,n}$ since $U(\bsl k_b)$ is identity matrix. The new basis under truncation $\{1,1\}$ becomes
\bea
\{\tilde\Psi^{\{1,1\}}\}&=&\{\tilde d^\dagger_{\bsl k_a,0}\tilde d^\dagger_{\bsl k_b,0}|0\rangle,\tilde d^\dagger_{\bsl k_a,2}\tilde d^\dagger_{\bsl k_b,1}|0\rangle ,\tilde d^\dagger_{\bsl k_a,1}\tilde d^\dagger_{\bsl k_b,2}|0\rangle   \}\\
&=&\{d^\dagger_{\bsl k_a,0}d^\dagger_{\bsl k_b,0}|0\rangle,d^\dagger_{\bsl k_a,1}d^\dagger_{\bsl k_b,1}|0\rangle ,d^\dagger_{\bsl k_a,2}d^\dagger_{\bsl k_b,2}|0\rangle   \}
\label{tildepsi012}
\eea
Comparing with the untruncated states in Eq.\eqref{statessel}, $\{\tilde\Psi^{\{1,1\}}\}$ keeps the first, fourth and fifth states. Therefore the new truncated Hamiltonian in basis $\{\tilde\Psi^{\{1,1\}}\}$ becomes identical to $H^{034}$ in Eq.\eqref{H012034}, and the lowest eigenstate becomes $\psi^{034}=(c',a',b')^T$. The density matrices in the $\tilde d^\dagger_{\bsl k,n}$ basis obtained from $\psi^{034}$ are:
\bea
\tilde\rho(\bsl k_a)=\left(\begin{array}{ccc}
|c'|^2 & 0 & 0  \\
 0& |b'|^2 & 0  \\
0 & 0 & |a'|^2 
\end{array}\right),\ \ \tilde\rho(\bsl k_b)=\left(\begin{array}{ccc}
|c'|^2 & 0 & 0  \\
 0& |a'|^2 & 0  \\
0 & 0 & |b'|^2 
\end{array}\right).
\eea

First we look at the case with $|a'|>|b'|$. In this case we get unitary matrices with the same form as Eq.\eqref{Uflip} since $|c'|^2>|a'|^2>|b'|^2$:
\be
\tilde U(\bsl k_a)=\left(\begin{array}{ccc}
1 & 0 & 0  \\
 0& 0 & 1  \\
0 & 1 & 0 
\end{array}\right),\ \ \tilde U(\bsl k_b)=\left(\begin{array}{ccc}
1 & 0 & 0  \\
 0& 1 & 0  \\
0 & 0 & 1 
\end{array}\right).
\label{tildeUflip}
\ee
These unitary matrices will flip $\tilde d^\dagger_{\bsl k_a,1}$ and $\tilde d^\dagger_{\bsl k_a,2}$ again. Therefore, after this iteration step the new basis $\tilde{\tilde d}^\dagger_{\bsl k,n}$ will be the same as $d^\dagger_{\bsl k,n}$, and we are back to the Hamiltonian $H^{012}$ that we started from. Hence during the iteration procedure with truncation parameter $\{1,1\}$, the ground state will oscillate back and forth between $\psi^{012}$ and $\psi^{034}$ without convergence.

Next we look at the case with $|a'|<|b'|$ while keeping $|a|>|b|$. Then the unitary matrices in Eq.\eqref{tildeUflip} become
\be
\tilde U'(\bsl k_a)=\left(\begin{array}{ccc}
1 & 0 & 0  \\
 0& 1 & 0  \\
0 & 0 & 1 
\end{array}\right),\ \ \tilde U'(\bsl k_b)=\left(\begin{array}{ccc}
1 & 0 & 0  \\
 0& 0& 1  \\
0 & 1 & 0 
\end{array}\right).
\label{tildeUflipp}
\ee
This transformation leads to a flip between band 1 and band 2 at $\bsl k_b$ rather than $\bsl k_a$. Then the basis in the next iteration step becomes
\bea
\{\tilde{\tilde{\Psi}}^{\{1,1\}}\}&=&\{\tilde{\tilde{d}}^\dagger_{\bsl k_a,0}\tilde{\tilde{d}}^\dagger_{\bsl k_b,0}|0\rangle,\tilde{\tilde{d}}^\dagger_{\bsl k_a,2}\tilde{\tilde{d}}^\dagger_{\bsl k_b,1}|0\rangle ,\tilde{\tilde{d}}^\dagger_{\bsl k_a,1}\tilde{\tilde{d}}^\dagger_{\bsl k_b,2}|0\rangle   \}\\
&=&\{\tilde d^\dagger_{\bsl k_a,0}\tilde d^\dagger_{\bsl k_b,0}|0\rangle,\tilde d^\dagger_{\bsl k_a,2}\tilde d^\dagger_{\bsl k_b,2}|0\rangle ,\tilde d^\dagger_{\bsl k_a,1}\tilde d^\dagger_{\bsl k_b,1}|0\rangle   \} \\
&=&\{ d^\dagger_{\bsl k_a,0} d^\dagger_{\bsl k_b,0}|0\rangle, d^\dagger_{\bsl k_a,1}d^\dagger_{\bsl k_b,2}|0\rangle , d^\dagger_{\bsl k_a,2}d^\dagger_{\bsl k_b,1}|0\rangle   \}
\label{ttildepsi}
\eea
Comparing with the untruncated states in Eq.\eqref{statessel}, the new Hamiltonian in the next iteration becomes
\bea
H^{021}=\left(\begin{array}{ccc}
h_{00} & h_{02} & h_{01}  \\
h_{20} & h_{22} & h_{21}  \\
h_{10} & h_{12} & h_{11}
\end{array}\right),
\eea
with lowest eigenstate given by
\be
\psi^{021}=(c,b,a)^T.
\ee
The density matrices computed from $\psi^{021}$ are:
\bea
\tilde{\tilde{\rho}}(\bsl k_a)=\left(\begin{array}{ccc}
|c|^2 & 0 & 0  \\
 0& |a|^2 & 0  \\
0 & 0 & |b|^2 
\end{array}\right),\ \ \tilde{\tilde{\rho}}(\bsl k_b)=\left(\begin{array}{ccc}
|c|^2 & 0 & 0  \\
 0& |b|^2 & 0  \\
0 & 0 & |a|^2 
\end{array}\right).
\eea
The unitary matrices are:
\be
\tilde{\tilde{U}}(\bsl k_a)=\left(\begin{array}{ccc}
1 & 0 & 0  \\
 0& 1 & 0  \\
0 & 0 & 1 
\end{array}\right),\ \ \tilde{\tilde{U}}(\bsl k_b)=\left(\begin{array}{ccc}
1 & 0 & 0  \\
 0& 0& 1  \\
0 & 1 & 0 
\end{array}\right).
\label{tildeUflippp}
\ee
This transformation will flip band 1 and 2 again at $\bsl k_b$, which will change the one-body basis from $\tilde{\tilde{d}}_{\bsl k,n}$ back to $\tilde d_{\bsl k,n}$, and change the truncated Hamiltonian back to $H^{034}$. Therefore, during the iteration procedure the ground state will oscillate between $\psi^{034}$ and $\psi^{021}$ without convergence when $|a'|<|b'|$. Combining with the previous conclusion for the case when $|a'|>|b'|$, we have shown that the iteration will not converge as long as Eq.\eqref{assumecab} is valid, no matter the relative magnitude between $a,b$ and $a',b'$.

This analytical example, inspired from the smallest system size exhibiting oscillations, shows that certain cases where the iteration with truncation parameter $\{1,1\}$ will oscillate between two ground states without convergence, i.e., when the conditions in Eq.\eqref{assumecab} are satisfied for the Hamiltonian in Eq.\eqref{Horig}. Note that we do not require this condition to be satisfied in the Hamiltonian before iteration. As long as the iteration procedure makes this condition valid for the Hamiltonian in the rotated basis, this analysis will be applicable.

\subsection{Convergence (or lack of) for truncations $\{2,1\},\{1,2\}$ and $\{2,2\}$ }

The model built from the five states in Eq.\eqref{statessel} to study the truncation $\{1,1\}$ can also reveal the reason why the oscillation in iteration also occurs for $\{1,2\}$ but not at other truncations like $\{2,1\}$ or $\{2,2\}$.  For truncation $\{2,2\}$, none of the five states in Eq.\eqref{statessel} is truncated, hence the iteration will converge after one step, and the oscillation will not happen. Therefore we mainly focus the comparison between truncation $\{1,2\}$ and $\{2,1\}$.

Both truncations $\{2,1\}$ and $\{1,2\}$ allow one more state in addition to the three states allowed by $\{1,1\}$ (defined in Eq.\eqref{psi012}). The truncation $\{2,1\}$ allows $d^\dagger_{\bsl k_a,1}d^\dagger_{\bsl k_b,1}|0\rangle$ and $\{1,2\}$ allows $d^\dagger_{\bsl k_a,2}d^\dagger_{\bsl k_b,2}|0\rangle$. We demonstrate the convergence of $\{2,1\}$ and the non-convergence of $\{1,2\}$ by considering an iteration procedure where the single-particle bands at the start of iteration are the same as those in $H_b$ in Eq.\eqref{eq:hb} (taking $H_a$ instead of $H_b$ will lead to the same conclusion). The $C_3$ eigenvalues of single particle bands 0,1,2 are $\omega^*,\omega,1$ at $\bsl k_a$ and $\omega^*,1,\omega$ at $\bsl k_b$. The basis at truncation $\{2,1\}$ is:
\bea
\{d^\dagger_{\bsl k_a,0}d^\dagger_{\bsl k_b,0}|0\rangle,d^\dagger_{\bsl k_a,2}d^\dagger_{\bsl k_b,1}|0\rangle,d^\dagger_{\bsl k_a,1}d^\dagger_{\bsl k_b,2}|0\rangle,d^\dagger_{\bsl k_a,1}d^\dagger_{\bsl k_b,1}|0\rangle\},
\label{basis4state}
\eea
which has $C_3$ eigenvalues $\omega,1,\omega^*,\omega$. The Hamiltonian in this basis is:
\bea
\overline H=\left(\begin{array}{cccc}
h_{00} &0 & 0 & h_{03}  \\
0 & h_{11} & 0& 0  \\
0 & 0 & h_{22}& 0 \\
h_{30} & 0 & 0& h_{33}\\
\end{array}\right),
\label{Hbar4}
\eea
where $h_{00}$ is the smallest diagonal element, and the locations of the zeros are required by $C_3$ symmetry.

We first review the non-convergence behavior in this example when taking truncation $\{1,1\}$. In this case the Hamiltonian becomes the first $3\times 3$ block of $\overline H$, with a ground state of the form $(1,0,0)^T$. However, in the numerical computation of the iteration process, the zero components of the Hamiltonian and the ground state cannot be exactly zero at machine precision, therefore the ground state wave function generically becomes $(1,\epsilon_1,\epsilon_2)^T$, where $\epsilon_1$ and $\epsilon_2$ are very small numbers. This leads to density matrices:
\bea
\rho(\bsl k_a)=\left(\begin{array}{ccc}
1 & 0 & 0  \\
 0& |\epsilon_2|^2 & 0  \\
0 & 0 & |\epsilon_1|^2 
\end{array}\right),\ \ \rho(\bsl k_b)=\left(\begin{array}{ccc}
1 & 0 & 0  \\
 0& |\epsilon_1|^2 & 0  \\
0 & 0 & |\epsilon_2|^2 
\end{array}\right).
\eea
The iteration procedure performs a unitary transform to make all density matrices diagonal with eigenvalues in descending order. Therefore, depending on the relative size of $|\epsilon_1|$ and $|\epsilon_2|$, the unitary transform at one $\bsl k$ point will flip band 1 and band 2, which leads to the oscillatory behavior. Indeed, since this wave function satisfies the conditions in Eq.\eqref{assumecab}, the iteration with $\{1,1\}$ will oscillate no matter $|\epsilon_1|>|\epsilon_2|$ or $|\epsilon_1|<|\epsilon_2|$.

Then we move on to consider truncation $\{2,1\}$ with the Hamiltonian given by $\overline H$. In this case the ground state is $(\cos \theta,\epsilon_1,\epsilon_2,\sin\theta e^{i\phi})^T$ for some parameters $\theta$ and $\phi$, and the density matrices become:
\bea
\rho'(\bsl k_a)=\left(\begin{array}{ccc}
\cos^2\theta & 0 & 0  \\
 0& |\epsilon_2|^2+\sin^2\theta & 0  \\
0 & 0 & |\epsilon_1|^2 
\end{array}\right),\ \ \rho'(\bsl k_b)=\left(\begin{array}{ccc}
\cos^2\theta & 0 & 0  \\
 0& |\epsilon_1|^2+\sin^2\theta & 0  \\
0 & 0 & |\epsilon_2|^2 
\end{array}\right).
\eea
Therefore, when $\theta<\frac{\pi}{4}$, i.e., the ground state is still dominated by the state $d^\dagger_{\bsl k_a,0}d^\dagger_{\bsl k_b,0}|0\rangle$, the density matrices at both $\bsl k_a$ and $\bsl k_b$ are already diagonal with eigenvalues in descending order, hence no rotation of basis is needed, and any further iteration will stay in the same basis. This is the case for the parameters of $H_b$ in Eq.\eqref{eq:hb}, where we find $\theta=0.72<\frac{\pi}{4}$. Therefore, the iteration with truncation $\{2,1\}$ has converged.

Next we consider truncation $\{1,2\}$, where the fourth state in the basis in Eq.\eqref{basis4state} is changed from $d^\dagger_{\bsl k_a,1}d^\dagger_{\bsl k_b,1}|0\rangle$ to $d^\dagger_{\bsl k_a,2}d^\dagger_{\bsl k_b,2}|0\rangle$. The $C_3$ eigenvalues in the basis remain the same, hence the Hamiltonian still takes the same form as Eq.\eqref{Hbar4} but with different values of $h_{30}$, $h_{03}$ and $h_{33}$. The ground state also has the same form $(\cos\theta',\epsilon'_1,\epsilon'_2,\sin\theta'e^{i\phi'})^T$. However, because the fourth state is now located in band 2 rather than band 1, the density matrices receives a large contribution at band 2:
\bea
\rho''(\bsl k_a)=\left(\begin{array}{ccc}
\cos^2\theta' & 0 & 0  \\
 0& |\epsilon'_2|^2 & 0  \\
0 & 0 & |\epsilon'_1|^2 +\sin^2\theta'
\end{array}\right),\ \ \rho''(\bsl k_b)=\left(\begin{array}{ccc}
\cos^2\theta' & 0 & 0  \\
 0& |\epsilon'_1|^2 & 0  \\
0 & 0 & |\epsilon'_2|^2 +\sin^2\theta'
\end{array}\right).
\eea
This makes the $\rho''_{22}>\rho''_{11}$ at both momentum points. Therefore, the unitary transformation in the iteration will flip band 1 and 2, contrary to the case in truncation $\{2,1\}$ where convergence is reached. Further iteration steps show that the ground state will oscillate between different values without convergence for truncation $\{1,2\}$.

\end{document}

%% file: reference.bbl
%